\documentclass[useAMS,usenatbib]{mn2e}
\usepackage{times,graphicx,aas_macros,amssymb}

\newcommand{\sau}{{\tt SAURON}}  					
\newcommand{\spit}{{\it Spitzer}}						

\newcommand{\um}{$\mu$m}

\newcommand{\msun}{M$_{\odot}$}					
\newcommand{\mstar}{M$_*$}						
\newcommand{\peryr}{yr$^{-1}$}					
\newcommand{\lr}{$\lambda_R$}					

\newcommand{\ha}{H$\alpha$}					
\newcommand{\hb}{H$\beta$}						
\newcommand{\hone}{H{\small I}}					
\newcommand{\htwo}{H{\small II}}					
\newcommand{\mgb}{Mg $b$}						
\newcommand{\oiii}{[O{\small III}]}					

\newcommand{\mhi}{M$_{\rm H{\small I}}$} 			
\newcommand{\mhtwo}{M$_{\rm H{\small II}}$} 		

\newcommand{\nspit}{48}							

\newcommand{\pivt}{Paper~IV}
\newcommand{\pvt}{Paper~V}

\newcommand{\pixt}{Paper~IX}

\newcommand{\pxiit}{Paper~XII}

\newcommand{\pxvit}{Paper~XVI}
\newcommand{\pia}{Paper~I}
\newcommand{\piia}{Paper~II}
\newcommand{\piiia}{Paper~III}
\newcommand{\piva}{Paper~IV}
\newcommand{\pva}{Paper~V}
\newcommand{\pvia}{Paper~VI}

\newcommand{\pviiia}{Paper~VIII}
\newcommand{\pixa}{Paper~IX}
\newcommand{\pxa}{Paper~X}
\newcommand{\pxia}{Paper~XI}
\newcommand{\pxiia}{Paper~XII}
\newcommand{\pxiiia}{Paper~XIII}
\newcommand{\pxvia}{Paper~XVI}

\newcommand{\piiip}{(Paper~III)}

\newcommand{\pvp}{(Paper~V)}
\newcommand{\pvip}{(Paper~VI)}

\newcommand{\pixp}{(Paper~IX)}

\newcommand{\pxiip}{(Paper~XII)}

\newcommand{\pxvip}{(Paper~XVI)}

\newcommand{\classone}{widespread star-forming}
\newcommand{\classtwo}{circumnuclear star-forming}

\title[The SAURON Project - XV]{The SAURON project - XV. Modes of star formation in early-type galaxies and the evolution of the red sequence}

\author[K. L. Shapiro et al.]{Kristen L. Shapiro,$^{1}$\thanks{E-mail: shapiro@astro.berkeley.edu} 
Jes\'us Falc\'on-Barroso,$^{2,3}$ Glenn van de Ven,$^{4}$\thanks{Hubble Fellow} 
P. Tim de Zeeuw,$^{5,6}$ 
\newauthor
Marc Sarzi,$^{7}$ Roland Bacon,$^{8}$ Alberto Bolatto,$^{9}$ Michele Cappellari,$^{10}$ 
Darren Croton,$^{11}$ 
\newauthor
Roger L. Davies,$^{10}$ Eric Emsellem,$^{5,8}$ Onsi Fakhouri,$^{1}$ Davor Krajnovi\'c,$^{10}$ 
\newauthor
Harald Kuntschner,$^{5}$ Richard M. McDermid,$^{12}$ Reynier F. Peletier,$^{13}$ 
\newauthor
Remco C. E. van den Bosch,$^{14}$ Guido van der Wolk$^{13}$ \\
$^{1}$UC Berkeley Department of Astronomy, Berkeley, CA 94720, USA \\
$^{2}$Instituto de Astrof\'isica de Canarias, Canarias, Via Lactea s/n, 38700 La Laguna, Tenerife, Spain \\
$^{3}$European Space and Technology Centre (ESTEC), Keplerlaan 1, Postbus 299, 2200 AG Noordwijk, The Netherlands \\
$^{4}$Institute for Advanced Study, Einstein Drive, Princeton, NJ 08540, USA \\
$^{5}$European Southern Observatory, Karl-Schwarzschild-Str. 2, D-85748 Garching, Germany \\
$^{6}$Sterrewacht Leiden, Universiteit Leiden, Postbus 9513, 2300 RA Leiden, The Netherlands \\
$^{7}$Centre for Astrophysics Research, University of Hertfordshire, Hatfield, Herts AL1 09AB, UK \\
$^{8}$Universit\'e Lyon 1, CRAL, Observatoire de Lyon, 9 av. Charles Andr\'e F-69230 Saint Genis Laval; CNRS, UMR 5574; ENS de Lyon, France \\
$^{9}$Deparment of Astronomy, University of Maryland, College Park, MD 20742, USA\\
$^{10}$Sub-department of Astrophysics, University of Oxford, Denys Wilkinson Building, Keble Road, OX1 3RH, UK \\
$^{11}$Centre for Astrophysics and Supercomputing, Swinburne University of Technology, Hawthorn, VIC 3122, Australia \\
$^{12}$Gemini Observatory, Northern Opertations Center, 670 N. A'ohoku Place, Hilo, HI 96720 USA \\
$^{13}$Kapteyn Astronomical Institute, Postbus 800, 9700 AV Groningen, The Netherlands \\
$^{14}$Department of Astronomy, University of Texas, Austin, TX 78712, USA
}

\begin{document}

\date{Accepted 2009 November 20.  Received 2009 October 28; in original form 2009 July 9}

\pagerange{\pageref{firstpage}--\pageref{lastpage}} \pubyear{}

\maketitle

\label{firstpage}


\begin{abstract}
We combine \sau\ integral field data of a representative sample of local early-type, red sequence galaxies with \spit/IRAC imaging in order to investigate the presence of trace star formation in these systems.  With the \spit\ data, we identify galaxies hosting low-level star formation, as traced by PAH emission, with measured star formation rates that compare well to those estimated from other tracers.  This star formation proceeds according to established scaling relations with molecular gas content, in surface density regimes characteristic of disk galaxies and circumnuclear starbursts.  We find that star formation in early-type galaxies happens exclusively in fast-rotating systems and occurs in two distinct modes.  In the first, star formation is a diffuse process, corresponding to widespread young stellar populations and high molecular gas content.  The equal presence of co- and counter-rotating components in these systems strongly implies an external origin for the star-forming gas, and we argue that these star formation events may be the final stages of (mostly minor) mergers that build up the bulges of red sequence lenticulars.  In the second mode of star formation, the process is concentrated into well-defined disk or ring morphologies, outside of which the host galaxies exhibit uniformly evolved stellar populations.  This implies that these star formation events represent rejuvenations within previously quiescent stellar systems.  Evidence for earlier star formation events similar to these in all fast rotating early-type galaxies suggests that this mode of star formation may be common to all such galaxies, with a duty cycle of roughly 1/10, and likely contributes to the embedded, co-rotating inner stellar disks ubiquitous in this population.
\end{abstract}





\section{Introduction}

Early-type (elliptical and lenticular) galaxies, as the most massive and most evolved components of the local Universe, are important test sites for our understanding of the growth of structure.  In the past, these systems likely hosted much more powerful star formation and nuclear activity than is found in present day assembling systems (``downsizing"; e.g. \citealt{Cow+96,Kri+07}).  However, these galaxies have long since ceased such activity and are now part of the ``red sequence" in the optical color-magnitude space, so called because of the red colors produced by old stellar populations.

The lack of star formation in these galaxies has been attributed to various mechanisms, including gas exhaustion in the major mergers that supposedly give early-type systems their bulge-dominated morphologies, gas stripping as these typically more clustered galaxies encountered the cluster environment, and gas heating by AGN or by shock-heating as the accreting halos cross a mass threshold above which the gas shocks upon entry.  Detailed studies of the gas content and of low-level star formation in early-type galaxies can therefore be a critical constraint on the relative importance of each of these mechanisms in regulating the interstellar media (ISM) and therefore star formation in these systems.  Such data are a window into the mechanisms responsible for bringing early-type galaxies from the blue cloud onto the red sequence \citep[e.g.][]{Fab+07} and keeping them there.

Over the past several decades, increasingly sensitive probes of the ISM in early-type galaxies have shown that these systems contain detectable amounts of atomic hydrogen, cold dust, and ionized gas \citep[see review by][also e.g. \citealt{Sad+00,Bre+06,TemBriMat07,Kan+08b}]{Sch87}.  More recently, \citet{Mor+06} and \citet[][hereafter \pva]{SAURONV} have shown that the majority of early-type galaxies in non-cluster environments contain atomic and ionized gas, with environment playing a critical role in the presence of an ISM.  In some cases, the gas content in these systems is large, with \mhi~=~10$^6 - $10$^9$ \msun\ and \mhtwo~=~10$^2 - $10$^5$ \msun\ (\citealt{Mor+06}; \pva).

This abundance of gas naturally raises the question of whether these systems, some of which have up to the same amount of atomic hydrogen as the Milky Way, also sustain a molecular component and on-going star formation.  Several surveys of CO emission in early-type galaxies have found substantial amounts of molecular gas in some systems (\citealt{Lee+91,You02}; see also the compilation by \citealt{BetGalGar03}).  More recently, millimeter interferometry has demonstrated that this gas is sometimes arranged in regularly rotating disks, whose size and orientation is consistent with that of the ionized gas and of the embedded stellar disks observed in this class of galaxies \citep{YouBurCap08}.  However, the trace amounts of star formation likely associated with these molecular reservoirs have historically been difficult to quantify and to disentangle from other gas excitation mechanisms.

In 2003, the launch of the {\it Spitzer Space Telescope} made possible sensitive observations of polycyclic aromatic hydrocarbons (PAHs) and of small hot dust grains, both ionized primarily by young stellar populations, in early-type galaxies.  Nearly simultaneously, the launch of the {\it GALEX} satellite enabled similarly sensitive observations in the ultraviolet, in which the stellar photospheric signatures of these young stars can be probed directly.  Data from these satellites have revealed that star formation is occurring in a significant fraction of early-type galaxies \citep{Yi+05,YouBenLuc09,TemBriMat09} and that this late-time star formation contributes 1$-$10\% of the current stellar mass \citep{Sch+07,Kav+07}.

The detection capabilities and spatial resolution available with these satellites now allows detailed comparisons within galaxies of star formation processes and other properties.  With a suitable sample and extensive multi-wavelength data, it is then possible to study what governs star formation in early-type systems and to probe why some of these galaxies continue to regenerate their stellar populations while others truly are ``red and dead."

In this paper,  we therefore present \spit\ data for our representative sample of early-type galaxies, the \sau\ survey.  In the context of this survey, these galaxies have all been observed with optical integral-field spectroscopy, which provides stellar kinematics and absorption line indices, as well as ionized gas fluxes and kinematics (\S\ref{SAURON}).  Such properties can be compared in detail with direct probes of star formation, as available in the mid-IR with \spit/IRAC (\S\ref{Spitzer}).  We use these data to locate regions of on-going star formation in the \sau\ galaxies and to probe the corresponding rates and efficiencies (\S\ref{Compare}).  We then combine the \sau, \spit, and auxiliary data sets to study the photometric and kinematic properties associated with star formation in early type galaxies (\S\ref{SF}) and to probe the origins, evolution, and fate of these systems (\S\ref{discussion}).  Finally, in \S\ref{Conclu}, we summarize our conclusions about the continuing evolution of these red sequence galaxies.


\section{The SAURON Survey}
\label{SAURON}

The \sau\ survey is a study of the two-dimensional kinematic, stellar population, and gas properties of representative samples of early-type galaxies and bulges, using integral-field observations obtained with the \sau\ instrument (\citealt{SAURONI}, hereafter \pia), mounted on the William Herschel Telescope in La Palma.  The sample was selected to be evenly divided between morphological classes (E, S0, Sa) and environment (field, Virgo cluster), and to representatively cover the ellipticity versus absolute magnitude parameter space (\citealt{SAURONII}, hereafter \piia).  In the context of this program, 48 elliptical and lenticular galaxies and 24 spiral bulges were observed out to one effective radius.  Here, we focus on the 48 early-type systems in the main \sau\ survey, for which the stellar kinematics (\citealt{SAURONIII}, hereafter \piiia), absorption line strengths (\citealt{SAURONVI}, hereafter \pvia), and emission line fluxes and kinematics \pvp\ have been studied in detail.  In these data, the pixel size is 0.8\arcsec\ and the median seeing FWHM is 1.7\arcsec\ \piiip.

\subsection{Survey Results: Stellar Kinematics}
\label{SAUkin}

Analysis and comparison of the stellar kinematics maps of the \sau\ sample has revealed that early-type galaxies are structurally quite complex.  In particular, \citet[][hereafter \pixa]{SAURONIX} showed that these systems naturally divide into two new categories: fast and slow rotators, as parametrized by their specific angular momentum within one effective radius, \lr.  Detailed analysis of the kinematics in these systems and dynamical modeling have revealed that the fast rotators, those systems with high \lr, are uniformly characterized by significant embedded disk components and oblate axisymmetry, while the slow rotators, those systems with low \lr, are fairly round and triaxial (\citealt{SAURONX}, hereafter \pxa; \citealt{SAURONXII}, hereafter \pxiia; van den Bosch et al. {\it in prep}).

Within these systems, the stellar velocity maps also reveal a variety of kinematic substructures, most notably the kinematically decoupled components (KDCs, \pxiia).  Moreover, \citet[][hereafter \pviiia]{SAURONVIII} have shown that the characteristics of the KDCs, where present, are linked to the rotational class of the host system.  KDCs in fast rotators are all compact ($<$500 pc) and are generally composed of young ($<$5 Gyr) stars, while those in the slow rotators are universally large ($>$1 kpc) and much older ($>$8 Gyr; \pviiia).  It is thus apparent that the assembly and star formation processes involved in the creation of an early-type system are intimately linked with the resulting kinematic substructure.

\subsection{Survey Results: Ionized Gas and Auxiliary Data}
\label{SAUSF}

Additionally, the \sau\ survey has located and spatially resolved likely regions of continuing evolution within early-type galaxies, where star formation appears to be on-going.  From the \sau\ maps, \pvt\ identified a subset of the sample in which the low \oiii/\hb\ emission-line ratios and settled gas systems can only be interpreted as sites of star formation activity.  In these systems, the ionized gas is arranged in a regular, disk-like configuration, with low velocity dispersion \pvp.  Stellar population estimates (\pvia; Kuntschner et al. {\it in prep}) and {\it GALEX} UV imaging \citep[][hereafter \pxiiia]{SAURONXIII} have further revealed that these systems contain the young stellar populations that must exist in the presence of on-going star formation.  Additional evidence for star formation in these galaxies comes from CO observations, which reveal the presence of molecular gas and show that, in some cases, this gas is organized into disks that are co-spatial and co-rotating with the ionized gas and the young stars \citep{ComYouBur07,YouBurCap08}.  However, it is important to note that despite this evidence for trace on-going star formation in this subset of the sample, wide field optical imaging of the \sau\ galaxies continues to confirm that all the galaxies in this sample reside solidly on the red sequence in optical colors (Falc\'on-Barroso et al. {\it in prep}).

These star-forming galaxies in the \sau\ sample are, however, the minority of the population, with the exact fraction depending on the tracer used (5/48: from \oiii/\hb\ emission line ratios, \pva; 9/34: from UV emission, \pxiiia).  The majority of the galaxies in the \sau\ survey do contain measurable ionized gas, but these systems display emission line ratios inconsistent with on-going star formation (\pva; Sarzi et al. {\it in press}, hereafter \pxvia).  Many of these systems also contain extensive reservoirs of neutral hydrogen (\mhi\ = 10$^6 - $10$^9$ \msun), which in some cases is organized into large, regularly rotating structures (\citealt{Mor+06}; \citealt{Wei+08}; Oosterloo et al. {\it in prep}).  The presence of this neutral gas and its potential processing into stars is therefore complicated, and critical, to understand.


\section{Spitzer Data}
\label{Spitzer}

The obvious questions raised by these \sau\ results are then: what are the origins of the large amounts of gas in some early-type galaxies, and what governs its conversion into stars?  To address these questions, we must probe the star-forming structures in the \sau\ galaxies in detail.

This requires a star formation tracer that is more direct than the \oiii/\hb\ ratio, which can be affected by AGN activity and metallicity gradients, and the presence of CO, which by itself is not evidence of on-going star formation.  While both the ultraviolet and the mid-infrared, in which reprocessed UV light is emitted, are suitable for this task, the infrared telescope \spit\ yields spatial resolution $\sim$3 times better than that of the UV space telescope {\it GALEX} and comparable to that of \sau.  \spit\ observations, therefore, provide the unique opportunity to probe the star formation properties of the \sau\ galaxies on the spatial scales of interesting stellar and ionized gas features, some of which lie below the {\it GALEX} spatial resolution.

To do this, we measure the mid-IR emission bands from polycyclic aromatic hydrocarbons (PAHs) in the \sau\ galaxies.  These large molecules, with sizes of few \AA\ and containing up to a few hundred carbon atoms, are each stochastically heated by far UV photons, which excite high-energy vibrational modes that, as they decay, emit primarily in broad emission bands centered at 3.3\um, 6.2\um, 7.7\um, 8.6\um, 11.2\um, and 12.7\um.  This emission is typically seen on the boundaries of H{\small II} regions, suggesting that late O and early B type stars provide the requisite UV photons and that harder radiation fields (e.g. within H{\small II} regions or from AGN) either destroy or chemically transform the molecules such that they no longer emit in these bands \citep[e.g.][]{Ces+96}.  We therefore use these mid-IR emission bands, probed by \spit/IRAC broad-band images and supplemented where possible with \spit/IRS spectroscopy, to study the intensity and morphology of star formation in the \sau\ galaxies.

\subsection{Observations and Archival Data}
\label{Archive}

For this study, we combine observations with data made publicly available through the Spitzer Science Center (SSC) archive, as originally acquired in the context of a number of different programs.  These archival data cover a significant fraction of the \sau\ sample.  The remaining systems were observed in the context of Program 50630 (PI: G. van der Wolk) during Cycle 5, such that \spit\ data was obtained for the full \sau\ sample of early-type galaxies.

When possible, the IRAC broad-band images are supplemented with archival IRS data, either with the optimal coverage of PAH emission bands achieved in observations with the SL module ($\lambda$ = 5.2$-$14.5\um) or with coverage of the longer-wavelength PAH bands with the SH module ($\lambda$ = 9.9$-$19.6\um).  Table \ref{Obs} describes the available data for each system.

\begin{table*}
\caption{\spit\ Data for the \sau\ Sample Galaxies}
\label{Obs}
\begin{center}
\begin{minipage}{14.5cm}
\begin{center}
\begin{tabular*}{14.5cm}{cccccl}
\hline
Galaxy & IRAC Exp. Time & IRS Spectral Range & $f_{3.6}$ & $f_{8.0}$ & \ \ \ \ Observers \\
 & (sec) & (\um) & (Jy) & (Jy) & (PIs, Program IDs) \\
\hline
\hline
NGC 474		& 150	& 9.9 - 19.6			& 0.072	& 0.018		& A. Zezas (20140) \\
NGC 524		& 150	& --					& 0.283	& 0.098		& G. van der Wolk (50630) \\
NGC 821		& 44		& 5.2 - 14.5			& 0.148	& 0.041		& G. Fabbiano (20371); J. Bregman (03535) \\
NGC 1023 	& 150	& 5.2 - 7.7				& 0.635	& 0.167		& G. Fazio (00069) \\
NGC 2549	& 150	& --					& 0.106	& 0.028		& G. van der Wolk (50630) \\
NGC 2685	& 48		& --					& 0.095$^{\dagger}$	& 0.050$^{\dagger}$	& G. Rieke (40936) \\
NGC 2695    	& 150	& --					& 0.067	& 0.017		& G. van der Wolk (50630) \\
NGC 2699    	& 150	& --					& 0.040	& 0.011		& G. van der Wolk (50630) \\
NGC 2768	& 500	& --					& 0.362	& 0.099		& G. Fazio (30318) \\
NGC 2974	& 500	& 5.2 - 14.5			& 0.175$^{\dagger}$	& 0.062$^{\dagger}$	& G. Fazio (30318); H. Kaneda (03619) \\
NGC 3032	& 480	& --					& 0.038	& 0.110		& S. Kannappan (30406) \\
NGC 3156 	& 210	& --					& 0.028	& 0.016		& J. Surace (03403) \\
NGC 3377	& 150	& 5.2 - 14.5			& 0.219	& 0.058		& G. Fazio (00069); J. Bregman (03535) \\
NGC 3379	& 60		& 5.2 - 14.5			& 0.604	& 0.150		& G. Fazio (00069); J. Bregman (03535) \\
NGC 3384	& 500	& --					& 0.298	& 0.088		& G. Fazio (30318) \\
NGC 3414	& 150	& --					& 0.142	& 0.044		& G. van der Wolk (50630) \\
NGC 3489	& 60		& 5.2 - 7.7; 9.9 - 19.6		& 0.227$^{\dagger}$	& 0.104$^{\dagger}$	& G. Fazio (00069); C. Leitherer (03674) \\
NGC 3608	& 500	& 5.2 - 14.5			& 0.145	& 0.035		& G. Fazio (30318); J. Bregman (03535) \\
NGC 4150	& 150	& 5.2 - 7.7; 9.9 - 19.6		& 0.061$^{\dagger}$	& 0.054$^{\dagger}$	& G. Fazio (00069); C. Leitherer (03674) \\
NGC 4262    	& 150	& --					& 0.086	& 0.022		& G. van der Wolk (50630) \\
NGC 4270	& 150	& 9.9 - 19.6			& 0.040	& 0.011		& A. Zezas (20140) \\
NGC 4278	& 60		& 5.2 - 14.5			& 0.282	& 0.087		& G. Fazio (00069); E. Sturm (03237) \\
NGC 4374	& 60		& 5.2 - 14.5			& 0.770	& 0.204		& G. Fazio (00069); G. Rieke (00082) \\
NGC 4382	& 150	& 5.2 - 14.5			& 0.880	& 0.243		& A. Zezas (20140); A. Bressan (03419) \\
NGC 4387    	& 150	& --					& 0.038	& 0.010		& G. van der Wolk (50630) \\
NGC 4458	& 500	& --					& 0.039	& 0.010		& G. Fazio (30318) \\
NGC 4459	& 60		& 9.9 - 19.6			& 0.294	& 0.156		& P. Cote (03649); C. Leitherer (03674) \\
NGC 4473	& 60		& 5.2 - 14.5			& 0.255	& 0.066		& P. Cote (03649); A. Bressan (03419) \\
NGC 4477	& 48		& 5.2 - 14.5			& 0.257	& 0.071		& G. Rieke (40936) \\
NGC 4486	& 150	& 5.2 - 14.5			& 1.386	& 0.402		& W. Forman (03228); G. Rieke (00082) \\
NGC 4526	& 60		& 5.2 - 7.7				& 0.492	& 0.286		& G. Fazio (00069) \\
NGC 4546    	& 150	& --					& 0.227	& 0.063		& G. van der Wolk (50630) \\
NGC 4550    	& 150	& 5.2 - 14.5			& 0.056	& 0.019		& G. van der Wolk (50630); A. Bressan (03419) \\
NGC 4552	& 240	& 5.2 - 14.5			& 0.401	& 0.104		& R. Kennicutt (00159) \\
NGC 4564	& 44		& 5.2 - 14.5			& 0.125	& 0.032		& G. Fabbiano (20371); A. Bressan (03419) \\
NGC 4570	& 60		& 5.2 - 14.5			& 0.129	& 0.033		& P. Cote (03649); A. Bressan (03419) \\
NGC 4621	& 60*	& 5.2 - 14.5			& 0.419	& 0.110		& P. Cote (03649); A. Bressan (03419) \\
NGC 4660	& 60		& 5.2 - 14.5			& 0.107	& 0.028		& P. Cote (03649); A. Bressan (03419) \\
NGC 5198	& 150	& --					& 0.056	& 0.014		& G. van der Wolk (50630) \\
NGC 5308	& 150	& --					& 0.065	& 0.016		& G. van der Wolk (50630) \\
NGC 5813	& 60		& 5.2 - 14.5			& 0.248	& 0.060		& G. Fazio (00069); J. Bregman (03535) \\
NGC 5831	& 210	& 5.2 - 14.5			& 0.097	& 0.026		& J. Surace (03403); J. Bregman (03535) \\
NGC 5838    	& 150	& --					& 0.168	& 0.061		& G. van der Wolk (50630) \\
NGC 5845	& 44		& --					& 0.042	& 0.013		& G. Fabbiano (20371) \\
NGC 5846	& 150	& 5.2 - 14.5			& 0.429	& 0.087		& A. Zezas (20140); J. Bregman (03535) \\
NGC 5982	& 210	& 					& 0.121	& 0.031		& J. Surace (03403) \\
NGC 7332    	& 150	& 5.2 - 14.5			& 0.086	& 0.024		& G. van der Wolk (50630); R. Rampazzo (30256) \\
NGC 7457	& 500	& --					& 0.098	& 0.027		& G. Fazio (30318) \\
\hline
\end{tabular*}
\end{center}
IRAC data noted with (*) are saturated in the 3.6\um\ band.  Aperture photometry is given over the galaxy effective radius; note that for galaxies indicated with ($^{\dagger}$) non-stellar 8.0\um\ emission extends beyond $R_e$.  In all cases, errors on the photometry are completely dominated by the assumed 3\%\ systematic error in the IRAC photometric calibration (see text); consequently, we do not explicitly list this systematic error here.
\end{minipage}
\end{center}
\end{table*}

\subsection{Data Processing}
\label{IRAC}

To reduce the IRAC data, we began with the Basic Calibration Data (BCD), as generated in the \spit\ pipeline.  Artifacts were removed from each exposure using the publicly available artifact mitigation code on the SSC website (courtesy of S.~Carey).  The exposures were processed and combined into final mosaics using MOPEX (v16.3.7) with pixel sizes of 1.22\arcsec.  These images were then sky subtracted in the standard way.

Aperture photometry was performed on all systems using three apertures: one of radius 14\arcsec, for comparison with CO data obtained with a beam of this equivalent radius (in \S \ref{SFinstab}), another corresponding to the effective radius of the galaxy ($R_e$), and a third encompassing all of the 8.0\um\ non-stellar emission (see \S \ref{spitzersf}) for use in measuring the total star formation rate (SFR) in systems where the observed radius of star formation exceeds $R_e$.  During this process, foreground and background objects and associated saturated pixels were masked out.  (The two systems most affected by this are NGC~2974, with a bright foreground star to the southwest, and NGC~4150, with a $z$=0.52 QSO identified by \citealt{LirLawJoh00} to the southeast.)  The SSC-recommended extended aperture corrections were applied in all cases for the relevant aperture sizes and wavebands \citep{Rea+05,Coh+07}.  Errors on the photometry were extracted from the same apertures on the uncertainty images, whose processing in the IRAC pipeline and post-processing was identical to that of the images.  In cases where an image was saturated (see Table~\ref{Obs}), the photometry was recorded as a lower limit.  These errors and upper limits were combined in quadrature with the photometric error of IRAC of 3\%\ \citep{Rea+05}.

For visual analysis of the mid-IR morphology of the \sau\ galaxies, we also use the IRAC images directly.  To re-scale the measured intensities in the images for extended sources, we multiply each image by the extended aperture correction appropriate for an infinite aperture \citep[][see also \citealt{Gor+08}]{Rea+05}.  In order to compare the 3.6\um\ images and the 8.0\um\ images for each galaxy (see \S\ref{spitzersf}), we then convolve the 3.6\um\ images with a kernel designed by \citet{Gor+08} to transform their PSF to that of the 8.0\um\ images.  In some cases, the convolution kernels altered the centering of the image by several tenths of a pixel; for these galaxies, the images were re-aligned using foreground stars for reference.

\begin{figure}
	\centering
	\includegraphics[angle=90,height=9.1cm,trim= 0cm 0cm 0cm 1cm,clip]{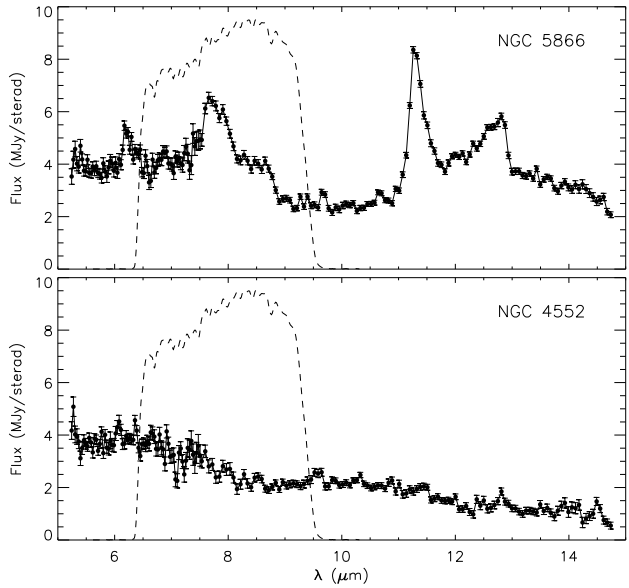}
	\caption{IRS spectra for an actively star-forming early-type galaxy ({\it top}) and a galaxy with no on-going star formation ({\it bottom}).  The star-forming galaxy emits significantly in the main PAH bands at 6.2, 7.7, 8.6, 11.3, and 12.7\um, which are excited by the FUV photons associated primarily with massive young stars.   In marked contrast, the quiescent galaxy's spectrum is featureless.  The spectra shown here are for NGC~5866, a galaxy observed with \sau\ but not in the context of the main sample, and NGC~4552, part of the \sau\ sample.  The spectra are taken from the SINGS survey \citep{Ken+03} data release.  For reference, the IRAC 8.0\um\ filter response provided by the SSC is overplotted in dashed lines, with arbitrary normalization.}
	\label{spectra}
\end{figure}

\subsection{Tracing Star Formation}
\label{spitzersf}

The presence of PAH emission in the resulting images and aperture photometry is best probed by the 8.0\um\ waveband, the spectral response function of which covers the strong 7.7\um\ and 8.6\um\ features (Figure~\ref{spectra}).  However, other emission processes can contribute to signal at 8.0\um, the most important of these being a non-negligible contribution from stars.

\begin{figure*}
	\centering
	\includegraphics[angle=90,width=16cm,trim=0cm 0cm 0cm 1cm]{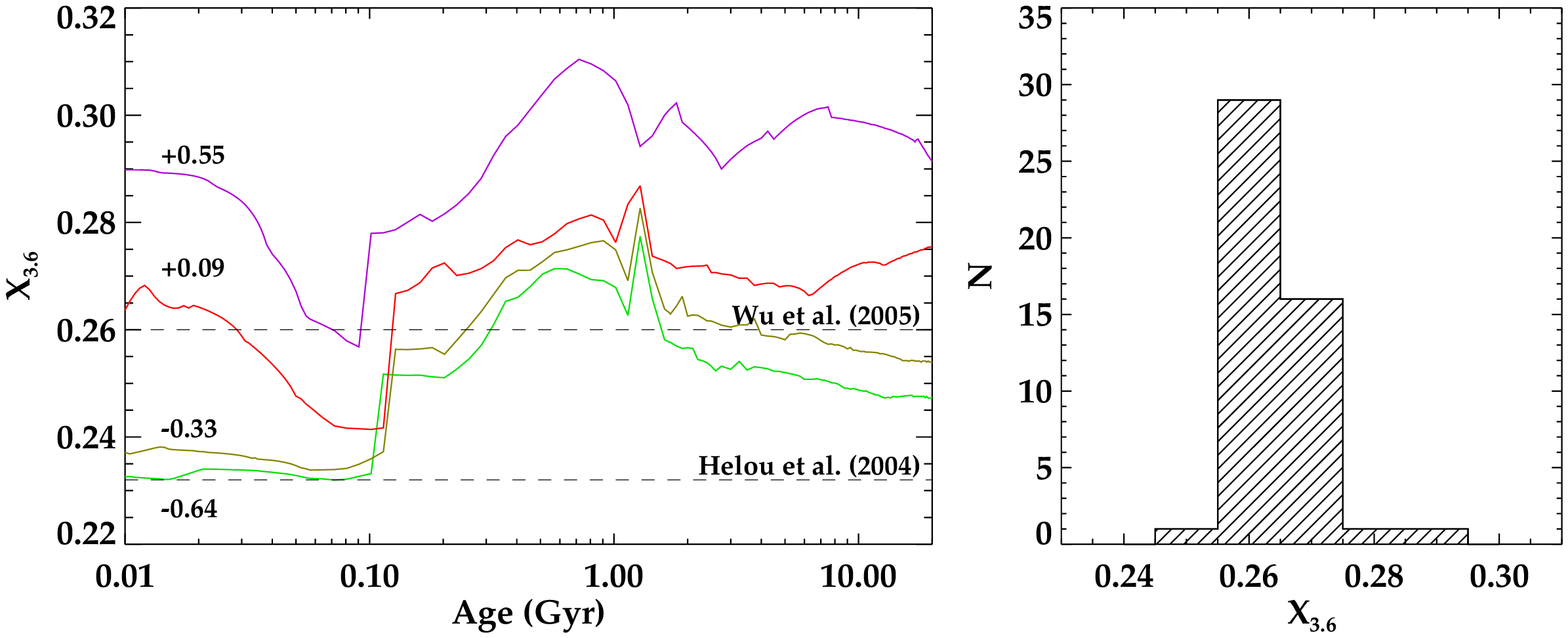}
	\caption{{\it Left}: Dependence of $X_{3.6}$ on a galaxy's age and metallicity, as predicted by \citet{BruCha03} SSP models.  Four different metallicities, ranging from sub-solar ({\it green}) to super-solar ({\it magenta}), are plotted and labeled.  The Starburst99 value found by \citet{Hel+04} and the empirical estimate by \citet{Wu+05} are overplotted with dashed lines.  {\it Right:} Histogram of best-fit values of $X_{3.6}$ for the SAURON sample galaxies.  See text for a description of the method.}
	\label{brucha}
\end{figure*}

The stellar contribution to the 8.0\um\ band can be estimated using the flux in the 3.6\um\ band, both of which are dominated by the Rayleigh-Jeans tail of the stellar light, via the simple formula $f_{8.0, stellar}$ = $X_{3.6} \times f_{3.6}$.   However, predictions of $X_{3.6}$ from \citet{BruCha03} single stellar population (SSP) models (see Figure~\ref{brucha}) and from Starburst99 continuous star formation models generated by \citet{Hel+04} reveal a wide range in this parameter depending on the age and metallicity of the stellar population \citep[see also][]{Cal+07}.  In late-type galaxies, for which a continuous star formation model is appropriate, \citet{Hel+04} find $X_{3.6}$ = 0.232; however, this value has limited applicability to our early-type galaxies, which are not well described by a continuous star formation model.  \citet{Wu+05} subsequently used images of quiescent galaxies to empirically estimate $X_{3.6}$ = 0.26, although there is little reason to apply this number to systems with stellar populations that deviate from being uniformly very evolved.

Given the large range in ages and metallicities found in the \sau\ galaxies (Kuntschner et al. {\it in prep}), it is preferable to directly measure $X_{3.6}$ in each of our sample galaxies, based on \citet{BruCha03} SSP models.  We begin by computing SSP models that finely sample age and metallicity parameter space.  For each galaxy, we then compare the observed stellar absorption line strengths (\hb, \mgb, Fe5015) to those predicted by the SSP models and locate the best-fit model in the maximum likelihood sense.  This model yields the predicted stellar spectral energy distribution in the mid-infrared, from which we measure $X_{3.6}$.  (Tests with more recent SSP libraries that include an updated treatment of AGB stars found similar results to those quoted here.)  The resulting range of $X_{3.6}$ for our sample galaxies is shown in the right panel of Figure~\ref{brucha}, with a sample median of 0.264.  We caution here that we find $X_{3.6}$ to differ significantly from 0.264 in some galaxies in our sample and especially in later-type systems (E/S0: Figure~\ref{brucha}; Sa: Falc\'on-Barroso, private communication), so the conversions used here should not be blindly applied to other galaxies or H{\small II} regions.  Using the derived values of $X_{3.6}$ for each galaxy, we estimate the non-stellar contribution to the 8.0\um\ flux via $f_{8.0, non-stellar}$~=~$f_{8.0} - X_{3.6}~f_{3.6}$.

In galaxies where a significant amount of non-stellar 8.0\um\ emission is detected, we inspect the pipeline-processed pBCD (post Basic Calibration Data) IRS spectra, where available, to differentiate between emission dominated by PAH features (Figure~\ref{spectra}) and emission dominated by strong non-thermal continuum emission powered by an AGN \citep[see e.g.][]{Bre+07}.

In cases where the 8.0\um\ non-stellar emission is due to PAH bands, the star formation rate needed to drive this PAH emission can be computed directly from $f_{8.0, non-stellar}$, using the relation from \citet{Wu+05},

\begin{equation}
{\rm SFR}\ ({\rm M}_{\odot}\ {\rm yr}^{-1}) = \frac{\nu L_{\nu} (8.0, non-stellar)}{1.39 \times 10^9\ L_\odot} .
\label{eq:sfr}
\end{equation}

\noindent
These authors have calibrated this relation from the \citet{YunRedCon01} relation between SFR and radio continuum (based on the SFR-FIR and radio-FIR relations), with correlation coefficient $\rho$~=~0.88 between the 8.0\um\ and the radio continuum emission.  \citet{Wu+05} have also calibrated the 8.0\um\ SFR estimator against the \citet{Ken98} SFR-\ha\ conversion, which lowers the estimated SFR by $\sim$13\%.  The SFRs derived from equation~\ref{eq:sfr} for the \sau\ galaxies are typically several tenths of a solar mass per year (Table~\ref{tbl:sfr}), which is within the regime in which \citet{Wu+05} report a linear relation between SFR and 8.0\um\ flux.  We note that non-linear relations between 8.0\um\ emission and SFR have been reported; however, this effect appears to be confined to galaxies with metallicities less than one third solar \citep{Eng+05, Cal+07} and likely does not apply to the metal-rich early-type galaxies discussed here.

The errors on our 8.0\um\ luminosities are derived in \S\ref{IRAC} and include the 3\%\ IRAC photometric calibration.  We then propagate the uncertainties in the relations between 8.0\um\ and radio luminosities \citep{Wu+05} and SFR and radio luminosity \citep{YunRedCon01} in order to estimate the total (statistical and systematic) error on our SFRs.  The final error estimates derived in this way are listed in Table~\ref{tbl:sfr}.

We measure SFRs for the 13 galaxies in our sample in which the 8.0\um\ non-stellar emission is significant (with respect to the statistical error bars) and is likely associated with star formation (as identified in \S \ref{SFproof} and shown in Figure~\ref{profiles}{\it a}).  For systems in which there is negligible 8.0\um\ non-stellar emission, we quote the r.m.s. errors on the non-detections as the upper limits in Table~\ref{tbl:sfr}.  For systems in which there is 8.0\um\ non-stellar emission that is likely not associated with star formation (as identified in \S\ref{SFother} and shown in Figure~\ref{profiles}{\it b}), we quote the derived SFRs from this emission as the upper limits.  These latter upper limits are often contaminated by AGN emission or other processes (see \S \ref{SFother}) and are consequently less stringent limits on the SFRs in these systems.

\begin{table*}
\caption{Observed SFRs and Related Properties for the \sau\ Sample Galaxies}
\label{tbl:sfr}
\begin{center}
\begin{minipage}{14cm}
\begin{center}
\begin{tabular*}{14cm}{cccccccccc}
\hline
Galaxy & Rotator & Type & R$_e$ & Distance & $X_{3.6}$ & SFR & {\it d}SFR & $R_{\rm SF}$ & ${\rm M}_{\rm H_2}$\\
 & (Fast/Slow) &  & (\arcsec) & (Mpc) & & (\msun\ \peryr) & (\msun\ \peryr) & (\arcsec) & (10$^8$ \msun)\\
\hline
\hline
 NGC 474	    	& F	& S0		& 28.0	& 26.06	& 0.264	& 0		& 0.0025	& 0		& $<$0.26 \\
 NGC 524	    	& F	& S0		& 35.4	& 23.99	& 0.274	& 0.0971 	& 0.0415 	& 13		&    1.11 \\
 NGC 821	    	& F	& E6		& 31.3	& 24.10	& 0.261	& 0 		& 0.0132 	& 0		& $<$0.39 \\
NGC 1023 	& F	& SB0	& 48.7	& 11.43	& 0.268	& 0		& 0.0012 	& 0		& $<$0.05 \\
NGC 2549	& F	& S0		& 13.9	& 12.65	& 0.277	& 0 		& 0.0008 	& 0		& $<$0.10 \\
NGC 2685	& F	& SB0	& 23.6	& 11.86	& 0.258	& 0.0950 	& 0.0406	& 50		&    0.12 \\
NGC 2695    	& F	& SAB0	& 18.7	& 32.36	& 0.261	& 0 		& 0.0031 	& 0		& $<$0.87\\
NGC 2699    	& F	& E		& 12.1	& 26.92	& 0.264	& 0 		& 0.0039 	& 0		& $<$0.31\\
NGC 2768	& F	& E6		& 68.0	& 22.39	& 0.258	& 0	 	& 0.0265 	& 0		&    0.74\\
NGC 2974	& F	& E4		& 28.3	& 21.48	& 0.271	& 0.1426 	& 0.0610 	& 70		& $<$0.39 \\
NGC 3032	& F	& SAB0	& 19.3	& 21.98	& 0.293	& 0.4017 	& 0.1717 	& 18		&    5.27 \\
NGC 3156 	& F	& S0		& 14.8	& 22.39	& 0.264	& 0.0349 	& 0.0149 	& 10		&    0.48 \\
NGC 3377	& F	& E5-6	& 38.3	& 11.22	& 0.258	& 0 		& 0.0012 	& 0		& $<$0.08 \\
NGC 3379	& F	& E1		& 44.9	& 10.57	& 0.264	& 0 		& 0.0036 	& 0		& $<$0.05\\
NGC 3384	& F	& SB0	& 28.5	& 11.59	& 0.271	& 0		& 0.0091 	& 0		& $<$0.12\\
NGC 3414	& S	& S0		& 32.0	& 25.23	& 0.258	& 0 		& 0.0385 	& 0		& $<$0.14\\
NGC 3489	& F	& SAB0	& 21.5	& 12.08	& 0.271	& 0.0619 	& 0.0265 	& 38		&    0.15 \\
NGC 3608	& S	& E2		& 33.6	& 22.91	& 0.264	& 0 		& 0.0060 	& 0		& $<$0.33 \\
NGC 4150	& F	& S0		& 15.9	& 13.74	& 0.271	& 0.0606 	& 0.0259 	& 16		&    0.59\\
NGC 4262    	& F	& SB0	& 10.6	& 15.42	& 0.258	& 0	 	& 0.0004 	& 0		& $<$0.10 \\
NGC 4270	& F	& S0		& 13.7	& 37.33	& 0.261	& 0	 	& 0.0053 	& 0		& $<$0.58 \\
NGC 4278	& F	& E1-2	& 30.6	& 16.07	& 0.261	& 0 		& 0.0294 	& 0		& $<$0.32 \\
NGC 4374	& S	& E1		& 70.2	& 18.45	& 0.261	& 0 		& 0.0124 	& 0		& $<$0.14 \\
NGC 4382	& F	& S0		& 94.4	& 17.86	& 0.264	& 0 		& 0.0298 	& 0		& $<$0.20 \\
NGC 4387    	& F	& E		& 11.0	& 17.95	& 0.261	& 0 		& 0.0004 	& 0		& $<$0.21 \\
NGC 4458	& S	& E0-1	& 19.9	& 16.37	& 0.252	& 0 		& 0.0010 	& 0		& $<$0.17 \\
NGC 4459	& F	& S0		& 41.0	& 16.07	& 0.264	& 0.1689 	& 0.0722 	& 12		&    1.68 \\
NGC 4473	& F	& E5		& 26.8	& 15.28	& 0.264	& 0 		& 0.0016 	& 0		& $<$0.10 \\
NGC 4477	& F	& SB0	& 46.5	& 16.67	& 0.261	& 0.0088 	& 0.0038 	& 7		&    0.34 \\
NGC 4486	& S	& E0-1	& 106.2	& 17.22	& 0.268	& 0 		& 0.0761 	& 0		& $<$0.12 \\
NGC 4526	& F	& SAB0	& 35.7	& 16.90	& 0.268	& 0.3700 	& 0.1581 	& 11		&    6.05 \\
NGC 4546    	& F	& SB0	& 22.0	& 14.06	& 0.261	& 0 		& 0.0055 	& 0		& $<$0.08 \\
NGC 4550    	& S	& SB0	& 11.6	& 15.49	& 0.258	& 0.0084 	& 0.0036 	& 6		&    0.14 \\
NGC 4552	& S	& E0-1	& 33.9	& 15.85	& 0.271	& 0 		& 0.0041 	& 0		& $<$0.16 \\
NGC 4564	& F	& E		& 19.3	& 15.85	& 0.268	& 0 		& 0.0014 	& 0		& $<$0.20 \\
NGC 4570	& F	& S0		& 12.8	& 17.06	& 0.261	& 0 		& 0.0008 	& 0		& $<$0.24 \\
NGC 4621	& F	& E5		& 46.0	& 14.93	& 0.261	& 0 		& 0.0004 	& 0		& $<$0.15 \\
NGC 4660	& F	& E		& 11.5	& 15.00	& 0.261	& 0 		& 0.0002 	& 0		& $<$0.17 \\
NGC 5198	& S	& E1-2	& 18.0	& 38.37	& 0.264	& 0 		& 0.0028 	& 0		& $<$0.61 \\
NGC 5308	& F	& S0		& 9.9		& 29.43	& 0.268	& 0 		& 0.0032 	& 0		& $<$0.55 \\
NGC 5813	& S	& E1-2	& 55.9	& 32.21	& 0.264	& 0 		& 0.0202 	& 0		& $<$0.43 \\
NGC 5831	& S	& E3		& 29.2	& 27.16	& 0.264	& 0 		& 0.0005 	& 0		& $<$0.62 \\
NGC 5838    	& F	& S0		& 20.6	& 24.66	& 0.271	& 0.0819 	& 0.0350 	& 6		& $<$0.38 \\
NGC 5845	& F	& E		& 4.3		& 25.94	& 0.274	& 0.0062 	& 0.0027 	& $<$3	& $<$0.28 \\
NGC 5846	& S	& E0-1	& 76.8	& 24.89	& 0.268	& 0 		& 0.0624 	& 0		& $<$0.52 \\
NGC 5982	& S	& E3		& 24.9	& 51.76	& 0.271	& 0 		& 0.0152 	& 0		& $<$0.57 \\
NGC 7332    	& F	& S0		& 9.2		& 23.01	& 0.268	& 0 		& 0.0022 	& 0		& $<$2.61\\
NGC 7457	& F	& S0		& 33.2	& 13.24	& 0.264	& 0 		& 0.0016 	& 0		& $<$0.11 \\
\hline
\end{tabular*}
\end{center}
Rotator class is as derived in \pixt.  Hubble type is taken from NED.  Distances and effective radii are taken from Falc\'on-Barroso et al. ({\it in prep}).  SFRs are measured here, with upper limits indicated by SFR = 0 and the upper limit given in the error ({\it d}SFR) column.  The spatial extent of the star formation activity $R_{\rm SF}$ and the ratio of the 8.0\um\ to 3.6\um\ stellar emission $X_{3.6}$ are derived in the text.  Molecular gas masses are from \citet{SchSco02}, \citet{ComYouBur07}, \citet{YouBurCap08}, \citet{Cro+08}, and Crocker et al. ({\it in prep}) and are adjusted to the distances assumed here.
\end{minipage}
\end{center}
\end{table*}

As a final caveat to the SFRs measured here, it should be noted that there is some debate as to what radiation fields can drive PAH emission.  In particular, \citet{Pee+04} have suggested that PAHs may be better tracers of B stars than of the harder radiation fields found around O stars.  Likewise, \citet{Cal+05} found evidence for a diffuse component of PAH emission unrelated to \htwo\ regions, suggesting that these molecules may be excited by UV photons in the general galactic radiation fields, possibly originating from B stars.  \citet{Pee+02} have also shown that post-AGB stars and planetary nebulae can produce radiation fields appropriate for PAH absorption and reprocessing.  To differentiate between those systems in which the 8.0\um\ non-stellar emission is associated with young stellar populations and those in which it is not, in \S \ref{SFproof}, we compare the \spit\ data with \sau\ and auxiliary data to find compelling evidence for and against star formation in individual galaxies.


\section{Star Formation in the SAURON Galaxies}
\label{Compare}

\begin{figure*}
	\centering
	\includegraphics[width=17cm]{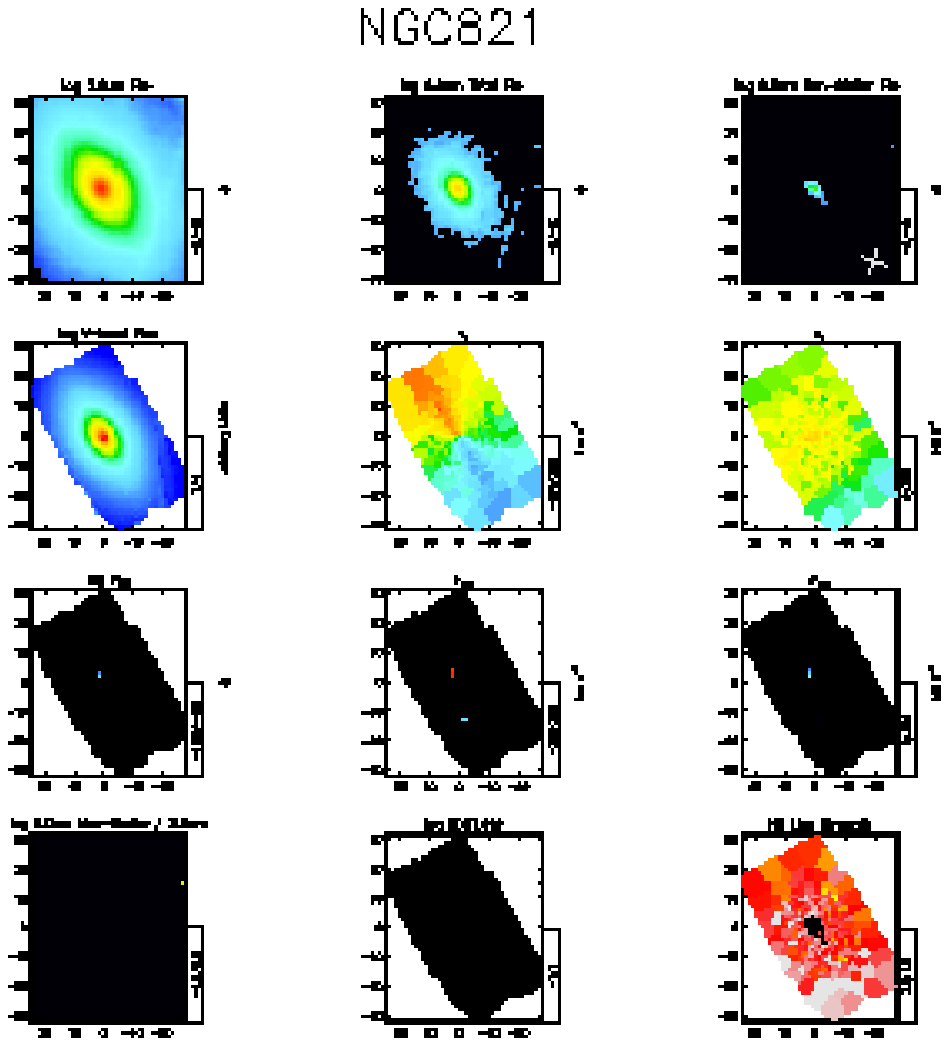}	
	\caption{Comparison of \sau\ maps and available \spit\ data for a representative galaxy with unremarkable infrared properties.  {\it From left to right. Top row:} IRAC 3.6\um\ image, IRAC 8.0\um\ image, and IRAC 8.0\um\ image with stellar contribution removed (see text for details).  The cross indicates the original axes of the IRAC CCD; high flux in a pixel (e.g. the galaxy center) can produce spurious structures along these axes.  {\it Second row:} Broad-band image from \sau\ integrated flux covering roughly the $V$-band, \sau\ stellar velocity map, and \sau\ stellar velocity dispersion map.  {\it Third row:} \sau\ \hb\ emission line flux, \sau\ ionized gas velocity map, and \sau\ ionized gas velocity dispersion map (both of the latter derived from the typically stronger \oiii\ emission).  {\it Bottom row:} IRAC 8.0\um\ equivalent width (8.0\um\ non-stellar / 3.6\um) map, \sau\ \oiii/\hb\ map, and \sau\ \hb\ line strength map (both of the latter with contours of the 8.0\um\ non-stellar map overplotted).  In all maps, North is up and East to the left, and axes are labelled in arcseconds.}
	\label{Main-noSF}
\end{figure*}

Figures \ref{Main-noSF} and \ref{Main-SF} present a detailed comparison of the \sau\ stellar and ionized gas properties with the IRAC data.  These data reveal a wide range in the mid-IR properties in the sample galaxies, from negligible 8.0\um\ non-stellar emission to highly-structured 8.0\um\ non-stellar emission.  Since many of the systems in the \sau\ sample are unremarkable in their infrared properties, we present a single representative galaxy from this group in Figure \ref{Main-noSF}.  In Figure~\ref{Main-SF}, we show all systems with marginal to significant infrared emission (also shown in Figure~\ref{profiles}).  Representative spectra from these two groups are shown in Figure~\ref{spectra}.

\subsection{Evidence for Star Formation in \boldmath 8.0\um\ Emission}
\label{SFproof}

Of the \nspit\ \sau\ galaxies, roughly half (Figure~\ref{Main-SF}, and reproduced in Figure~\ref{profiles}) have 8.0\um\ non-stellar maps with significant power.  Since this 8.0\um\ non-stellar emission can be powered by a variety of mechanisms, we combine the \spit\ data with \sau\ and auxiliary data to differentiate between these emission processes.  In our sample, we see evidence for several 8.0\um\ emission mechanisms; here we discuss the galaxies in each class in turn.

\subsubsection{Clear Signatures of Star Formation}
\label{yesSF}

\begin{figure*}
	\centering
	\includegraphics[width=15cm]{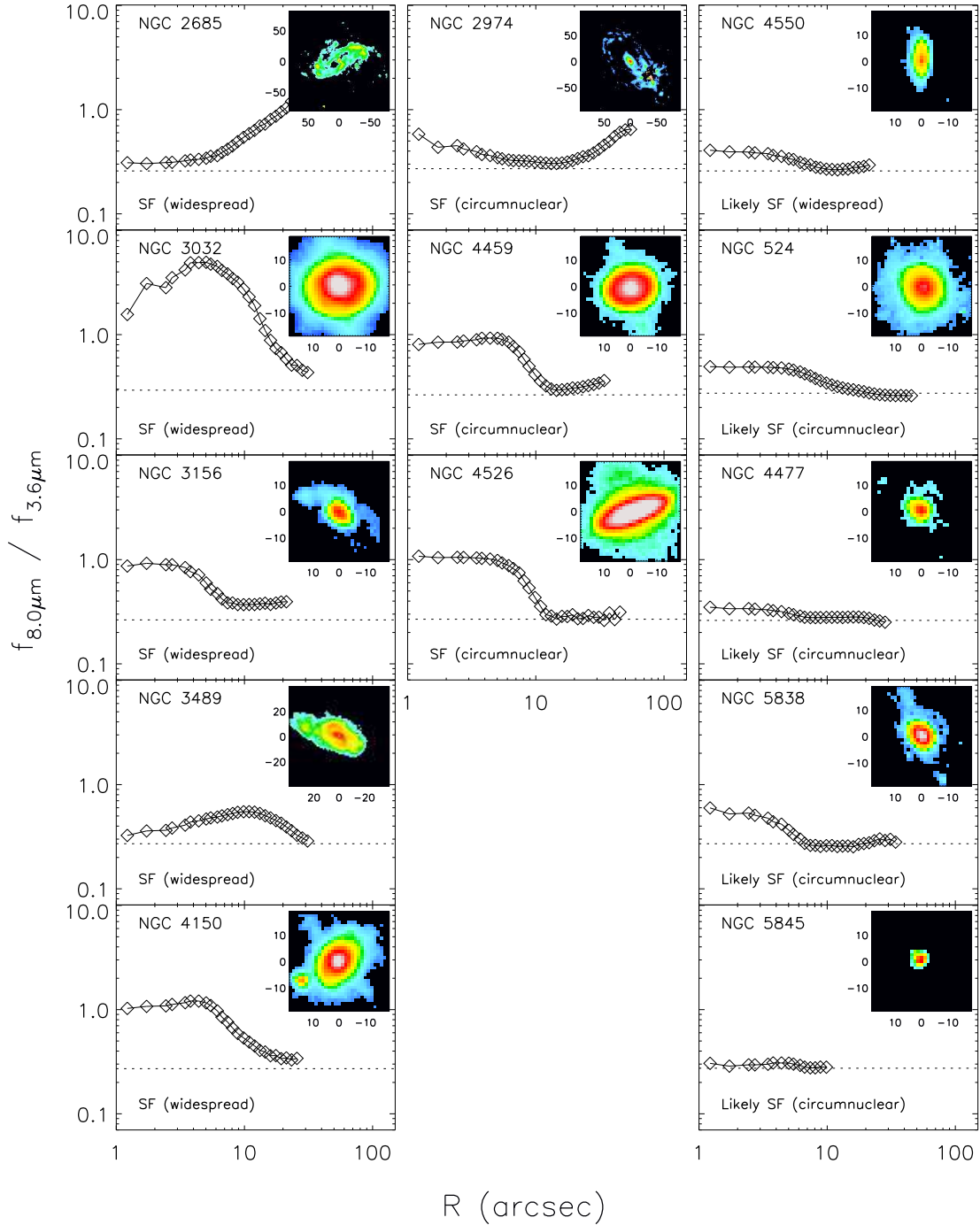}
	\caption{({\it a}) Radial profiles of ratio of 8.0\um\ and 3.6\um\ fluxes and images of the 8.0\um\ non-stellar emission ({\it inset}) for a subset of the sample galaxies: clearly star-forming galaxies ({\it first and second columns from left}) and possibly star-forming galaxies ({\it third column}).  The presence of star formation in each galaxy is indicated in its panel, along with the physical mechanism likely responsible for the 8.0\um\ non-stellar emission (see text in \S\ref{Compare} and \S\ref{SF}).  Horizontal lines indicate the flux ratio expected of a purely stellar profile, $X_{3.6}$, for each galaxy (see text).  Profiles were measured using the method of \citet{Mun-Mat+09} in which the growth curve of the 3.6\um\ and 8.0\um\ images were constructed, the proper extended aperture correction was applied at each radius, and the flux at a given radius was computed by subtracting the flux of the adjacent inner aperture.  Note that extended linear structures in these images (e.g. NGC~5838) are artifacts caused by high fluxes in galaxy centers (see Figure~\ref{Main-SF} for more details).}
	\label{profiles}
\end{figure*}

In 8 of the \nspit\ \sau\ galaxies, there is strong evidence that the 8.0\um\ non-stellar flux reflects PAH transitions excited by young stars.  These systems are shown in the left and middle columns of Figure~\ref{profiles}{\it a}; all eight display strong 8.0\um\ structures, and their spectra, where available, contain prominent PAH emission features.  Furthermore, Figure~\ref{profiles}{\it a} highlights the marked difference in spatial distribution between the 8.0\um\ and 3.6\um\ emission in these galaxies.  This provides key evidence that the 8.0\um-emitting dust is not produced and excited by the stellar mass loss of an older FUV-bright population (e.g. the AGB stars responsible for the so-called ``UV upturn," Bureau et al. {\it in prep}), since the spatial distribution of such populations is expected to closely follow that of the 3.6\um-emitting old stellar populations.  The radial profiles in Figure~\ref{profiles}{\it a} are also useful in quantifying the extent of the star formation $R_{\rm SF}$ (Table~\ref{tbl:sfr}), which we define here as the radius at which $f_{8.0\mu m} / f_{3.6\mu m}$ decreases to $<$10\%\ its peak height above $X_{3.6}$.

\begin{figure*}
	\centering
	\includegraphics[width=16cm]{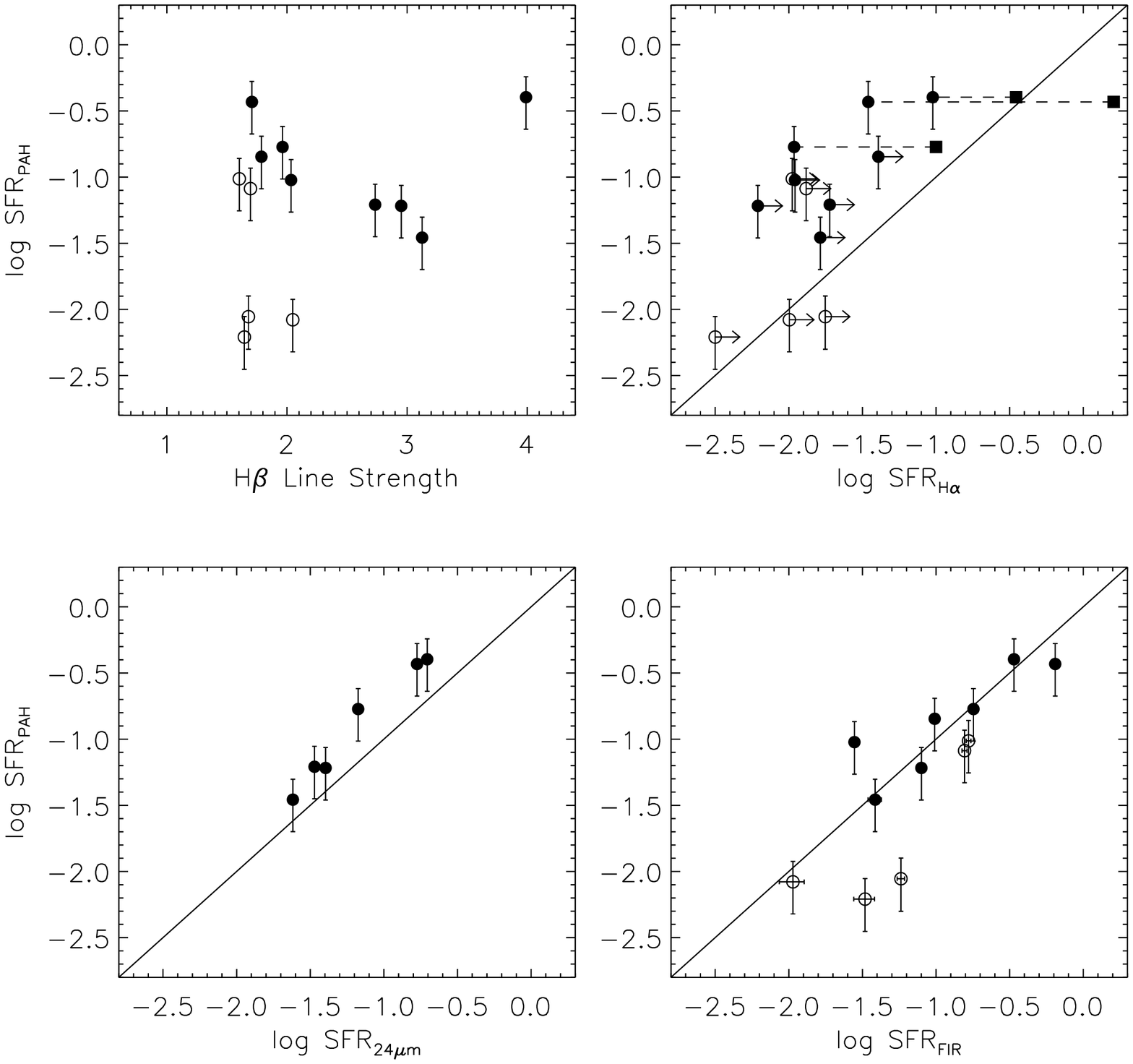}
	\caption{SFRs derived from PAH emission for the star-forming galaxies in our sample compared with other age and star formation indicators: \hb\ absorption line strength, as measured from the \sau\ data over an effective radius ({\it top left}); SFRs derived from \ha\ emission, as estimated from the \sau\ \hb\ emission and assuming no extinction ({\it top right}); SFRs derived from 24\um\ emission by \citet{TemBriMat09} for 6/13 of our star-forming galaxies ({\it bottom left}); and SFRs derived from FIR emission, estimated from the 60\um\ and 100\um\ {\it IRAS} fluxes collected from the NASA Extragalactic Database by \citet{ComYouBur07} for 12/13 of our star-forming galaxies ({\it bottom right}).  Galaxies with clear signs of star formation are indicated with filled circles, while those with suggestive signs of star formation are indicated with open circles (see text for details).  For three of our galaxies, direct \ha\ observations have been obtained \citep{GalHunTut84,You+96}; those measurements, adjusted to the distances and SFR-\ha\ conversion assumed here \citep{Ken98}, are plotted in the top right panel with squares.  In all panels, SFR is given in \msun\ \peryr.}
	\label{sfrcompare}
\end{figure*}

Additional evidence for young stellar populations in these galaxies comes from the \sau\ integral field data, in which the PAH-emitting regions are found to be coincident with higher \hb\ line strengths (Figures~\ref{sfrcompare}~and~\ref{Main-SF}; see also \citealt{TemBriMat09}), indicative of the presence of younger stars (\pvia; see \S\ref{SF} for details).  These findings are also supported by auxiliary {\it GALEX} imaging, in which all five of the star-forming systems identified here that are included in that sample (NGC~2974, NGC~3032, NGC~4150, NGC~4459, NGC~4526) display blue $UV-V$ colors consistent with star formation at the same radii at which they exhibit non-stellar 8.0\um\ emission (\citealt{Jeo+07}; \pxiiia). 

Observations of the interstellar media of these eight systems are also consistent with the presence of low-level star formation.  In \pvt, ionized gas was detected in all of these systems, and some also display the low \oiii/\hb\ ratios that are associated with on-going star formation.  Furthermore, the \hb\ emission can be used to estimate the star formation rate via the \citet{Ken98} SFR-\ha\ relation \pvp.  These estimates are compared to the PAH-derived SFRs in Figure~\ref{sfrcompare} and are lower limits, since the extinction that likely accompanies star formation in these regions cannot be quantified with only the narrow \sau\ spectral range (\pva; \citealt{TemBriMat09}).  Indeed, the few direct literature measurements of \ha-derived SFRs for these systems \citep{GalHunTut84,You+96} are systematically higher than the \hb\ estimates and more consistent with the PAH-derived SFRs (Figure~\ref{sfrcompare}).  \citet{TemBriMat09} have also measured 24\um -derived SFRs for these eight galaxies.  In the six galaxies with centrally-located and concentrated star-forming regions (excluding NGC~2685 and NGC~2974), the 24\um -derived SFRs are broadly consistent with, though slightly lower than, the PAH-derived SFRs (Figure~\ref{sfrcompare}).

\begin{figure*}
	\centering
	\includegraphics[width=15cm]{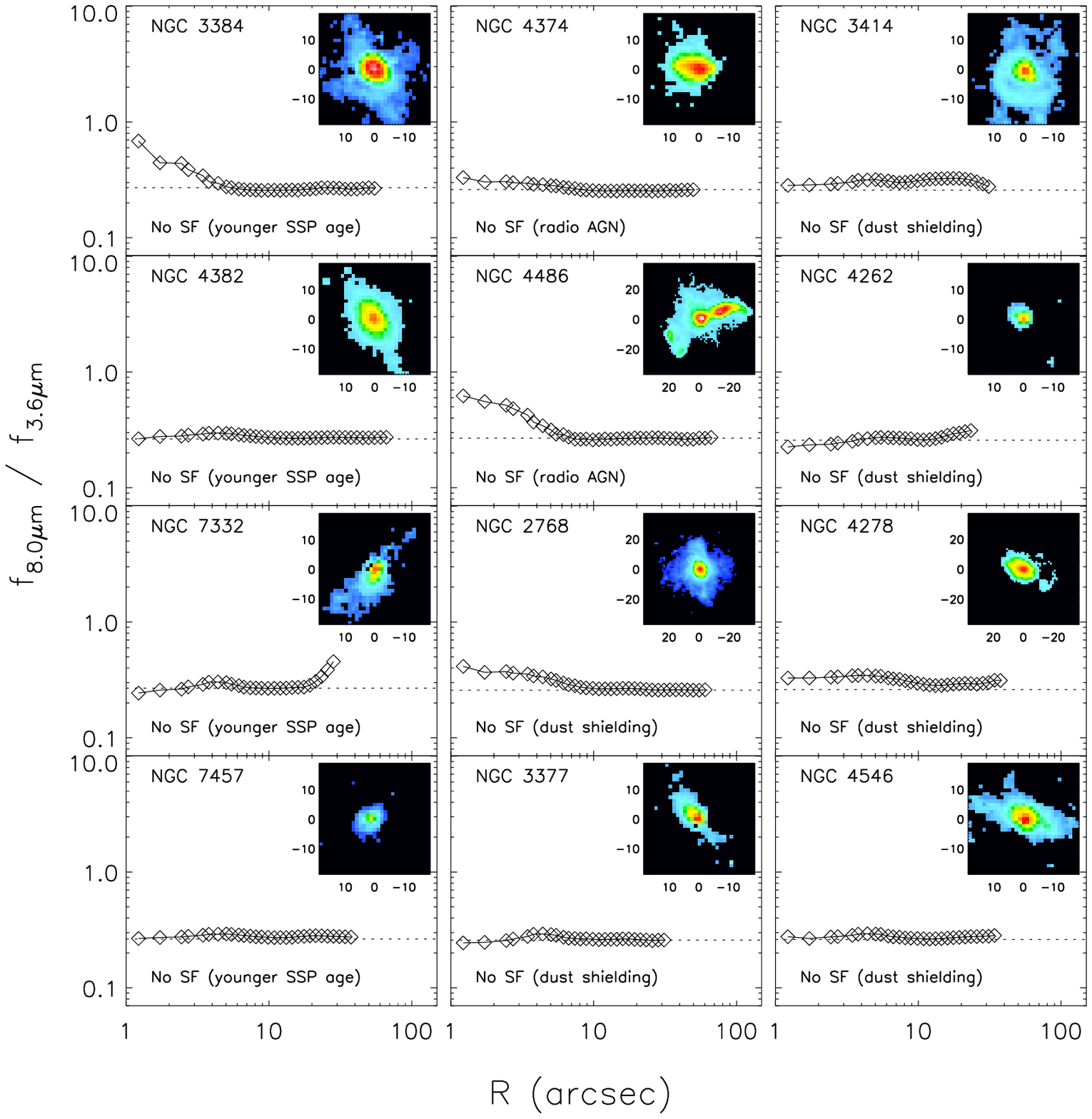}
	\begin{center}
{\bf Figure~\ref{profiles}.} ({\it b}) Same as ({\it a}), for non-star-forming galaxies with 8.0\um\ non-stellar emission.
	\end{center}
\end{figure*}

Furthermore, the molecular content of the \sau\ sample has been studied with single-dish CO observations \citep{ComYouBur07}, and six of the eight of the systems identified in the \spit\ data as hosting star formation are also detected in CO emission (Table~\ref{tbl:sfr}).  The two remaining galaxies are NGC~2685 and NGC~2974, in which the CO beam of \citet{ComYouBur07} only marginally overlaps the regions of star formation at high radii; the former has been detected with observations at larger radii by \citet{SchSco02}.  Follow-up CO interferometry has revealed molecular disk-like structures in several of the star-forming galaxies \citep{YouBurCap08}, whose spatial extents and morphologies are excellent matches to the structures seen in PAH emission.  \citet{ComYouBur07} also compared their observed molecular gas masses to {\it IRAS} FIR fluxes and estimated SFRs.  As a further test of the star formation origin of our observed PAH emission, we compare these FIR-derived SFRs to our PAH-derived SFRs in Figure~\ref{sfrcompare} and find excellent agreement.  This result is reassuring, if unsurprising, given that the PAH-derived SFRs are calibrated against the SFR-FIR relation (\S \ref{spitzersf}).

Finally, the atomic content of the interstellar media in the \sau\ galaxies has also been observed for a subset of the full sample \citep{Mor+06,Wei+08}.  In the three star-forming galaxies in this subset (NGC~2685, NGC~2974, NGC~4150), \hone\ is detected in the same regions as the PAH emission (and beyond).  However, \citet{Mor+06} also find significant \hone\ reservoirs in early-type systems without star formation, so the link between this component of the ISM and star formation activity is, as with spiral and dwarf galaxies, less obvious.

\subsubsection{Suggestive Signatures of Star Formation}
\label{maybeSF}

In addition to the 8 galaxies with significant multi-wavelength evidence for on-going star formation, there are 5 (of \nspit) additional galaxies whose 8.0\um\ non-stellar emission is suggestive of star formation but whose \sau\ and auxiliary data render these cases inconclusive.  These systems (NGC~524, NGC~4477, NGC~4550, NGC~5838, NGC~5845) are shown in the right column of Figure~\ref{profiles}{\it a}; as with the eight star-forming galaxies, these systems show prominent 8.0\um\ structures. Likewise, in four of these five galaxies, the radial profiles reveal clear differences in the spatial distributions of the old stars (as traced by 3.6\um) and the 8.0\um\ emission, which must therefore have a contribution not traceable to evolved stars.  In these four systems, we can therefore measure $R_{\rm SF}$; in the remaining galaxy, we obtain only an upper limit.

The regular morphologies of the 8.0\um\ non-stellar emission in these galaxies suggest an analogy to some of the star-forming galaxies in Figure~\ref{profiles}{\it a}; however, the SFRs measured in these five systems are several times lower.  This difference may explain why the signatures of star formation in these systems are less apparent in the \sau\ data.  In four of these five systems, the \hb\ line strengths are very low (Figure~\ref{sfrcompare}) and so do not obviously point to a young stellar population \pvip.  Likewise, the UV broad-band emission reveals bluer $UV-V$ colors in the 8.0\um-emitting regions in only the galaxy with slightly higher \hb\ absorption (NGC~4550; \pxiiia).

The interstellar media of these systems similarly contribute to this complex picture.  Although all five galaxies contain ionized gas, much of it is barely detected, and the \oiii/\hb\ emission-line ratios, where measurable, are neither low enough to be consistent with excitation only from star formation nor high enough to be consistent with other excitation mechanisms \pvp.  Nevertheless, the kinematics of the ionized gas are regular and are consistent with rotating disks, a picture that is also supported by prominent, dust disks visible in unsharp-masked $HST$/WFPC2 $V$-band images \pvp.  In three of these systems (NGC~524, NGC~4477, NGC~4550), sensitive CO interferometry has identified molecular gas disks coincident with the dust (\citealt{Cro+09}; Crocker et al. {\it in prep}), while the remaining two (NGC~5838, NGC~5845) were undetected with single-dish observations \citep{ComYouBur07}.  Despite this absence, the FIR emission measured by \citep{ComYouBur07} in all five systems produces SFR estimates broadly consistent with those from the PAHs.  \citet{TemBriMat09} have observed two of these five galaxies (NGC~4477, NGC~5845) in 24\um\ emission and do not detect star formation; however, the PAH-derived SFRs of these two galaxies are the lowest in our sample by nearly an order of magnitude (Table~\ref{tbl:sfr}; NGC~4550 has a similar SFR but was not observed by \citealt{TemBriMat09}), suggesting that they may have escaped detection at 24\um.

It is consequently difficult to assess whether the 8.0\um\ non-stellar emission in these systems represents on-going star formation.  The analogy to some of the galaxies with evident star formation is highly suggestive, but the multi-wavelength data of these systems are somewhat contradictory and do not converge on this picture.  On the one hand, the 8.0\um\ non-stellar emission in these systems may not be tracing star formation.  On the other hand, however, it may be that the very small amounts of star formation measured in 8.0\um\ are not visible in other tracers, due to its low level and to the much more massive underlying old stellar populations.

\subsubsection{Other Sources of 8.0\um\ Emission}
\label{SFother}

In contrast to the two classes of galaxies discussed above, in which all or some of the evidence points to on-going or recently ceased star formation, there are 12 (of \nspit) additional systems in which 8.0\um\ emission is observed but cannot be easily linked to star formation activity.  These galaxies are shown in Figure~\ref{profiles}{\it b}.

In one of these systems, NGC~4486 (M87), the 8.0\um\ features are coincident with the well-known radio jet, seen prominently and ubiquitously in ultraviolet, optical, and infrared imaging, and is thus almost certainly synchrotron emission \citep[see also][]{Per+07}.  Another, NGC~4374 (M84), is also a well-known radio galaxy, although in this case the 8.0\um\ emission is aligned with the dust filaments seen in the optical that are perpendicular to the radio jet.  In at least these two cases, the 8.0\um\ non-stellar emission can be attributed to nuclear activity rather than to star formation.

In six systems (NGC~2768, NGC~3377, NGC~3414, NGC~4262, NGC~4278, NGC~4546), the 8.0\um\ non-stellar emission is patchy.  Five of these galaxies have no detected molecular gas (\citealt{ComYouBur07}; Crocker et al. {\it in prep}), and the one system with detected CO emission (NGC~2768, \citealt{Cro+08}) reveals little spatial overlap between this gas and the 8.0\um\ emission.  Likewise, the stellar populations of these galaxies show no evidence of recent star formation and range from fairly to very evolved (8$-$12 Gyr; Kuntschner et al. {\it in prep}).  As a group, therefore, these six 8.0\um\ emitters are unlikely to host star formation activity.

The 8.0\um\ emission in these galaxies does correlate loosely with optically observed asymmetric dust lanes seen in all of these galaxies in \pvt\ and therefore may be highlighting the densest regions of these structures.  The IRS spectra for these systems, where available, reveal strong neutral PAH features at 11.3\um\ and only marginally detected ionized PAH 7.7\um\ emission, the latter of which is the primary PAH feature associated with star formation.  The dominance of neutral PAHs has been observed in a number of early-type galaxies and attributed to softer interstellar radiation fields, as would be expected of older stellar populations in the absence of AGN \citep[e.g.][but see also \citealt{Bre+08}]{Kan+08a}.

\pxvit\ discusses the ionizing mechanisms in the \sau\ galaxies in detail and, for these quiescent galaxies with 8.0\um\ non-stellar emission, compares the spatial distribution of this emission to the ionization of the galaxies, as parametrized by \oiii/\hb.  In fact, the spatial coincidence between these two features is remarkable, in that regions of relatively lower \oiii/\hb\ are the regions where 8.0\um\ non-stellar emission is detected (see Figure~\ref{Main-SF}).  Within the hard radiation fields (seen as high \oiii/\hb) in these galaxies, there are embedded regions of softer radiation fields, in which lower \oiii/\hb\ and 8.0\um\ non-stellar emission is measured.  The lower ionization in these regions may be due to the denser ISM found in these amorphous dust lanes, which absorbs some of the ionizing photons, partially shielding the gas from the hard ionization field.  In this manner, PAH molecules could survive in these dusty regions and produce the observed emission.  This mechanism may likewise shield the very small grains that produce 24\um\ emission, as detected in these galaxies by \citet{TemBriMat09}.  For details, we refer the reader to \pxvit\ and do not further discuss this class of non-SF galaxies here.

In addition to the eight galaxies discussed above, four more have weak 8.0\um\ non-stellar emission detections.  The IRAC images show the emission in these systems to be relatively weak, suggesting that perhaps this emission is associated with trace excitation of PAHs by B stars in these galaxies (NGC~3384, NGC~4382, NGC~7332, NGC~7457), which all have younger stellar populations.

\subsection{Efficiency of Star Formation}
\label{SFinstab}

\begin{figure}
	\centering
	\includegraphics[width=10cm]{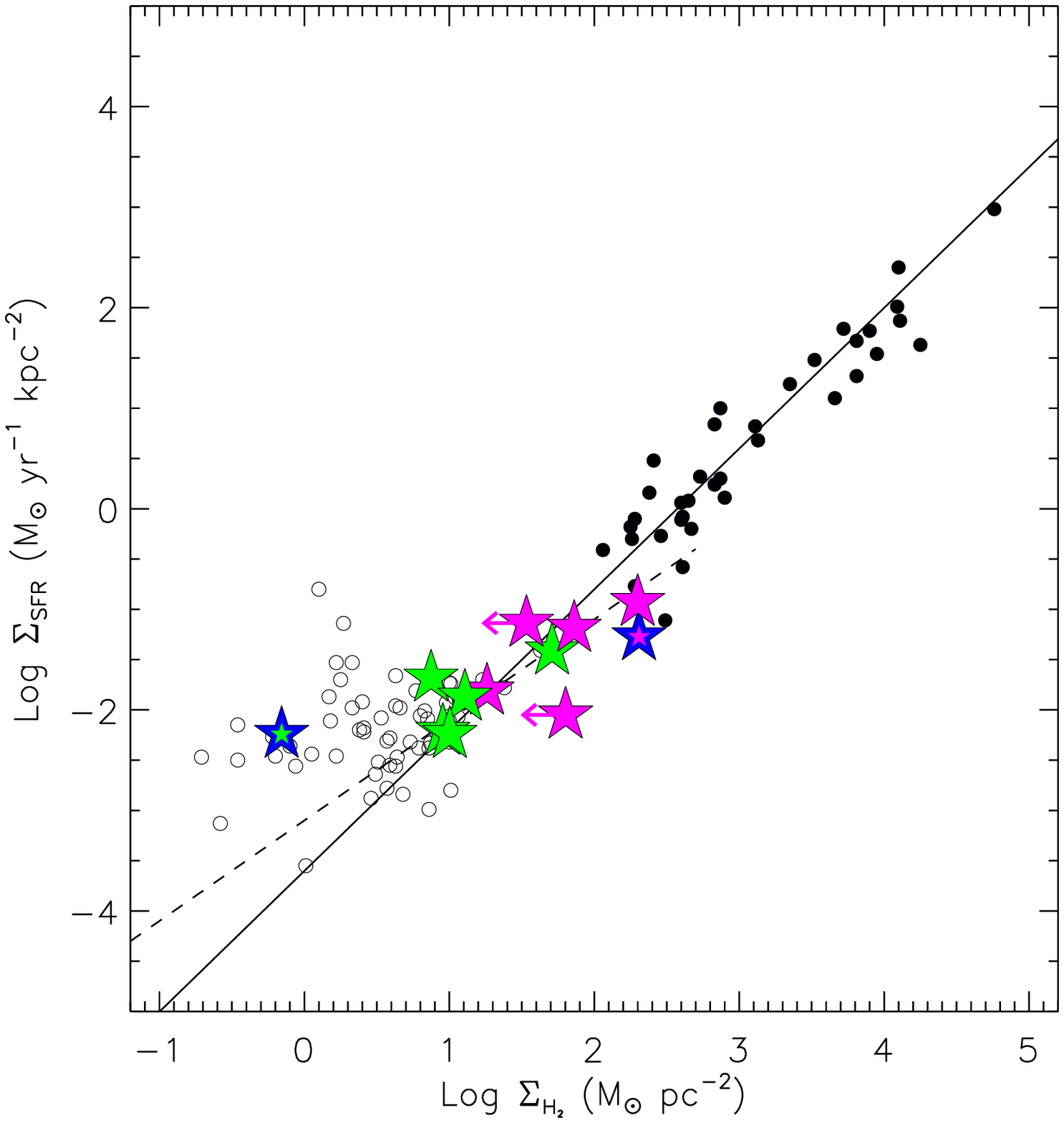}
	\caption{Schmidt-Kennicutt relation between gas and star formation densities ({\it solid line}), with data from \citet{Ken98} overplotted (open circles for spiral galaxies, filled circles for circumnuclear starbursts).  The recent analysis of \citet{Big+08} and \citet{Ler+08} with higher resolution data reveal a linear relation between these quantities ({\it dashed line}) in the spiral galaxy regime, although they note that a steepening of the relation probably occurs at higher molecular gas densities.  The star formation rates for early-type galaxies measured in this work are combined with existing CO measurements and overplotted with stars, colored according to the mode of star formation as described in \S\ref{SFdyn} (error bars on SFRs are smaller than the symbols).  Only 12 of the 13 star-forming galaxies are shown here, since the outer regions of NGC~2974, in the vicinity of the star formation activity, have not been observed in CO emission.}
	\label{ks}
\end{figure}

For the 13 galaxies identified in \S\ref{yesSF} and \S\ref{maybeSF} that are probably or potentially hosting star formation, we compare our measured SFRs to the molecular gas mass measurements of \citet{SchSco02}, \citet{ComYouBur07}, \citet{YouBurCap08}, \citet{Cro+08} and Crocker et al. ({\it in prep}) in order to probe the efficiency of star formation in the \sau\ early-type galaxies.  Our SFRs are measured over the star-forming regions, and we assume that all of the (mostly single-dish) CO emission originates in the same regions, as has been explicitly demonstrated for a subset of these galaxies with follow-up interferometry by \citet{YouBurCap08}.  Defining our apertures in this way, we can directly compare with the results of \citet{Ken98}, in which the aperture size is determined by the extent of the \ha\ disk (for normal spiral galaxies) or of the central molecular disk (for starbursts).  Our comparison to these data is shown in Figure~\ref{ks}; for reference, the results of more recent work by \citet{Big+08} are also shown.

An earlier version of this plot has been presented by \citet{ComYouBur07}, who compared their single-dish CO observations with the {\it IRAS} FIR luminosities discussed in \S\ref{SFproof}.  However, their single-dish data and the large {\it IRAS} beam did not allow them to constrain the size of the star-forming regions and to normalize their SFRs and H$_2$ luminosities to this aperture.  As a result, their star formation and gas surface densities were artificially lowered by their much larger aperture sizes ($R_{25}$), which are revealed in the \spit\ data (and CO interferometry, e.g. \citealt{YouBurCap08}) to only contain star formation activity in smaller, central regions.

Integrating over only the star-forming region, we find that early-type galaxies mainly inhabit the same regions of the plot as typical star-forming spiral galaxies (Figure~\ref{ks}).  Interestingly, however, several of the galaxies fall closer to the circumnuclear starburst regime on these diagrams (filled circles), with gas surface densities approaching and exceeding 100 \msun$/$pc$^2$ (preferentially magenta).  This implies that, whenever even trace amounts of star formation are present, the same physical processes that govern the rate and efficiency of star formation in late-type galaxies and starbursts are also at work in early-type galaxies.


\section{Modes of Star Formation in Early-Type Galaxies}
\label{SF}

In the following sections, we compare the properties of the 13 galaxies in the \sau\ survey that show signs of star formation in the infrared to those of the sample as a whole.  Since the \sau\ sample is representative of local elliptical and lenticular galaxies, this provides insight into how star formation is connected to the structure and evolution of early-type galaxies.  From the comparisons in this section, two distinct modes of star formation in these galaxies emerge, and the properties of galaxies in these classes are observed to sample those of the early-type population in independent ways.

\subsection{Broadband Properties}
\label{sec:cmd}

By definition, the \sau\ galaxies are drawn from a sample of local red sequence galaxies, whose very tight adherence to the optical red sequence is demonstrated by Falc\'on-Barroso et al. ({\it in prep}).  For this reason, it is intriguing to find on-going or recent star formation in a number of these ``red and dead" systems.

\begin{figure}
	\centering
	\includegraphics[height=7cm,angle=90,trim=0cm 0cm 1cm 2cm]{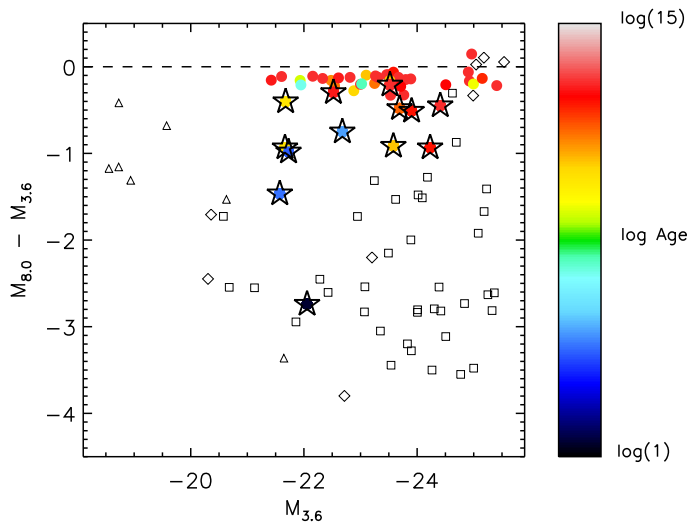}
	\caption{Infrared color-magnitude diagram for the \sau\ galaxies, measured over apertures of one effective radius.  Galaxies are indicated with points colored by their SSP ages (in Gyr; Kuntschner et al. {\it in prep}), such that younger stellar populations are blue.  Galaxies with on-going star formation are indicated with stars.  In the infrared bands, these systems are redder, since they appear more prominently in the redder 8.0\um\ (PAH) emission.  In the optical bands, which probe stellar photospheric emission, star-forming populations are typically bluer.  For straightforward comparison to optical CMDs, in which the (bluer) star-forming galaxies are lower on the vertical axis, the color index here is defined such that the (redder) star-forming galaxies are also lower on the vertical axis.  For reference, the $R_{25}$ colors and magnitudes of the SINGS galaxy sample are overplotted (E/S0: {\it diamond}, Sa-Sd: {\it square}, Sm-Im: {\it triangle}), although their slightly larger aperture $R_{25}$ increases their luminosities (and thus decreases their magnitudes, moving them to the right) relative to the \sau\ sample (photometry measured over $R_e$).  An $M_{8.0}-M_{3.6}$ color of zero is indicated with the dashed line; the magnitudes quoted here are Vega magnitudes (zero-points given by the SSC), so this line corresponds roughly to the color of a purely stellar spectrum.}
	\label{cmd}
\end{figure}

However, the $V-K$ colors used to define the optical red sequence are relatively insensitive to low mass fraction young stellar populations and can be strongly affected by dust.  We therefore construct an infrared color-magnitude diagram (CMD), which has the advantage of being directly sensitive to star formation (Figure~\ref{cmd}).  In this CMD, star-forming galaxies have large 8.0\um\ fluxes and are therefore redder, although for clarity and comparison with optical CMDs, we reverse the color index such that redder objects (with higher SFR) are lower in the diagram.

To examine the extent to which our star-forming galaxies differ from quiescent systems in the infrared color-magnitude space, we compare the \sau\ data to that from the SINGS galaxy sample \citep{Ken+03,Dal+05}, in which galaxies with a wide range of properties were observed with \spit.  From this comparison (Figure~\ref{cmd}), it is evident that the star-forming \sau\ galaxies have infrared colors spanning the full range observed in star-forming spiral galaxies, although the majority have colors intermediate between star-forming and quiescent galaxies.  In the infrared CMD, the \sau\ red sequence star-forming galaxies are thus readily distinguished from quiescent early-type systems \citep[see also the UV CMD results of][]{Yi+05,Kav+07}.

\begin{figure*}
	\centering
	\includegraphics[width=17cm]{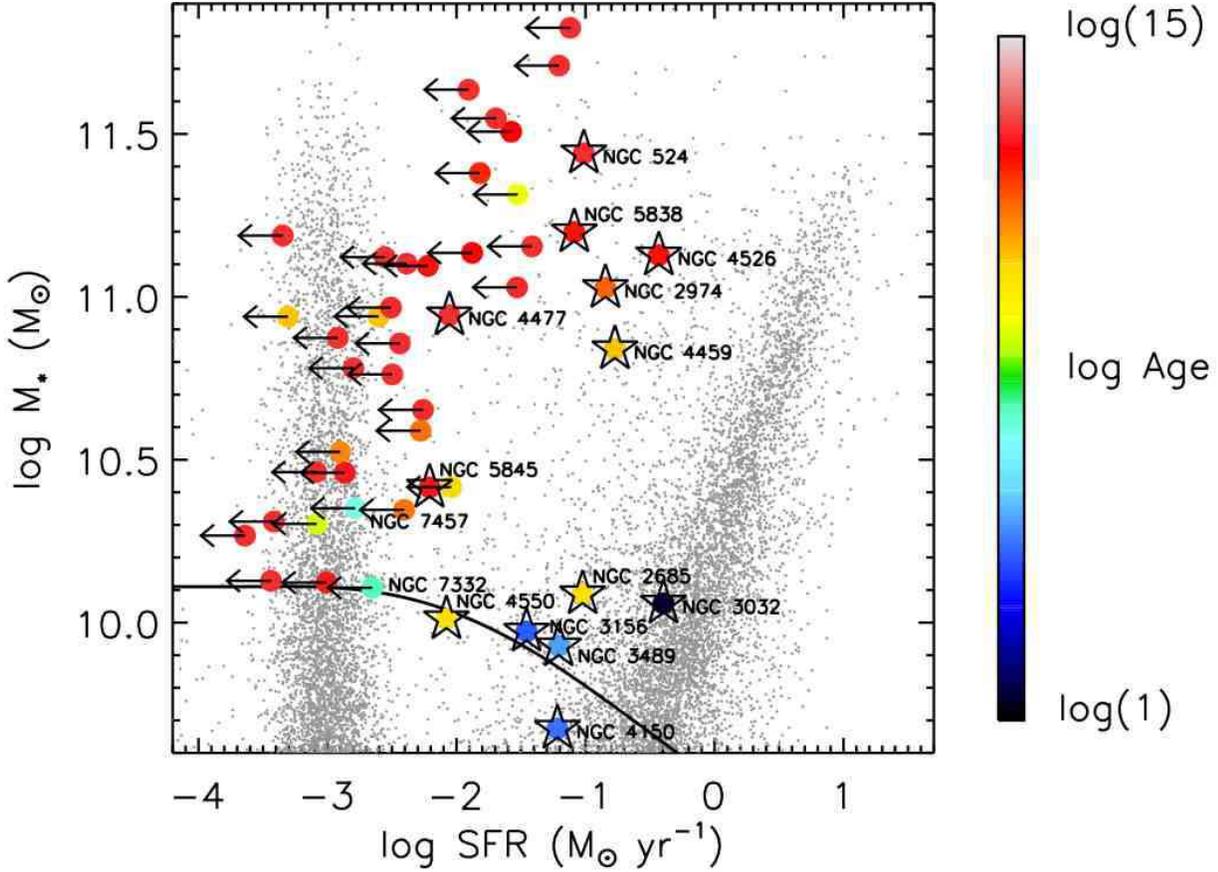}
	\caption{Relation of SFR and \mstar\ for the Millennium Simulation at $z$ = 0 ({\it grey dots}) and the \sau\ early-type galaxies ({\it large, colored circles}).  The \sau\ galaxies are colored by their SSP age (in Gyr, measured over $R_e$), such that young stellar populations are blue, while older stellar populations are red.  Those systems with clear evidence of star formation in the \spit\ images are indicated with stars; error bars are $\sim$1$-$2 times the size of the symbols and are omitted for clarity.  The $M_B \leq -$18 selection criterion for the \sau\ sample is evaluated in this parameter space using the simulated galaxies and plotted here with the black line.}
	\label{sfrmstar}
\end{figure*}

It is therefore instructive to study more fundamental properties of these star-forming galaxies than can be probed in a broadband CMD.  In Figure~\ref{sfrmstar}, we plot the \sau\ galaxies in the SFR-\mstar\ parameter space, a more physical analog to the color-magnitude diagram. To accomplish this, we first estimate the stellar mass of the \sau\ galaxies using the ``observed virial" masses M$_{vir} =$~5$R_e \sigma_e^2 / G$, where $\sigma_e$ is the luminosity-weighted second velocity moment within 1~$R_e$ (\citealt{SAURONIV}, hereafter \piva; \pixa).  The factor~5 scales the dynamical mass within one effective radius to the total baryonic mass of the galaxy; \pivt\ shows that this mass estimator contains an average dark matter contribution of $\sim$~30\%, although this percentage varies at the low and high mass extremes of the sample and depends strongly on the mass fraction of young stars in a galaxy (through the stellar population mass-to-light ratio).  We therefore approximate M$_* \approx$~0.7~M$_{vir}$, with the caveat that the stellar masses of individual galaxies may deviate from this estimate, and we plot these masses against the SFR measurements derived here (Table~\ref{tbl:sfr}).

These results are compared to simulated galaxies generated by the Munich semi-analytic models \citep{Cro+06,DeLucBla07} for the Millennium simulation \citep{Spr+05}.  For inclusion on this diagram, all simulated galaxies with no star formation are shifted to SFR = 10$^{-3}$ \msun\ \peryr, with an imposed scatter of 0.2 dex.  This reveals both a ``red" sequence of zero star formation (left) and a ``blue" sequence of the well-known SFR-\mstar\ relation for star-forming galaxies (right).  Additionally, the simulated broadband galaxy luminosities estimated in these models (including dust extinction) enable the \sau\ selection criterion of $B$-band absolute magnitude $\leq -$18 to be robustly identified in this parameter space (black line indicates the mass above which 90\%\ of galaxies with a given SFR fulfill this criterion).

Figure~\ref{sfrmstar} confirms the impression from the infrared CMD that nearly all of the \sau\ star-forming galaxies fall in the ``transition" region between actively star-forming and quiescent galaxies.  The only exception to this may be NGC~3032, which lies on the cusp of the SFR-\mstar\ relation and thus is consistent with a still actively star-forming galaxy.  Before proceeding further, it is important to reiterate that the \sau\ star-forming galaxies are detected as such in Figures~\ref{cmd}~and~\ref{sfrmstar} only due to the sensitivity of these parameter spaces to low-level star formation.  All of these systems are traditional early-type galaxies, with red optical colors and bulge-dominated morphologies, which nevertheless do form stars at significant rates.

\subsection{The Dynamics of Star Formation}
\label{SFdyn}

With the \sau\ data, it is possible to delve more deeply into the properties of early-type galaxies than can be accomplished with broadband data alone.  Of particular interest are the stellar and gas kinematics associated with star formation in early-type galaxies.

\begin{figure*}
	\centering
	\includegraphics[width=7cm]{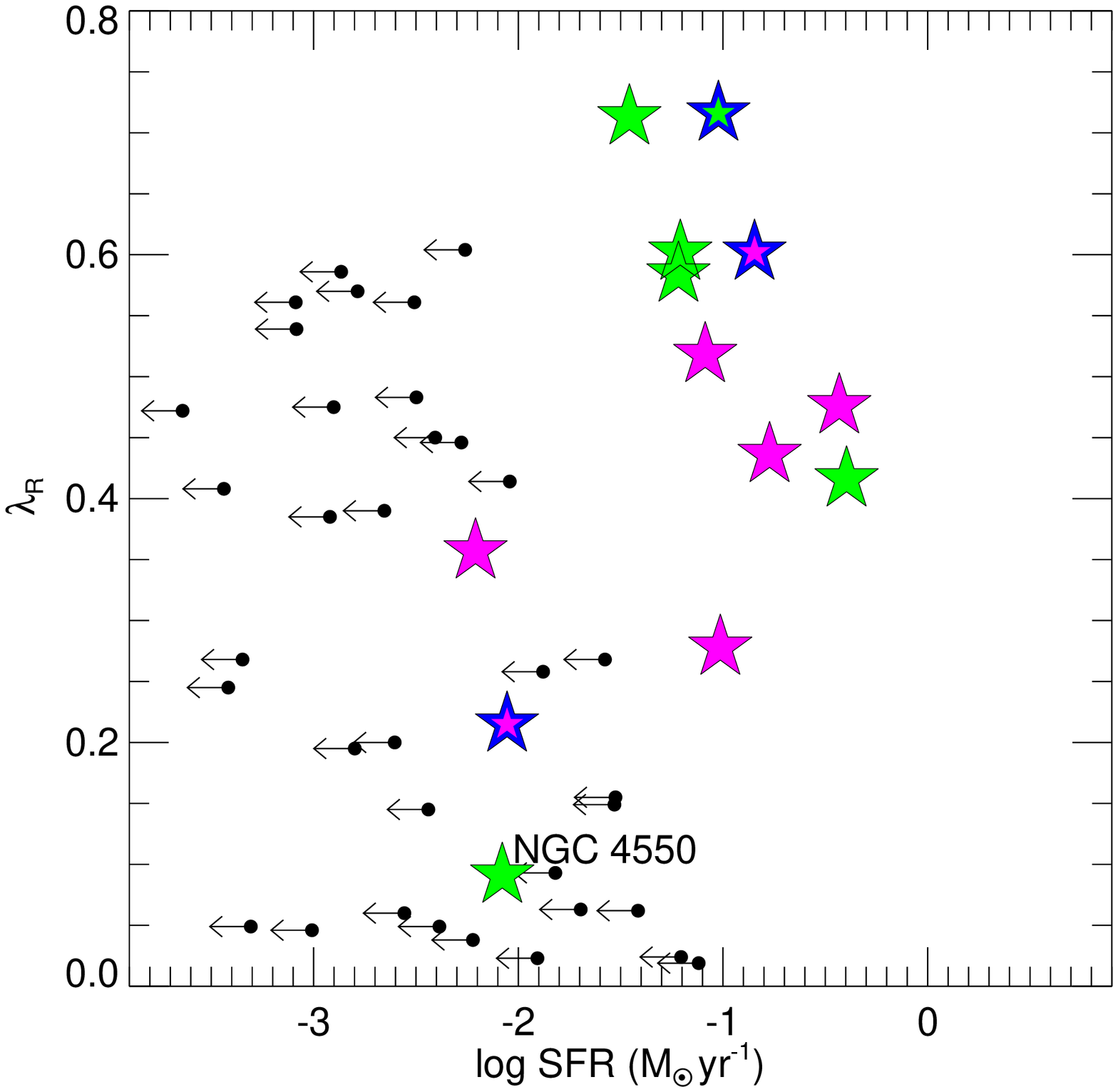}
	\includegraphics[width=7cm]{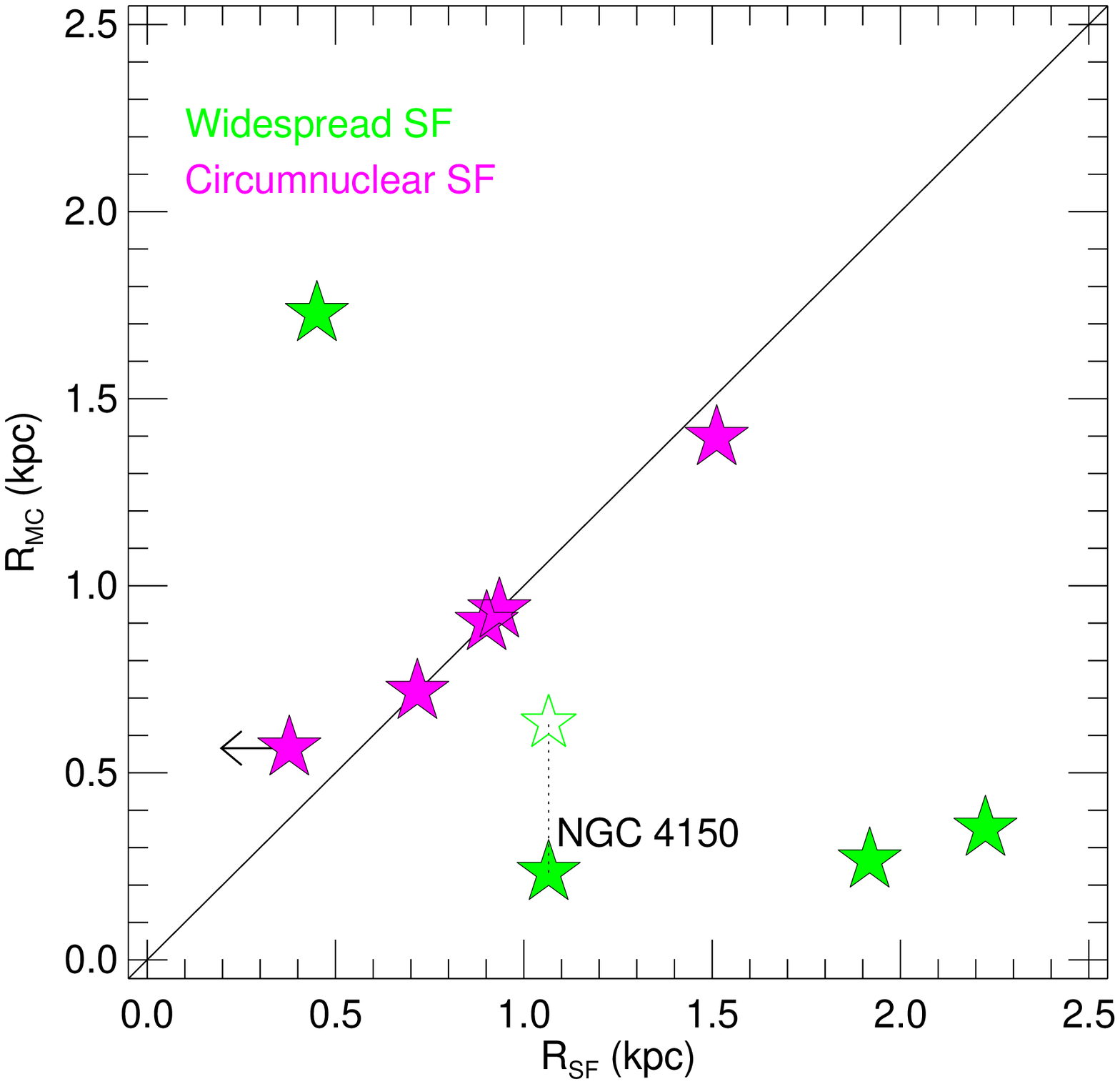}
	\includegraphics[width=7cm]{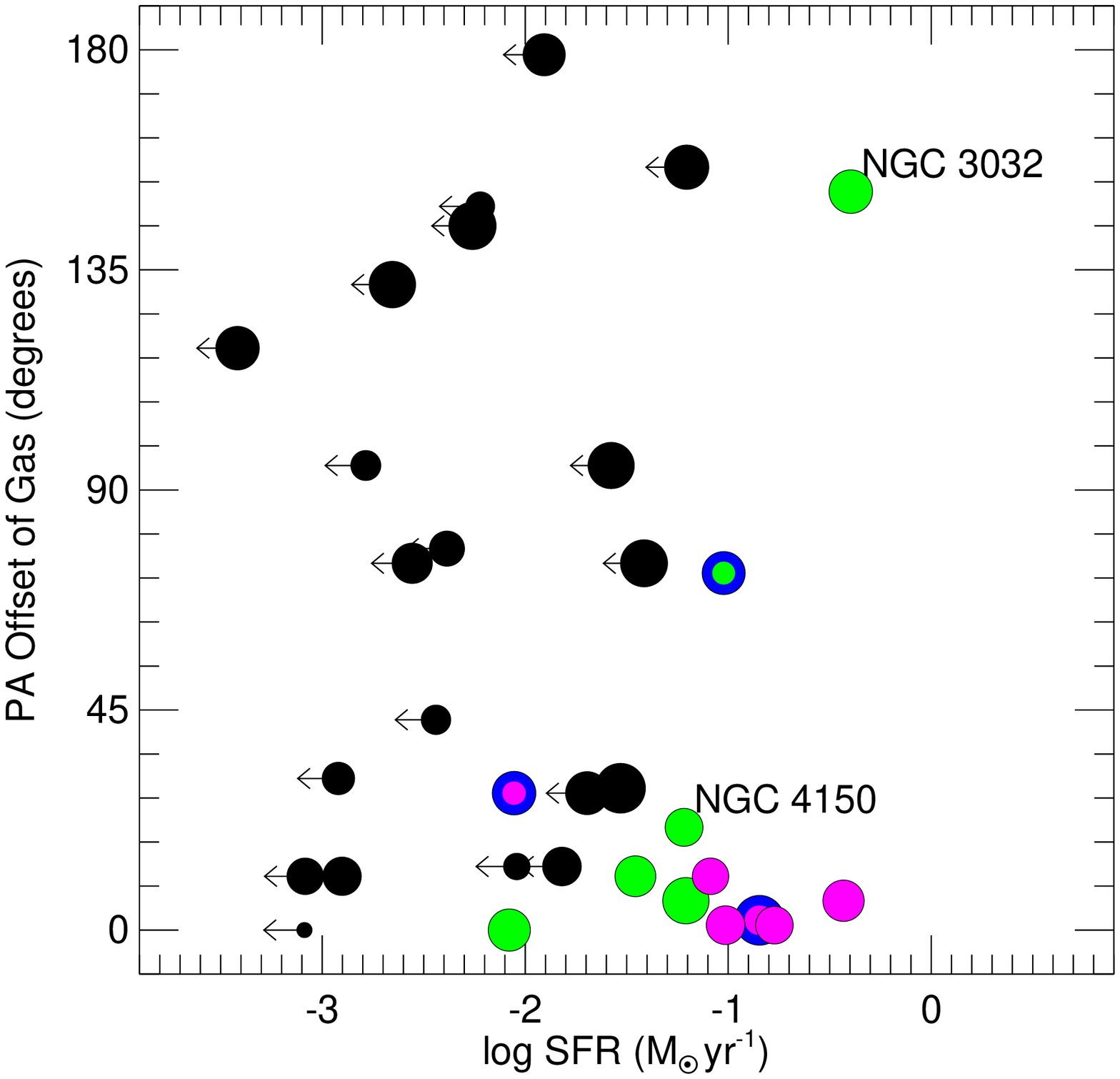}
	\includegraphics[width=7cm]{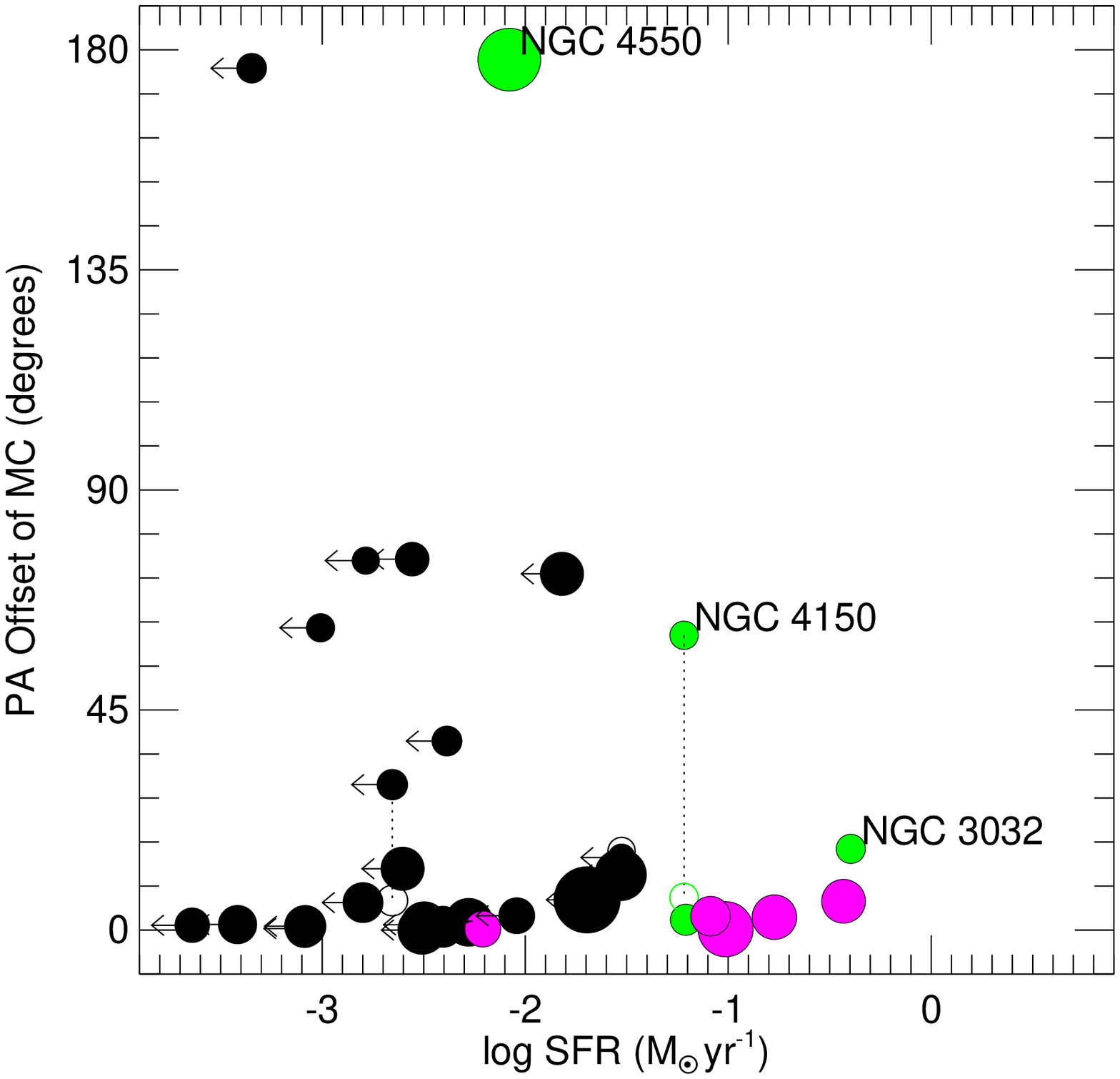}
	\caption{{\it Top left}: Comparison of (projected) specific angular momentum \pixp\ with SFR in the \sau\ galaxies.  {\it Bottom left}: Comparison of position angle offset between the gas and stellar kinematic major axes \pvp\ with SFR, including the 12/13 star-forming galaxies with sufficient ionized gas to make this measurement.  The sizes of the points correspond to the estimated masses of ionized gas in the galaxies (see \pva).  {\it Top right}: Comparison of the radial extent of distinct kinematic components $R_{\rm MC}$ \pxiip\ with radius of star formation activity $R_{\rm SF}$ for the star-forming galaxies with distinct kinematic substructures (see text).  {\it Bottom right}: Comparison of position angle offset between stellar kinematic components for galaxies with more than one component (from \pxiia).  The sizes of the points indicate the sizes of the MCs (in kpc).  In the latter two plots, the counter-rotating stellar disk in NGC~4550, which \pxiit\ notes is not detected in kinemetric analysis, is included, as is the marginally detected co-rotating MC in NGC~524.  In galaxies with two MCs~(=~three kinematic components), the outer MC is indicated with an open symbol.  In all plots, systems with star formation are indicated with colored symbols ({\it green}: widespread star-forming, {\it magenta}: circumnuclear star-forming, {\it blue outline}: outliers), with the specific locations of some galaxies noted (see text for discussion).  Error bars on the SFRs of these galaxies are 1$-$2 times the sizes of the symbols.}
	\label{starsgas}
\end{figure*}

Most apparent in the \sau\ kinematic data is that all star-forming galaxies are so-called ``fast rotators," with relatively high specific angular momenta (\lr $>$0.1).  The top left panel of Figure~\ref{starsgas} illustrates that some of the star-forming galaxies are even among the most rapidly rotating galaxies in the sample.  There is a single exception to this, NGC~4550.  This galaxy is a pathological ``slow rotator," in that it is composed of two rapidly counter-rotating disks (see stellar velocity maps in Figure~\ref{Main-SF}), whose angular momenta largely cancel each other; however, in terms of general properties, it belongs in the fast rotator family (\pixa; \pxa).

In \pxiit, fast rotators were shown to universally be composed of a large bulge and one or more stellar disks of varying thickness and spatial extent.  Since, in many cases, these stellar disks rotate more rapidly  and display different age and metallicity properties than their host galaxies (i.e. the bulges that dominate early-type systems), these structures may be connected to the more rapidly rotating gaseous component via recent star formation.  In the top right panel of Figure~\ref{starsgas}, we compare the spatial extent of star formation detected here with the spatial extent of kinematically distinct structures (counter-rotating cores, embedded stellar disks, etc), if present, as measured in \pxiit.  

It is apparent that in some cases (magenta stars), the kinematically distinct structure (or multiple component, MC) is directly related to the on-going star formation activity.  In these galaxies, the MCs are embedded stellar disks whose kinematic position angle is identical to that of the main galaxy body and of the gas (bottom panels in Figure~\ref{starsgas}).  These disks are also quite thin, as seen directly in edge-on systems (e.g. NGC~4526) or in optical images of the corresponding dust disks (e.g. NGC~4459).  The star formation morphologies in these galaxies are highly regular; all star formation is confined to a disk morphology that is spatially coincident with the ionized gas and dust disks, with variations in stellar populations, and with embedded stellar disks if present.

However, there are also star-forming galaxies in the sample with MCs whose extents are not obviously linked to the star formation activity (green stars in Figure~\ref{starsgas}).  This disconnect is also apparent in the kinematic alignments in these galaxies; roughly half of these systems (3/5) contain counter-rotating stars, gas, or both.  In many of these galaxies, the morphology of the star-forming region is markedly irregular (Figure~\ref{profiles}) and extends at low levels to large radii, in keeping with the ubiquity of ionized gas, with the universally young stellar populations (\S\ref{modeI}), and with the lack of distinct stellar kinematic features associated with the (widespread) star formation.

We can therefore divide the star-forming galaxies into two distinct classes using the top right panel of Figure~\ref{starsgas}; we refer to those galaxies with signs of star formation out to large radii (green stars) as ``widespread" star-forming galaxies and those galaxies in which the star formation is confined to a central disk morphology (magenta stars) as ``circumnuclear" star-forming galaxies, in analogy to the two modes of star formation seen in late-type galaxies \citep{Ken98}.  Four additional star-forming galaxies are not included in this panel due to their lack of distinct kinematic structures in the \sau\ data.  The first, NGC~3156, shows PAH and ionized gas emission with slightly irregular morphologies out to large radii, has no kinematic structure associated with the star formation event, and contains universally young stellar populations.  In all properties, this galaxy is therefore consistent with belonging to the widespread star-forming class (green), and we consider it a member of this group.  The remaining three galaxies do not fit neatly into either category.  Two of these three (NGC~2685, NGC~2974) contain large-scale rings of star formation and the third (NGC~4477) is highly disturbed by a strong bar.  It is therefore not immediately obvious how these three galaxies (shown with blue outlines in Figure~\ref{starsgas}; inner color determined from discussion in \S\ref{modeIII}) fit into the simple two-class division given above.

In the remainder of this section, we compare the properties of the five \classone\ galaxies (NGC~3032, NGC~3156, NGC~3489, NGC~4150, NGC~4550) to those of the five \classtwo\ galaxies (NGC~524, NGC~4459, NGC~4526, NGC~5838, NGC~5845) in our sample and find these two classes to be fundamentally distinct, in more than the morphology and kinematics of their star formation events.  We then describe the properties of the ``outliers" (NGC~2685, NGC~2974, NGC~4477) and find each of these galaxies to be consistent with one of the two modes of star formation, when a more nuanced approach is taken.

It is worth noting, before moving on, that there is one deep similarity between all of the star-forming galaxies, namely that the star formation is occurring in a thin disk or ring morphology.  This is directly apparent in the one \classone\ galaxy with a dynamically relaxed interstellar medium (NGC~3032), in the \classtwo\ galaxies (all of which contain thin dust disks), and in the two galaxies with star-forming rings.  In the remainder of the \classone\ galaxies and in the third ``outlier" galaxy (NGC~4477), the presence of a thin disk is implied by the rapid rotation velocities in the stars and by the low velocity dispersions in the ionized gas.  Similar results have been found for star formation in the bulges of Sa galaxies in \citet[][\pxia]{SAURONXI}.  Together, this evidence suggests that a thin disk/ring morphology may be the most common mode of star formation in early-type galaxies and bulges.

\subsection{Widespread Star-Forming Fast Rotators}
\label{modeI}

In Figure~\ref{sfrmstar}, the \classone\ galaxies (NGC~3032, NGC~3156, NGC~3489, NGC~4150, NGC~4550) are seen to be those star-forming early-type galaxies with lower stellar masses.  Given the comparable molecular content of all star-forming galaxies in our sample (see Table~\ref{tbl:sfr}), the \classone\ galaxies are consequently also characterized by higher gas fractions and specific star formation rates.  As a result, the current star formation event dominates the single stellar population (SSP) ages, which are measured to be young throughout these systems ($<$1$-$3 Gyr; Kuntschner et al.~{\it in prep}).  This is consistent with the location of the \classone\ galaxies in Figure~\ref{ks}; these systems have molecular gas and star formation surface densities similar to those found in late-type galaxies undergoing global star formation.  That the star formation event is widespread is also apparent in its morphology and in the absence of stellar kinematic structures with spatial extents related to those of the star formation (by definition; see \S\ref{SFdyn}).  Nevertheless, distinct stellar and gas kinematic structures do exist in these systems; galaxies in this class are evenly split between objects with counter-rotating stars, gas, or both (3/5) and those with purely co-rotating stellar and gaseous components (2/5).

Within this class, interesting trends in galaxy properties with SFR are found.  In particular, galaxies with higher SFRs (e.g. NGC~3032; SFR~=~0.4 \msun\ \peryr) have, unsurprisingly, more prominent signs of on-going star formation, including younger SSP ages, higher molecular gas masses, and lower \oiii/\hb\ ionized gas emission line ratios.  Those \classone\ systems with lower SFRs (e.g. NGC~3156, NGC~3489, NGC~4150; SFR~=~0.03$-$0.06 \msun\ \peryr) have SSP ages older by $\sim$1$-$2 Gyr, an order of magnitude less molecular gas, and high \oiii/\hb\ ratios that are inconsistent with significant on-going star formation \pxvip.  Lower SFRs in a \classone\ galaxy also correlates with an increasingly irregular star formation and gas distribution, from a regular, disk-like morphology (NGC~3032) to disturbed ionized and molecular gas distributions (NGC~3489; Figure~\ref{Main-SF} and Crocker et al.~{\it in prep}) to highly asymmetric molecular gas distributions (NGC~4550; \citealt{Cro+09}).  These increasingly irregular morphologies additionally correspond to more centralized star formation at lower SFR (compare NGC~3032, NGC~3156, and NGC~4550 in Figure~\ref{profiles}{\it a}).

In the absence of a replenishing gas reservoir, \classone\ galaxies will experience declining SFRs, causing them to move to the left in Figure~\ref{sfrmstar}; the system with the most intense star formation (NGC~3032) will therefore be expected to eventually move towards the other \classone\ galaxies.  Likewise, there is evidence that the \classone\ galaxies with lower SFRs previously had more intense star formation events that extended to larger radii, implying earlier positions to the right of their current location in Figure~\ref{sfrmstar}.  In particular, the young SSP ages (1$-$3 Gyr in NGC~3156, NGC~3489, NGC~4150; $\sim$6 Gyr in NGC~4550; Kuntschner et al.~{\it in prep}) found throughout these galaxies, beyond the star-forming regions observed in PAH emission, indicate that star formation activity occurred out to larger radii in the past than in the present.  Only in the galaxies' centers do the final traces of significant star formation, along with the youngest stellar populations, remain.

With this evidence for the movement of \classone\ galaxies towards lower SFR with time (to the left in Figure~\ref{sfrmstar}), along with the trends in galaxy properties with SFR, we can associate star formation events with the global production of young stars in rapidly rotating kinematic structures (Figures~\ref{starsgas}~and~\ref{Main-SF}).  Additionally, star formation in this class appears to cease in an ``outside-in" manner that is coincident with a consumption, disruption, and/or heating of the molecular reservoir.  It is interesting to note, however, that throughout this process, the last vestiges of star formation are still proceeding according to established SFR-H$_2$ scaling relations (Figure~\ref{ks}).

In \S\ref{discussI} below, we analyze the mechanisms that may be responsible for creating the \classone\ galaxies described here and for governing the cessation of their star formation.

\subsection{Circumnuclear Star-Forming Fast Rotators}
\label{modeII}

The \classtwo\ galaxies (NGC~524, NGC~4459, NGC~4526, NGC~5838, NGC~5845) appear to represent a completely different physical state than the \classone\ systems described above.  In Figure~\ref{sfrmstar}, these galaxies are seen to be those star-forming systems with higher masses and therefore lower molecular gas fractions and specific star formation rates.  Consequently, the current star formation events do not dominate the SSP ages, all of which are measured to be intermediate to old in the regions of star formation.  Unlike in the \classone\ galaxies, these regions of star formation are spatially very distinct, with all of the star formation, younger SSP ages, ionized gas, molecular gas, dust, and stellar kinematic structures being spatially coincident in well-defined central disks (see the definition of this class in \S\ref{SFdyn}).  Within these disks, the surface densities of star formation and molecular gas approach those seen in circumnuclear starbursts in late-type systems (Figure~\ref{ks}).  Additionally, the stellar and gas kinematic structures universally co-rotate with one another and with the old stellar populations in the quiescent regions of the host galaxies.

Unlike in \classone\ galaxies, none of the galaxies in this class have migrated all the way to the SFR-\mstar\ relation (Figure~\ref{sfrmstar}), likely because the star formation only occupies a small region of the galaxy.  Within these \classtwo\ systems, a range of properties is seen.  In some systems (NGC~4459, NGC~4526), all star formation indicators (PAHs, CO, \oiii/\hb, UV, low ionized gas velocity dispersion), where available, paint a consistent picture of on-going star formation in the region of the central disk.  In contrast, the presence of on-going star formation in other \classtwo\ galaxies (NGC~524, NGC~5838, NGC~5845) is more tenuous.  These galaxies have more tentative PAH detections (\S\ref{maybeSF}) and only marginally detected younger SSP ages and metallicity gradients.  Nevertheless, the PAH emission remains spatially coincident with thin, regular dust disks \pvp\ and with embedded stellar kinematic structures (Figure~\ref{starsgas}).  These systems may have reached the end of their star formation epochs, since only one (NGC~524; Crocker et al. {\it in prep}) is detected in CO and the other tracers of star formation (PAH, \oiii/\hb) can be affected by a B star population that has not yet died.  Nevertheless, as in \classone\ galaxies, even systems with trace amounts of molecular gas and PAH emission form their stars according to SFR-H$_2$ scaling relations (Figure~\ref{ks}).

In the following (\S\ref{discussII}), we speculate on the origins of gas responsible for this spatially distinct star formation event and on the mechanisms that spatially confine it into disks/rings.

\subsection{Outliers}
\label{modeIII}

While the two classes described above neatly bring together two groups of early-type galaxies that have a large number of similar properties, several ``outlier" galaxies in our sample (NGC~2685, NGC~2974, NGC~4477) resist such straightforward classification.  Here, we discuss the characteristics of each of these unique systems in turn and relate them to the two main modes of star formation.

\begin{figure*}
	\centering
	\includegraphics[width=17cm]{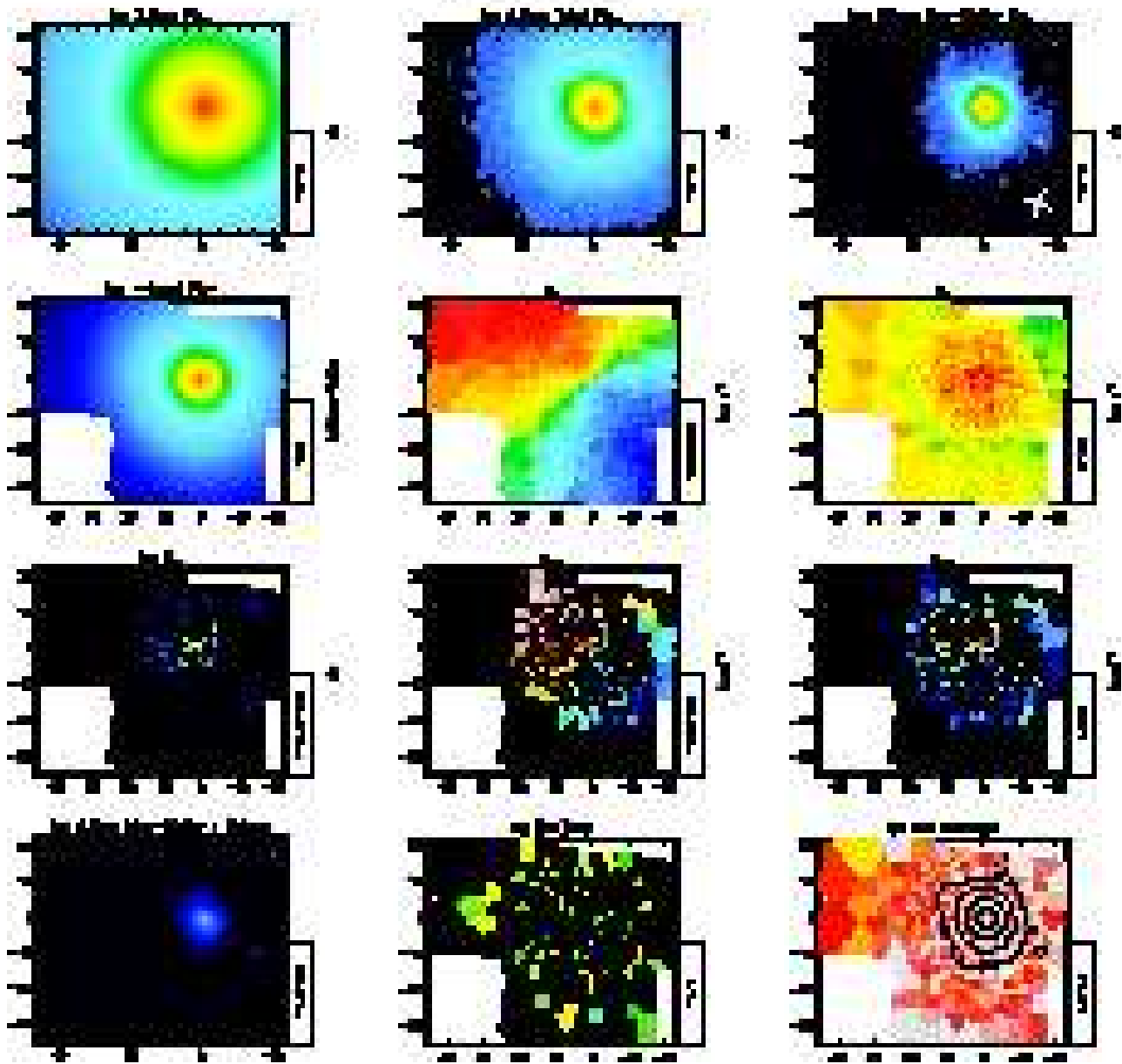}
	\caption{As with Figure~\ref{Main-noSF}, for all sample galaxies with significant infrared emission.}
	\label{Main-SF}
\end{figure*}

NGC~2685 is a prototypical polar ring galaxy, as is evident in a comparison of the morphology of the star-forming ring to the position angle of the galaxy and the stellar kinematics (Figure~\ref{Main-SF}; see the molecular and atomic gas maps of \citealt{SchSco02,Mor+06}; but see also \citealt{JozOosMor09}).  This rare configuration is often attributed to the accretion of gas or a gas-rich satellite at a very specific orientation (see \S\ref{discussI}), but the relevant aspect for this discussion is that the resulting dynamics of the gas are unassociated with those of the stars, as in \classone\ galaxies.  Moreover, NGC~2685 also resembles this class of star-forming galaxies in that it contains ubiquitous younger stellar populations (SSP age of 6 Gyr; Kuntschner et al. {\it in prep}) and ionized gas, the spatial distributions of which are unassociated with the current extent of the star formation event.  In the following discussion, we therefore consider this system to be related to the \classone\ galaxies (see blue-green symbols in Figures~\ref{ks}~and~\ref{starsgas}); we elaborate on this relationship in \S\ref{discussI}.

NGC~2974 also contains a large-scale star-forming ring, detected in PAHs, broadband UV emission, and \hone\ \citep[Figure~\ref{profiles}{\it a};][]{Jeo+07,Wei+08}.  In this system, the ring is morphologically and kinematically aligned with the stars \citep[Figure~\ref{Main-SF};][]{Wei+08}.  A probable interaction between this gas and a putative large-scale bar is responsible for generating the ring morphology \citep{Kra+05,Jeo+07}, suggesting that here, as with \classtwo\ galaxies, there is a deep connection between stellar kinematic structures and the spatially distinct morphology of star formation.  This galaxy resembles \classtwo\ systems in other ways, including the co-rotation of the gas and the stars as well as the old SSP ages (11 Gyr; Kuntschner et al. {\it in prep}), and we consequently include NGC~2974 in our discussion of the \classtwo\ galaxies in \S\ref{discussII} (see blue-magenta symbol in Figure~\ref{starsgas}).

Finally, NGC~4477, though unremarkable in the morphology of its star-forming region (Figure~\ref{profiles}{\it a}), is notable for being the most strongly barred of the star-forming galaxies in our sample (see Figure~\ref{Main-SF}), and the effects of this bar on the dynamics of the system are significant.  The ionized gas kinematics are driven by the bar, producing a spiral inflow structure that creates a central star-forming disk kinematically aligned with the bar (Figure~\ref{Main-SF}).  The resulting strong misalignment between the stars and the ionized gas is another expected consequence of the bar, since the gas is confined to the bar's circular, non-intersecting orbits.  In this sense, NGC~4477 resembles the \classtwo\ galaxies in a fundamental way - the stars, ionized gas, and star formation activity are strongly related dynamically - despite more superficial differences.  In addition, the stellar populations in NGC~4477, as with those in \classtwo\ galaxies, are very evolved (SSP age of 11 Gyr; Kuntschner et al. {\it in prep}).  This renders the detection of star formation in this galaxy slightly uncertain, although PAHs, molecular gas, dust, and age and metallicity gradients are detected in the central regions of this system (see \pva; Kuntscher et al. {\it in prep}; Crocker et al. {\it in prep}).  Given these similarities to \classtwo\ galaxies, we discuss NGC~4477 alongside those galaxies in \S\ref{discussII} (see blue-magenta symbols in Figures~\ref{ks}~and~\ref{starsgas}).

Despite being individually quite unique, these three ``outlier" galaxies nevertheless each have important similarities to one of the two main modes of star formation.  In the following sections, we speculate on these more nuanced connections and on the implications for the origin and evolution of the two main modes of star formation.


\section{The Evolution of the Red Sequence}
\label{discussion}

Our \spit\ data of the \sau\ early-type galaxies makes it possible to identify two different modes of star formation in these systems and to study the cessation of star formation in each regime.  We cannot, however, make quantitative statements from these data about the origins of the star-forming gas or about the mechanisms responsible for the shut-down of the star formation.  In this section, we nevertheless speculate on these topics, taking advantage of our multi-wavelength suite of data and the insights of \S\ref{SF}, in order to probe the role of the widespread and circumnuclear modes of star formation in the evolution of fast rotator early-type galaxies.  Afterwards, we briefly comment on slow rotators, none of which show any indication in our data for on-going or recent star formation; the formation and evolution of these systems is likely distinct from that of the fast rotators.

\subsection{Widespread Star-Forming Fast Rotators}
\label{discussI}

In \S\ref{SF}, \classone\ galaxies were found to have an equal presence of co- and counter-rotating stellar and gaseous structures, a distribution of kinematic alignments that is strongly suggestive of an external origin for the star-forming gas \pvp.  Indeed, this has been explicitly demonstrated for two of the three systems with counter-rotating structures (NGC~3032: \citealt{YouBurCap08}; NGC~4550: \citealt{Cro+09}; see also \citealt{TemBriMat09}) and for the related ``outlier" galaxy NGC~2685 by virtue of its polar ring \citep[e.g.][]{SchSco02,JozOosMor09}.  The \classone\ galaxies are therefore more likely caused by the accretion of external gas than by secular evolution or rejuvenated star formation via internal mass loss (see also \citealt{Sar+07}; Oosterloo et al. {\it in prep}).  Given the high gas mass fractions and specific SFRs resulting from these accretion events, it is perhaps more correct to call such events gas-rich mergers, with the accreted system being either a galaxy or cold gas.

Furthermore, the majority of these mergers are likely minor mergers, since all of the \classone\ galaxies live in field or small group environments (with the possible exception of NGC~4150, which may be interacting with the outskirts of the Virgo cluster).  It is difficult to justify invoking recent major mergers as the sole cause of the star formation in all six of these galaxies in our local (typical distance of a \sau\ sample galaxy $\sim$20 Mpc) sample.  While a major merger was likely responsible for one of these systems (NGC~4550; \citealt{Cro+09}), the more frequent minor mergers (see e.g. theoretical results of \citealt{FakMa08}) are more logical causes for the star formation seen in most of the \classone\ galaxies.  Similar conclusions have been reached by \citet{Kav+09}, who showed that the fraction of star-forming early-type galaxies, as detected in NUV, can be reproduced well by the rate of gas-rich minor mergers in cosmological simulations.

These minor mergers involve a large amount of gas that results in significant star formation throughout the system, bringing the remnants all the way to the SFR-\mstar\ relation.  The main progenitors of these mergers may have been quiescent early-type systems whose star formation has been rejuvenated; however, it is also possible that the main progenitors were spiral galaxies and that the minor merger is responsible for moving these galaxies off of the blue cloud, with the associated widespread star formation event representing the tail end of this process.  Indeed, in the mass range of these galaxies, 60\% of galaxies are late-type systems (from the mass functions of \citealt{Bel+03} and \citealt{Bal+04}), resulting in sufficient availability of such progenitors.  Additionally, recent simulations have shown that S0-like galaxies can be produced by spiral galaxies subjected to single minor mergers (4:1 to 10:1; \citealt{BouJogCom07}).  The \classone\ galaxies, all morphologically classified as S0, are thus good candidates for the products of recent minor mergers.

\begin{figure*}
	\centering
	\includegraphics[width=5.5cm]{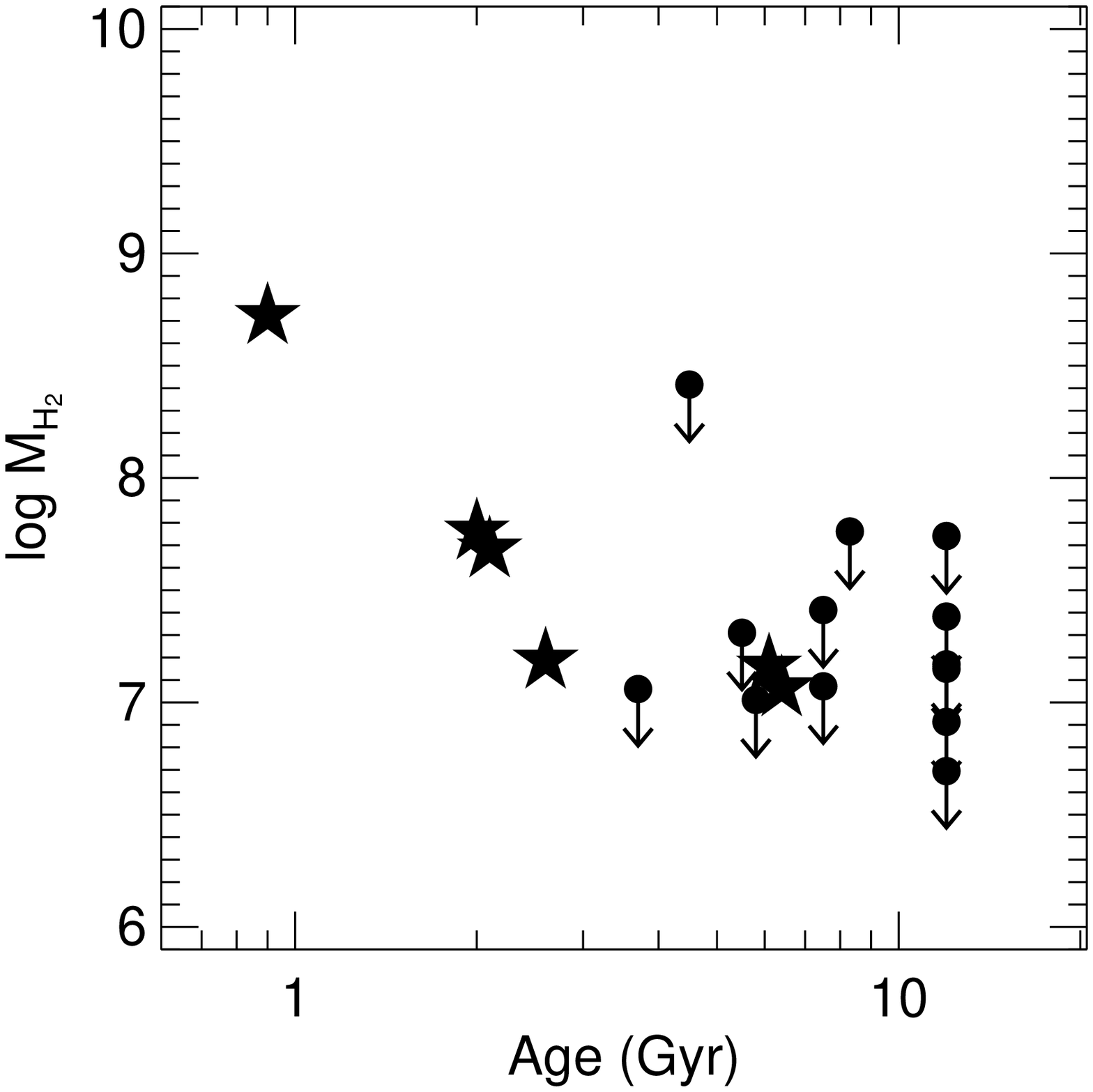}
	\includegraphics[width=5.5cm]{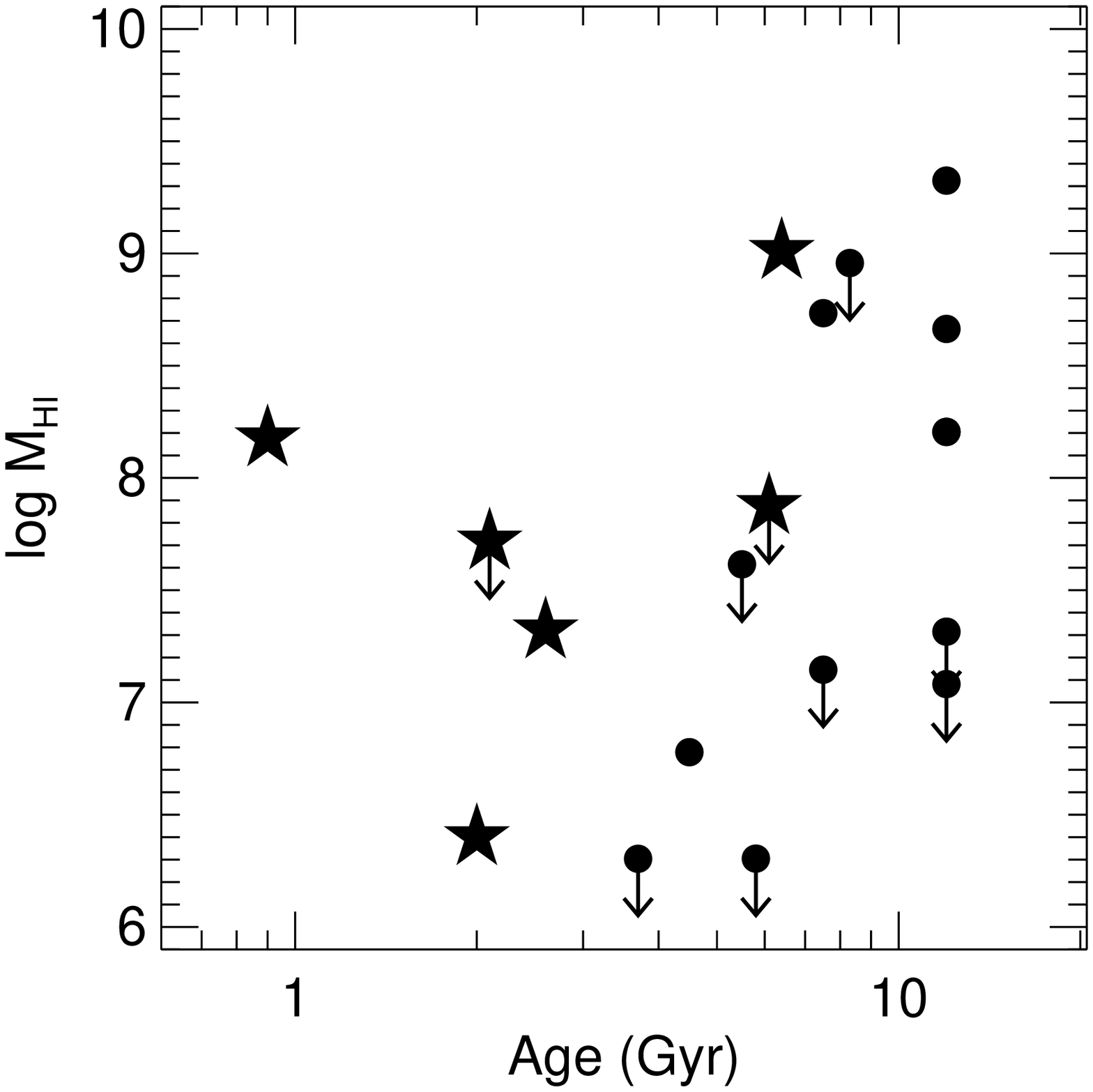}
	\includegraphics[width=5.5cm]{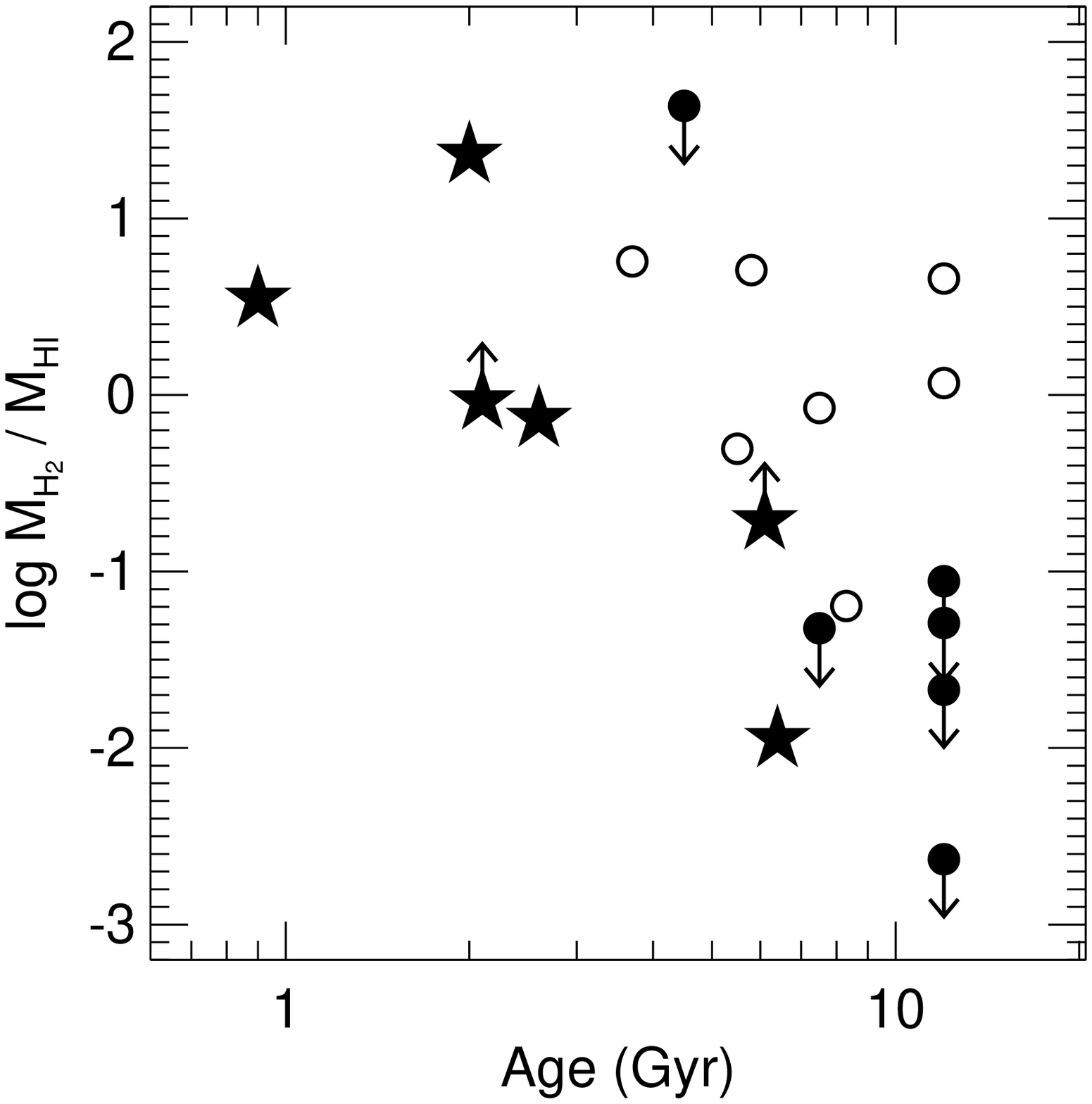}
	\caption{H$_2$ mass, H{\tiny I} mass, and the ratio of the two, as a function of SSP age measured over $R_e$ (Kuntschner et al.~{\it in prep}).  The galaxies plotted are all \sau\ systems classified as lenticular, excluding the \classtwo\ star-forming galaxies where applicable.  For the H$_2$/H{\tiny I} ratio, systems for which both the H$_2$ and H{\tiny I} masses are constrained only with upper limits are shown with open circles.  In all plots, \classone\ galaxies are indicated with stars.}
	\label{lowmass}
\end{figure*}

In the wake of such (potentially formative) mergers, galaxies appear as \classone\ systems, which may in fact trace out the post-merger cessation of star formation in detail.  Galaxies making this transition may initially resemble NGC~3032, the system that is forming stars very close to the SFR-\mstar\ sequence and that is the youngest and most gas rich galaxy in this class.  As the molecular gas is depleted and/or disrupted, the star formation will cease first at large radii (\S\ref{modeI}).  Indeed, the galaxies in which this is observed (NGC~3156, NGC~3489, NGC~4150) also have very high \oiii/\hb\ emission line ratios that \pxvit\ associates with excitation by planetary nebulae in post-starburst galaxies.  The residual star formation in these galaxies will fade to very low levels (as in NGC~4550) and eventually cease, after which \classone\ galaxies are expected to migrate onto the sequence of quiescent galaxies in SFR-\mstar\ space.  This evolutionary process is somewhat modified in the polar ring galaxy NGC~2685, whose post-merger evolution (disruption of the gas and ``outside-in" cessation of star formation) was halted by the special orientation of the interaction.  The cessation of star formation in the resulting, rare polar ring feature consequently may not follow a similar process to that in the other \classone\ galaxies, but the probable result, a quiescent lenticular galaxy, is the same.


When the star formation has ceased, galaxies will have intermediate SSP ages, with remnant B stars in the galaxies' centers that may produce marginal PAH detections (\S\ref{SFother} and Figure~\ref{profiles}{\it b}) and no corresponding molecular reservoir (\citealt{ComYouBur07}; Crocker et al. {\it in prep}).  Candidate quiescent descendants of \classone\ galaxies may then be S0 systems with widespread intermediate SSP ages (5$-$10 Gyr), several of which have been identified in the \sau\ sample by Kuntschner et al. ({\it in prep}).  Two of the youngest of these are NGC~7332 and NGC~7457 (SSP age of $\sim$3$-$5 Gyr; Kuntschner et al.~{\it in prep}), both of which contain clearly disrupted interstellar media and very high \oiii/\hb\ emission line ratios.  NGC~7332, in particular, is a likely descendant of an accretion event, caused by an interaction between this system and its companion gas-rich spiral galaxy.  Atomic and ionized extra-planar gas structures connect these two galaxies and may be responsible for the central KDC visible in NGC~7332 \citep{Fal+04,Mor+06}.  However, the intermediate SSP ages in this galaxy and NGC~7457, along with the lack of detection of on-going star formation in the \sau\ and \spit\ data, suggest that these systems are in fact quiescent relics, perhaps of \classone\ galaxies.

The cessation of star formation, which turns \classone\ galaxies into quiescent early-type systems, was shown in \S\ref{modeI} to be coincident with the heating or exhaustion of the available molecular reservoir.  Testing which of these processes is at work in \classone\ galaxies is not straightforward, but we can examine the state of the ISM to zeroth order in Figure~\ref{lowmass}, using H$_2$ masses measured by \citet{SchSco02}, \citet{ComYouBur07}, \citet{YouBurCap08}, \citet{Cro+08}, and Crocker et al. ({\it in prep}) and \hone\ masses compiled by \citet{Rob+91} and measured by \citet{Mor+06}, all corrected to the distances assumed here.   We find that the ratio of molecular to atomic gas declines steeply with SSP age; this decline reflects both a decreasing molecular component and a slightly increasing atomic component with age.  However, this Figure contains many upper limits and clearly requires more data, so we suffice to conclude that the observed phase of the ISM is not inconsistent with older systems having hotter interstellar media (more \hone\ than H$_2$).  Similarly, the data are also consistent with the end of star formation resulting from the gradual consumption of a finite reservoir of molecular gas (decreasing H$_2$ in the presence of constant \hone).

The picture sketched in this section can thus be summarized as follows: A star-forming spiral galaxy or a quiescent lenticular galaxy undergoes a major or minor gas-rich merger, resulting in a bulge-dominated remnant.  After the merger, the remaining gas settles into a plane, where it is gradually consumed or heated and therefore forms stars at ever decreasing rates.  The star formation ceases in an ``outside-in" manner, in which the star formation at large radii ends first, followed by that in the galaxy center.  During this process, the galaxy is moving from the blue cloud, through the green valley, and eventually onto the red sequence.  The residual star formation in this stage has created a young population that dominates the galaxy's stellar light and that rotates rapidly because of its (low velocity dispersion) gaseous origin.  This process may therefore be connected to the production of the fastest rotating early-type galaxies.

\subsection{Circumnuclear Star-Forming Fast Rotators}
\label{discussII}

For the \classtwo\ galaxies, in contrast, it is less straightforward to deduce the gas origins from the observations.  Given the co-rotation of the gas and the stars in all \classtwo\ systems (excepting NGC~4477, in which the gas kinematics are driven by the strong bar), a simple explanation would be that the gas originated within the galaxies, as a result of mass loss from evolved stellar populations.  \citet{TemBriMat09} use a simple analytic calculation to demonstrate this scenario and show that the amount of gas generated by post-AGB stars is more than sufficient for late-time star formation in early-type galaxies \citep[see also e.g.][]{Ho09}.  Likewise, \pxvia\ demonstrates that these stars dominate the ionized gas distribution and excitation in the \sau\ sample, implying that the mass loss associated with this phase of stellar evolution plays a key role in regulating the ISM in early-type galaxies.

However, in one of the ``outliers" associated with this class of galaxies (NGC~2974), the distribution of the star-forming gas in an outer ring is almost certainly caused by the interaction of accreted gas with a bar \citep{Kra+05,Jeo+07}.  Some galaxies in the \sau\ sample indeed show evidence of on-going gas accretion, in one of which (NGC~1023) the gas seems to be settling into co-rotation with the stars \citep{Mor+06}; nevertheless, it remains unclear what the ultimate fate of this gas will be.  With the available data, we therefore cannot rule out an external origin for the star-forming gas in at least some of the \classtwo\ objects.

Although definite statements cannot be made about the origins of the gas in \classtwo\ galaxies from the data presented here, their star formation can clearly be identified as a rejuvenation inside previously quiescent stellar systems.  This is evident in the combination of the regular, confined star-forming structures with the presence of uniformly old stellar populations (SSP age $\gtrsim$10 Gyr; Kuntschner et al.~{\it in prep}) outside of the star-forming regions in these galaxies.  If the gas causing this rejuvenation has internal origins, the similar stellar mass loss rates in all galaxies imply that this mode of star formation may be common to the majority of fast rotators.  Likewise, if the gas comes from wet minor mergers, it is expected to have been replenished at least once in the past 10 Gyr in these massive galaxies \citep{FakMa08}.  In either case, all fast rotating galaxies would be equally eligible for such star formation events.

In this scenario, the only difference between those fast rotating galaxies with on-going circumnuclear star formation and those without is the amount of gas in the system.  In the Toomre criterion for the stability of a system against star formation, the stability $Q$ depends on the epicycle frequency, the sound speed of the gas, and the surface density of the gas.  For the \sau\ galaxies, we have explicitly tested that the epicycle frequency radial profiles show no systematic differences between galaxies with and without circumnuclear star formation.  Consequently, gas of similar surface densities embedded in all of the fast rotators would result in similar $Q$ values and thus similar star formation rates and morphologies.

The above results hint that this mode of star formation may be a transient and recurrent phenomenon in all fast rotators; it is therefore possible to estimate the duty cycle of this process.  Since the \sau\ sample is roughly complete for local early-type galaxies with stellar masses exceeding $\sim$~10$^{11}$~\msun, we restrict our estimate to this regime.  One approximation of the duty cycle is simply the occurrence rate of on-going star formation in a central disk morphology; here, we observe this to be 2/9 for fast-rotating galaxies above our mass threshold (excluding the 3 \classtwo\ galaxies that are below this mass threshold and the 2 \classtwo\ galaxies that have only suggestive detections of on-going star formation).  Another approximation comes from the lifetime of the star formation event relative to that of the system; the mean SSP age of $\sim$10 Gyr of massive galaxies and the gas exhaustion timescales in these galaxies of $\sim$1 Gyr (SFR $\sim$ several tenths of \msun\ \peryr\ and molecular reservoirs of several $\times$ 10$^8$ \msun) yield a duty cycle of 1/10.

In the $\sim$1/10 fast rotators in which the necessary gas (with sufficiently low mass fraction) is present, the star formation typically occurs in a well-defined central disk structure.  This structure suggests a sharp radial limit exists in these galaxies.  One natural explanation is via bar resonances, as are responsible for the large-scale ring galaxy NGC~2974 \citep{Jeo+07}, for the gas inflow in the strongly barred galaxy NGC~4477, and for the circumnuclear disks seen in the bulges of late-type galaxies \citep{KorKen04}.  The role of bars in these systems is particular apparent in NGC~4477, in which the bar is driving a spiral structure in the gas and is funneling it towards the inner Lindblad resonance (ILR).  Together with the lack of strong signatures of star formation (in SSP age or \oiii/\hb) in this galaxy, these observations suggest that this galaxy may be in an initial stage of circumnuclear star formation.  The weaker bars detected in some of the other \classtwo\ galaxies may have already driven the majority of the gas into the ILR and now function to maintain the observed radial limit.  However, bars are not the only possible explanation for the radial limit of circumnuclear star formation events.  An alternative explanation may be a balance between the background radiation within the galaxy and the cooling time, which decreases rapidly in the central regions, perhaps resulting in a sharply defined region in which the gas can cool and form stars.

When this star formation ceases, the descendant galaxies will likely not be readily identifiable via gradually increasing SSP ages, unlike as in the case of \classone\ galaxies.  This is due to the difficulty inherent in detecting a low mass ratio, intermediate age population on top of a prominent background of stars with SSP ages $\gtrsim$10 Gyr; even in the \classtwo\ systems with on-going star formation, the measured SSP ages are quite high (several Gyr to 10 Gyr) because of this effect.  However, residuals from this process may in fact be visible via the resulting increased metallicity in the younger population; indeed, in \pvia, more than half of the fast rotators are found to contain ``\mgb\ disks," flattened, higher metallicity structures with similar spatial scales to that seen in PAH emission in \classtwo\ galaxies (see also Kuntschner et al.~{\it in prep}).  Moreover, the stellar disks that ought to be created from secondary star formation events are also observed to be ubiquitous in fast rotators; \pxiit\ notes that the presence of these embedded, flattened and rapidly rotating structures is in fact what defines fast rotating early-type galaxies as a population.  This mode of recurrent and transient star formation may contribute to these components in all fast rotators.

The picture emerging here can thus be summarized as follows: A quiescent, fast rotating early-type galaxy acquires a new gas reservoir, either through internal stellar mass loss or through accretion of external gas.  The gas settles into co-rotating orbits in the equitorial plane and forms stars, making the galaxy observable as a \classtwo\ system.  When the star formation ceases, over the course of $\sim$1 Gyr, the galaxy returns from being transiently star-forming back to passively evolving.  The rekindled star formation from this event has replenished the galaxy's stellar disk, adding a younger, thinner, faster rotating, and more metal-rich (though low mass fraction) component to this structure.  In this manner, the kinematically distinct stellar disks in fast rotators are maintained.

\subsection{Slow Rotators}
\label{discussSR}

Much of the discussion so far in this paper has focused on the evolution of fast rotator galaxies into and out of quiescent states.  However, the \sau\ sample of local early-type galaxies also contains a number of slow rotators, whose properties differ dramatically from those of the fast rotators.  Of particular relevance to this discussion is the complete absence of star formation, as traced by PAH emission, in these systems.  Instead, slow rotators can have significant X-ray halos of hot gas, unlike their rapidly rotating kin \citep{Sar+07,Kor+09}.  These halos naturally explain the lack of star formation activity in the slow rotators, in which the ISM is simply too hot for gas to cool and condense.

The hot halos in slow rotators may be maintained by radio AGN, which are detected in some of the slow rotators both in the radio \pxvip\ and in 8.0\um\ emission.  \citet{Cro+06} have demonstrated that the presence of such AGN, operating in only the highest mass halos, is effective at replicating observed properties of red sequence galaxies in simulations.  However, other mass-dependent ``maintenance modes" not relying on AGN activity \citep[e.g.][]{DekBir06,Naa+07} can also reproduce the characteristics of this population and maintain hot halos.

Whatever the cause, it is likely that the slow rotators in the \sau\ sample have been quiescent for much of their histories.  These galaxies do not host on-going star formation, and their SSP ages are uniformly $\gtrsim$10 Gyr (Kuntschner et al.~{\it in prep}).  This is consistent with the cosmological (major and minor) mergers thought to be necessary to the formation of these objects \citep{Naa+06,Bur+08}.  This galaxy population is therefore distinct in that it does not appear to participate in the continuing dissipational evolution observed here in the fast rotators.


\section{Conclusions}
\label{Conclu}

Using the synthesis of \sau\ integral field spectroscopy and \spit\ mid-IR imaging, we have probed star formation processes in a representative sample of early-type galaxies.  We find compelling evidence for star formation in 8 of \nspit\ early-type galaxies and potential evidence for star formation in an additional 5 of \nspit\ systems.  However, since the \sau\ sample was designed to cover parameter space rather than to statistically sample it, the observed frequency of star formation in early-type galaxies should be used with caution, especially in the lower mass range where the \sau\ sample is statistically incomplete.

Nevertheless, the representative nature of the \sau\ sample does suggest that these results can be used to understand the properties of early-type galaxies as a population.  With this sample and the associated auxiliary data, we find that star formation proceeds with the same surface densities and efficiencies as in spiral galaxies and circumnuclear starbursts.  This suggests that the low-level star formation observed in early-type galaxies is essentially a scaled down version of similar processes in more vigorously star-forming galaxies.  This analogy is also valid kinematically; we observe star formation to only be present in fast rotating early-type galaxies and to occur in flattened disk or ring morphologies.

Moreover, we find that star formation in these galaxies happens in one of two regimes: in the first, star formation is widespread and has no obvious link to stellar and gas kinematic substructures, while in the second, star formation is a spatially distinct process, often circumnuclear, whose extent corresponds well to those of the observed ionized gas, molecular gas, dust, and embedded stellar disks.  Analysis of the galaxies in each regime and their relation to the early-type population as a whole indicates that these two classes in fact represent very different stages in the evolution of red sequence galaxies.

In \classone\ galaxies, the star formation events have relatively high mass fractions, perhaps due to the lower masses of these galaxies, causing them to dominate the galaxies' stellar light and kinematics.  Together with the equal distribution of co- and counter-rotating kinematic structures in these systems that strongly indicates external origins for the gas, this suggests that \classone\ galaxies have recently undergone gas-rich minor mergers.  Simulations show that such minor mergers are sufficient to transform star-forming spiral galaxies into red lenticular galaxies, and we suggest that the \classone\ galaxies are either in the final stages of such a transformation or represent a reinvigoration of star formation in previously quiescent lenticular galaxies.  In either case, the eventual cessation of star formation in these galaxies is observed to occur in an ``outside-in" manner as the molecular gas is heated or consumed and the galaxies become quiescent.

In contrast, \classtwo\ galaxies are characterized by star formation with relatively low mass fractions, perhaps due to the higher masses of these galaxies.  These galaxies are experiencing an epoch of renewed star formation, as evident in the evolved stellar populations outside the star-forming regions, from gas that was acquired either through internal stellar mass loss or very minor mergers.  It is likely that this process, which has a duty cycle of $\sim$1/10 Gyr and can occur equally in all fast rotators, is important in reinforcing the embedded stellar disks that are found in this galaxy population.  The \classtwo\ star formation may therefore be a transient star formation event during the lives of all fast rotators, as they temporarily experience renewed star formation, resulting in additional thin, rapidly rotating, and metal-rich stellar disks.

A complete picture of early-type galaxies also includes the slow rotator population, which shows no evidence for on-going or recent star formation in our data.  In these systems, it is likely that a mass-dependent quenching mechanism, active only in these most massive galaxies, is continuously heating the interstellar media and preventing cooling and star formation.  Consequently, the stars in these galaxies are uniformly old.

The scenarios of galaxy formation and evolution presented here will naturally need to be confirmed and refined.  Future investigations with statistical samples will be crucial in probing the true nature of the two regimes of star formation proposed here, both of which are observed in only a handful of galaxies in our sample.  Another avenue with great promise is augmenting the multi-wavelength coverage of the \sau\ sample to include high resolution \hone\ observations, in order to relate star formation processes to the morphology and kinematics of atomic gas reservoirs in these galaxies and thus further constrain the gas origins in both \classone\ and \classtwo\ galaxies.  Finally, extending the SSP modeling to allow multiple stellar populations can provide greater insight into the star formation histories of early-type galaxies.  The results from these and similar studies will doubtlessly continue to improve our understanding of the formation and evolution of elliptical and lenticular galaxies.


\section*{Acknowledgments}

We would like to thank Fran\c{c}oise Combes, Phil Hopkins, and Jacqueline van Gorkom for interesting discussions, and Martin Bureau, Alison Crocker, Nicholas Scott, and Lisa Young for providing access to their data pre-publication and for useful comments and suggestions.  We are very grateful to the referee for comments that improved the quality of this manuscript.  KLS and JFB also acknowledge the repeated hospitality of the Institute for Advanced Study, to which collaborative visits contributed greatly to the quality of this work.

The \sau\ observations were obtained at the William Herschel Telescope, operated by the Isaac Newton Group in the Spanish Observatorio del Roque de los Muchachos of the Instituto de Astrof\'isica de Canarias.  This project is made possible through grants 614.13.003, 781.74.203, 614.000.301, and 614.031.015 from NWO and financial contributions from the Institut National des Sciences de l'Univers, the Universit\'e Lyon I, the Univerisities of Durham, Leiden, and Oxford, the Programme National Galaxies, the British Council, PPARC grant `Observational Astrophysics at Oxford 2002-2006,' support from Christ Church Oxford, and the Netherlands Research School for Astronomy NOVA.  KLS acknowledges support from Spitzer Research Award 1359449.  GvdV acknowledges support provided by NASA through Hubble Fellowship grant HST-HF-01202.01-A awarded by the Space Telescope Science Institute, which is operated by the Association of Universities for Research in Astronomy, Inc., for NASA, under contract NAS 5-26555.  MC acknowledges support from a STFC Advanced Fellowship (PP/D005574/1).  RLD is grateful for the award of a PPARC Senior Fellowship (PPA/Y/S/1999/00854) and postdoctoral support through PPARC grant PPA/G/S/2000/00729.

        
\bibliography{biblio}
 
\bsp

\clearpage

\clearpage
\begin{figure*}
	\centering
	\includegraphics[width=17cm]{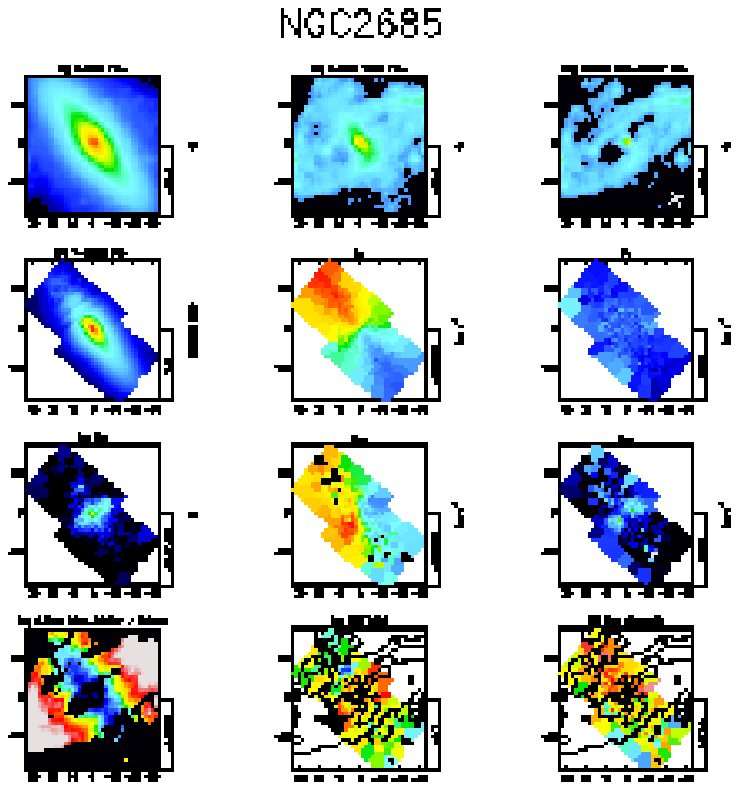}
	\begin{center}
{\bf Figure~\ref{Main-SF}.} {\it continued}
	\end{center}
\end{figure*}
\clearpage
\begin{figure*}
	\centering
	\includegraphics[width=17cm]{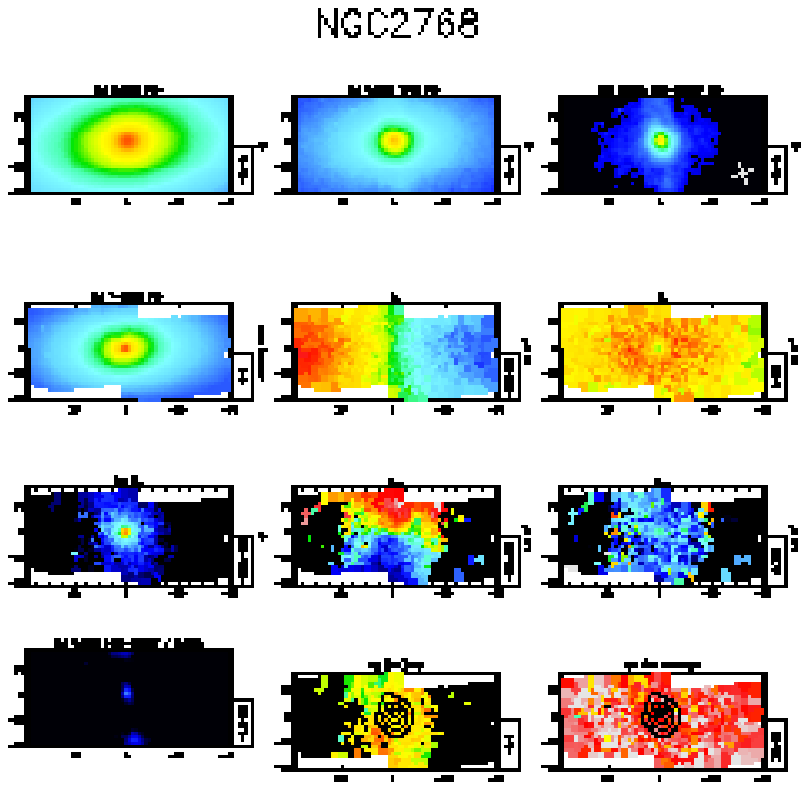}
	\begin{center}
{\bf Figure~\ref{Main-SF}.} {\it continued}
	\end{center}
\end{figure*}
\clearpage
\begin{figure*}
	\centering
	\includegraphics[width=17cm]{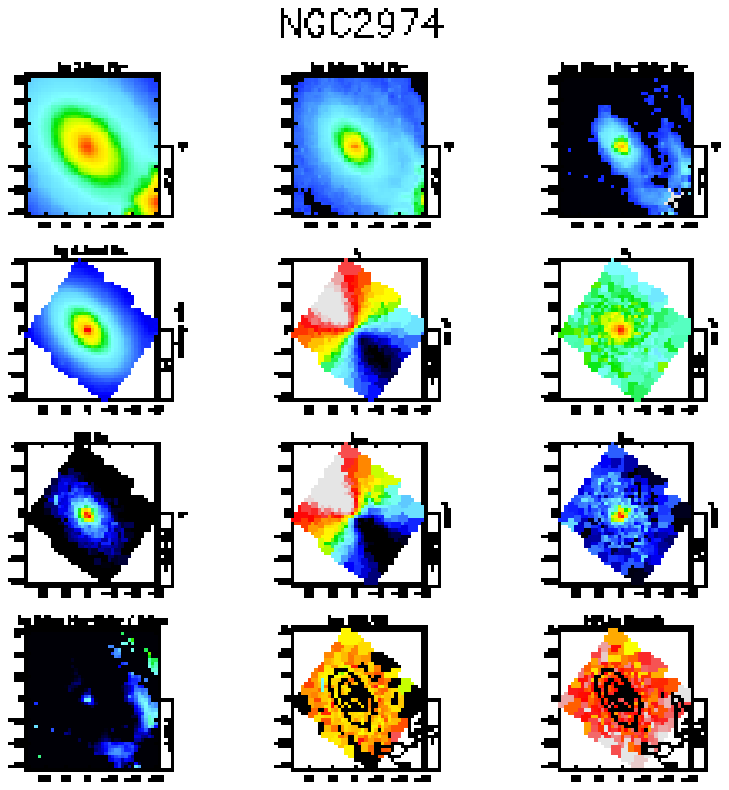}
	\begin{center}
{\bf Figure~\ref{Main-SF}.} {\it continued}  A foreground star is visible in the southwest corner of the \spit\ data.
	\end{center}
\end{figure*}
\clearpage
\begin{figure*}
	\centering
	\includegraphics[width=17cm]{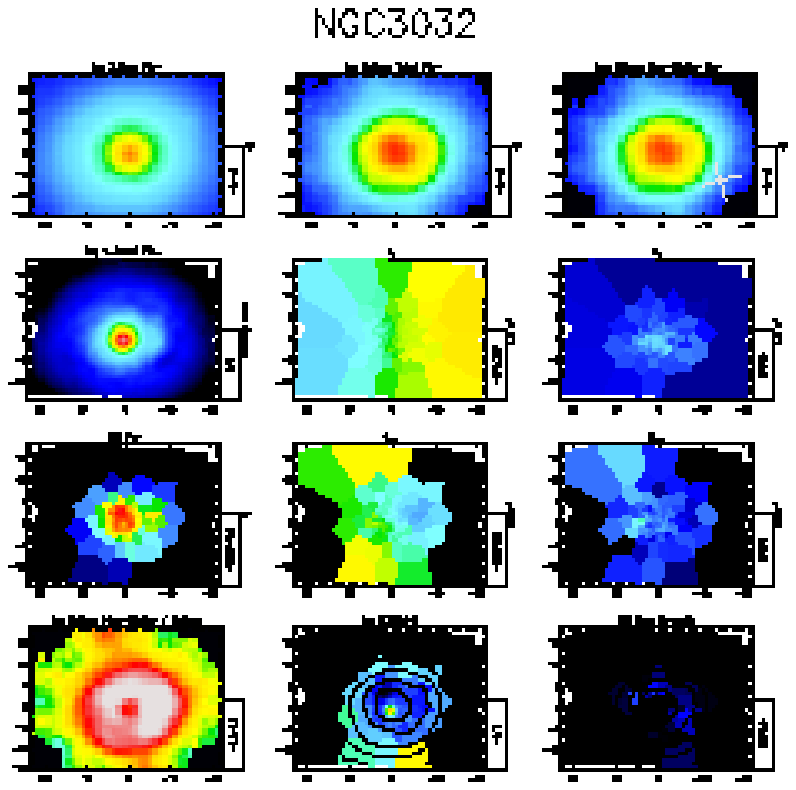}
	\begin{center}
{\bf Figure~\ref{Main-SF}.} {\it continued.}
	\end{center}
\end{figure*}
\clearpage
\begin{figure*}
	\centering
	\includegraphics[width=17cm]{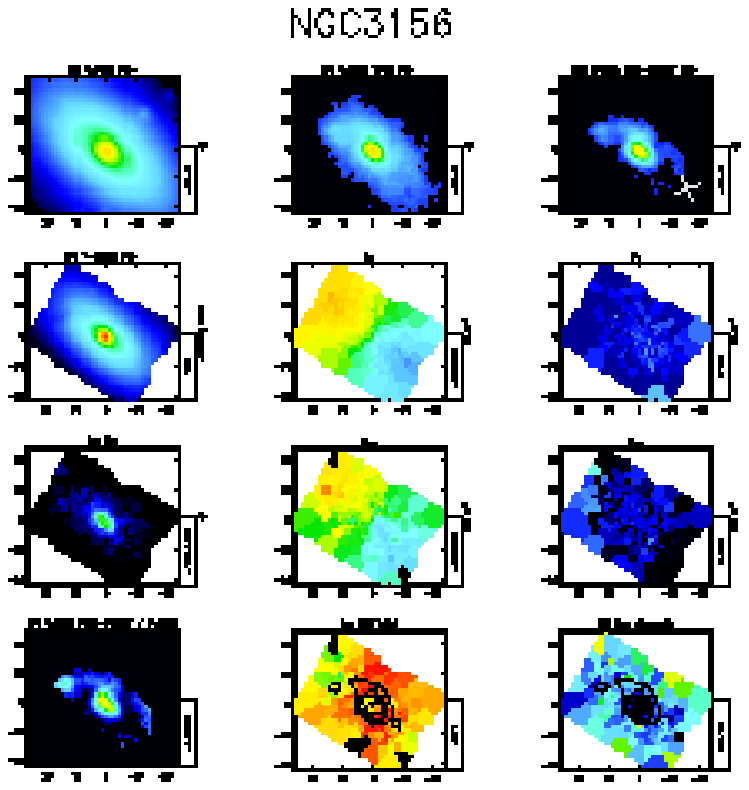}
	\begin{center}
{\bf Figure~\ref{Main-SF}.} {\it continued}
	\end{center}
\end{figure*}
\clearpage
\begin{figure*}
	\centering
	\includegraphics[width=17cm]{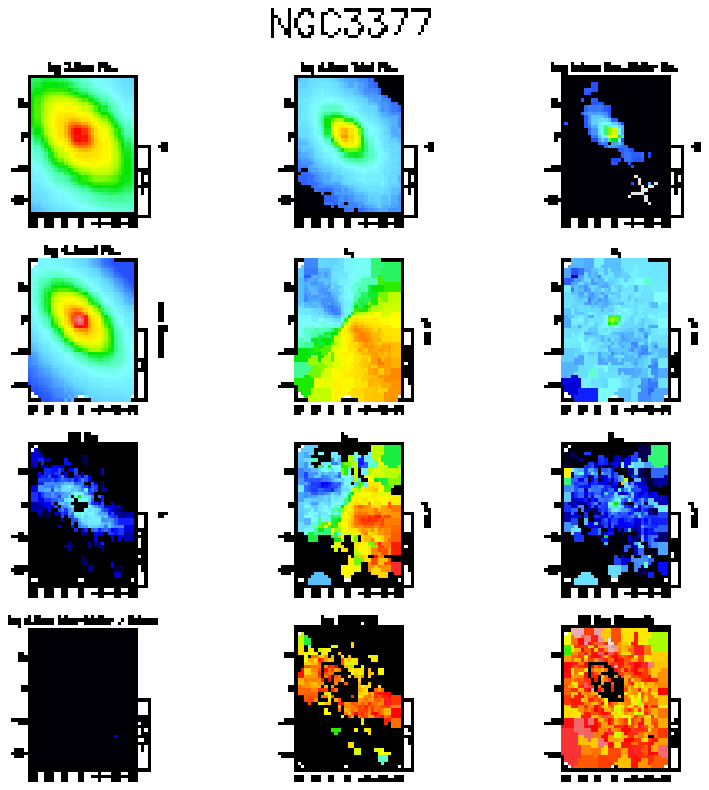}
	\begin{center}
{\bf Figure~\ref{Main-SF}.} {\it continued}
	\end{center}
\end{figure*}
\clearpage
\begin{figure*}
	\centering
	\includegraphics[width=17cm]{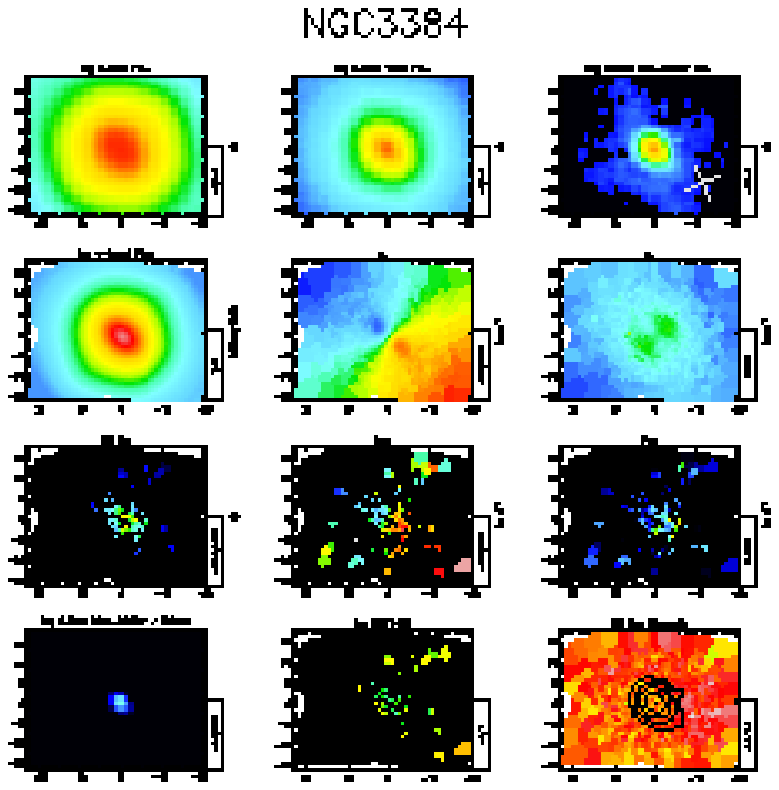}	
	\begin{center}
{\bf Figure~\ref{Main-SF}.} {\it continued}
	\end{center}
\end{figure*}
\clearpage
\begin{figure*}
	\centering
	\includegraphics[width=17cm]{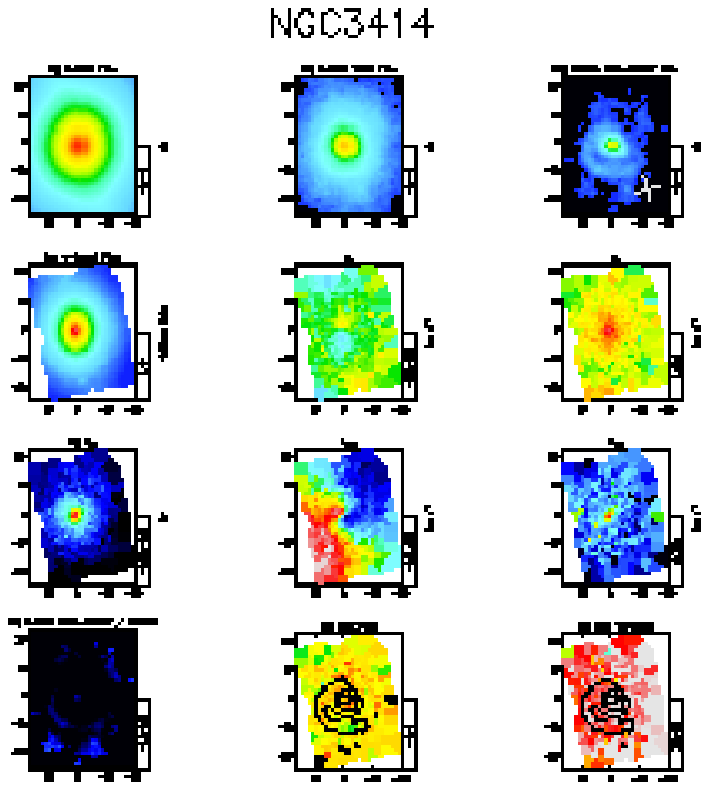}
	\begin{center}
{\bf Figure~\ref{Main-SF}.} {\it continued}
	\end{center}
\end{figure*}
\clearpage
\begin{figure*}
	\centering
	\includegraphics[width=17cm]{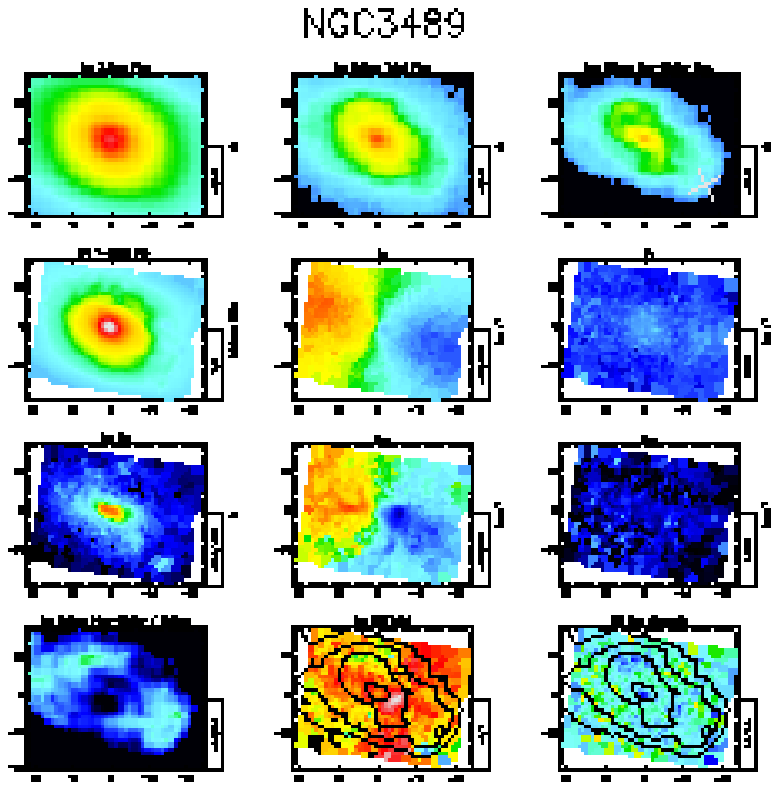}
	\begin{center}
{\bf Figure~\ref{Main-SF}.} {\it continued}
	\end{center}
\end{figure*}
\clearpage
\begin{figure*}
	\centering
	\includegraphics[width=17cm]{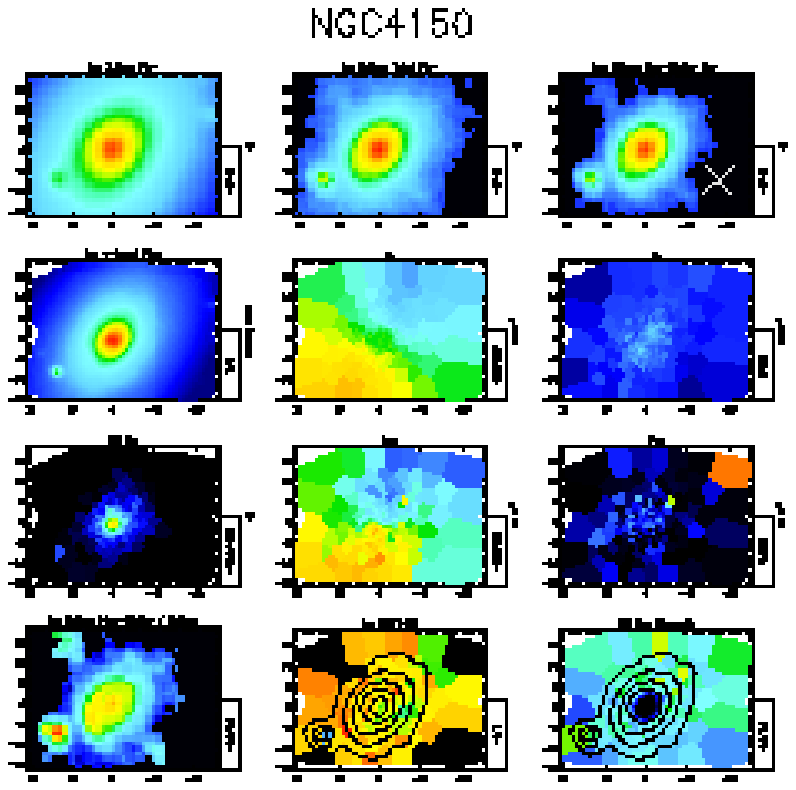}
	\begin{center}
{\bf Figure~\ref{Main-SF}.} {\it continued}  A quasar at $z$=0.52 \citep[see][]{LirLawJoh00} is visible to the southeast of the galaxy.
	\end{center}
\end{figure*}
\clearpage

\begin{figure*}
	\centering
	\includegraphics[width=17cm]{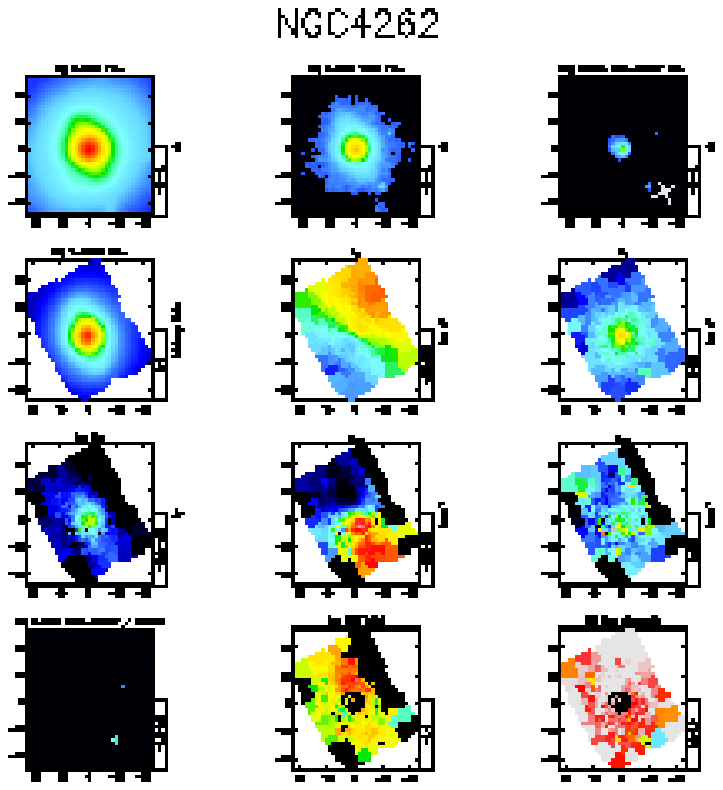}
	\begin{center}
{\bf Figure~\ref{Main-SF}.} {\it continued.}
	\end{center}
\end{figure*}
\clearpage
\begin{figure*}
	\centering
	\includegraphics[width=17cm]{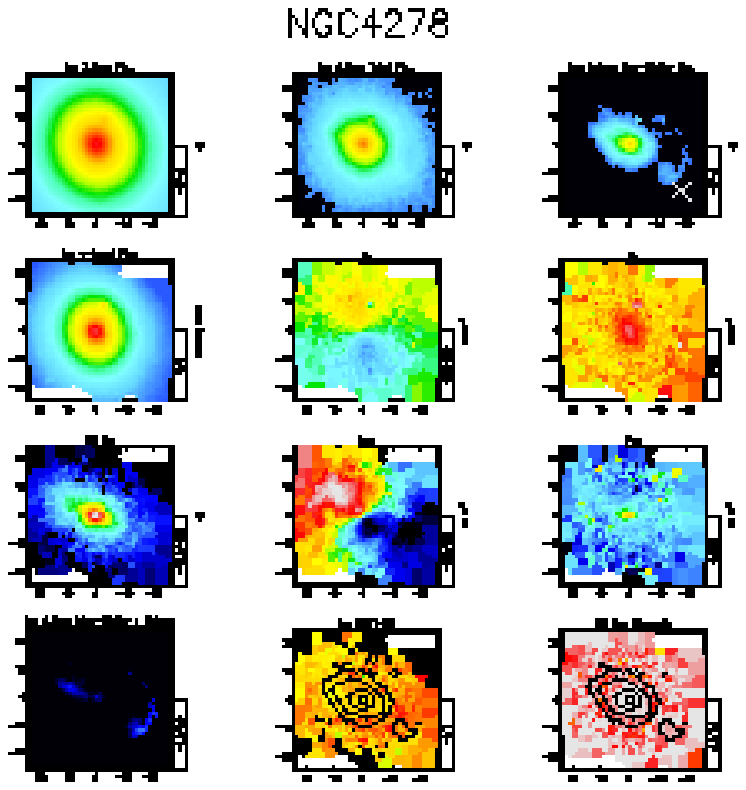}
	\begin{center}
{\bf Figure~\ref{Main-SF}.} {\it continued}
	\end{center}
\end{figure*}
\clearpage
\begin{figure*}
	\centering
	\includegraphics[width=17cm]{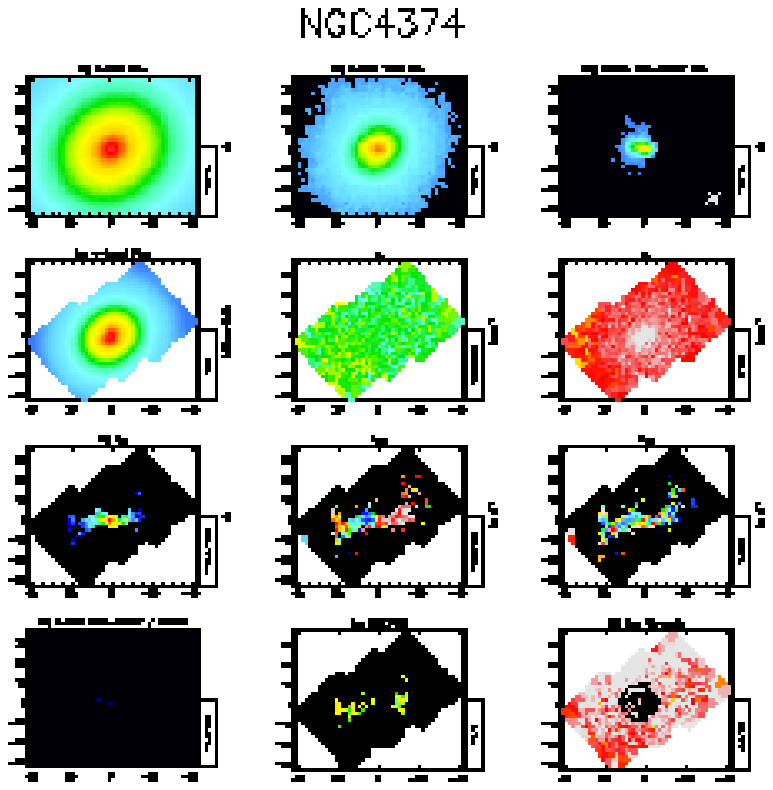}	
	\begin{center}
{\bf Figure~\ref{Main-SF}.} {\it continued}
	\end{center}
\end{figure*}
\clearpage
\begin{figure*}
	\centering
	\includegraphics[width=17cm]{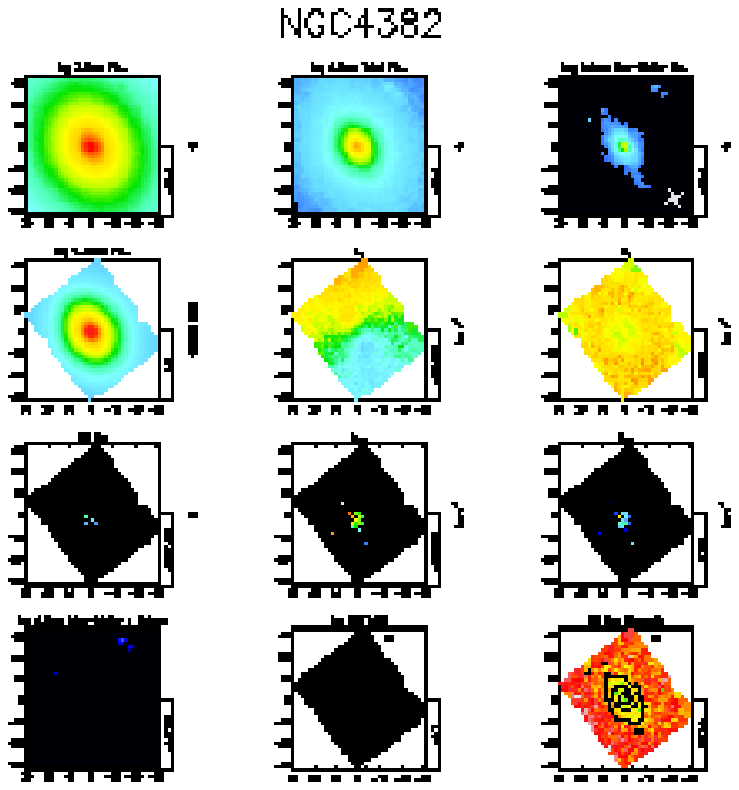}
	\begin{center}
{\bf Figure~\ref{Main-SF}.} {\it continued}
	\end{center}
\end{figure*}
\clearpage
\begin{figure*}
	\centering
	\includegraphics[width=17cm]{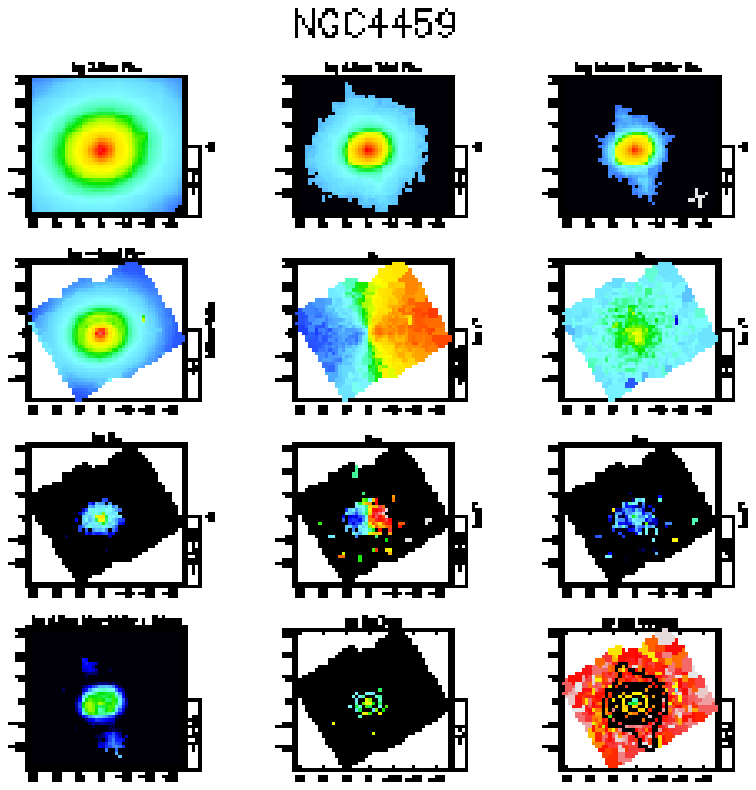}
	\begin{center}
{\bf Figure~\ref{Main-SF}.} {\it continued}
	\end{center}
\end{figure*}
\clearpage
\begin{figure*}
	\centering
	\includegraphics[width=17cm]{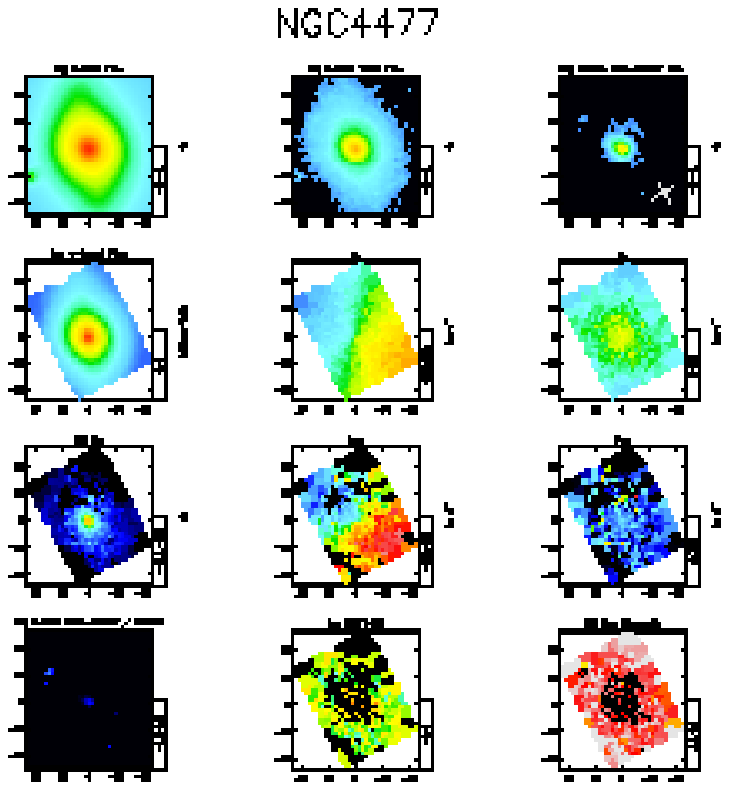}
	\begin{center}
{\bf Figure~\ref{Main-SF}.} {\it continued}
	\end{center}
\end{figure*}
\clearpage
\begin{figure*}
	\centering
	\includegraphics[width=17cm]{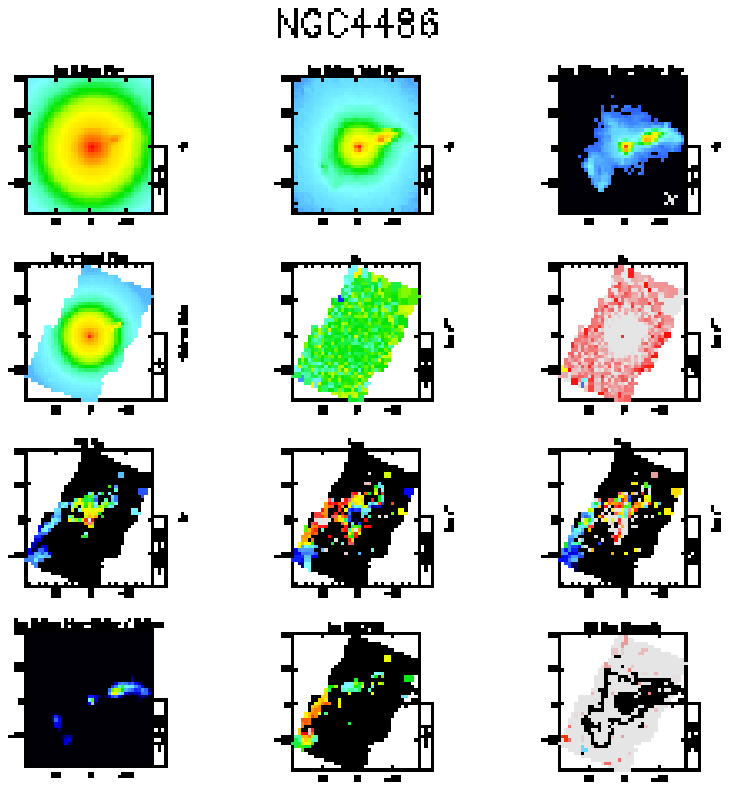}
	\begin{center}
{\bf Figure~\ref{Main-SF}.} {\it continued}
	\end{center}
\end{figure*}
\clearpage
\begin{figure*}
	\centering
	\includegraphics[width=17cm]{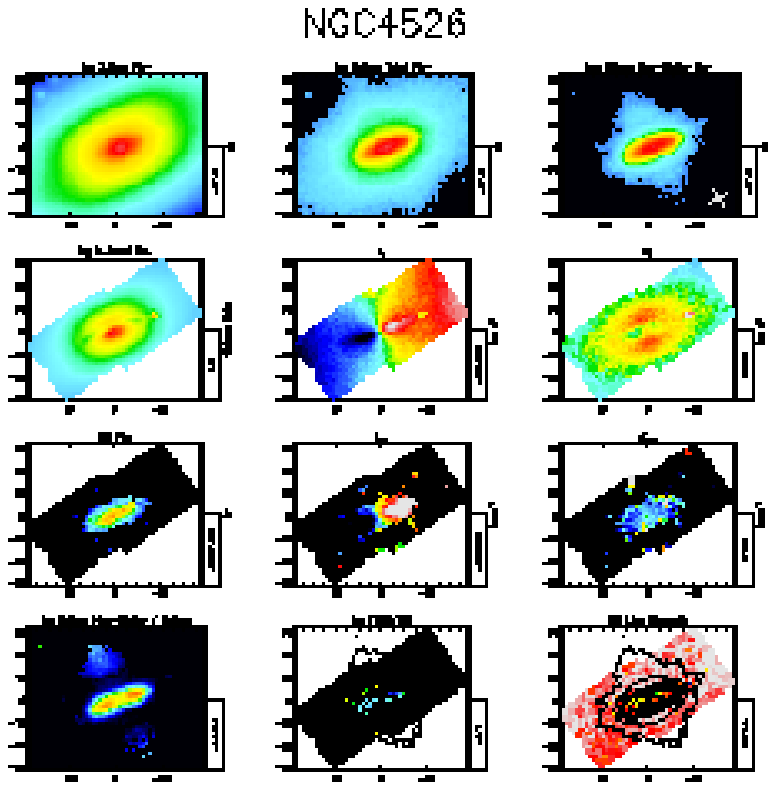}
	\begin{center}
{\bf Figure~\ref{Main-SF}.} {\it continued}
	\end{center}
\end{figure*}
\clearpage
\begin{figure*}
	\centering
	\includegraphics[width=17cm]{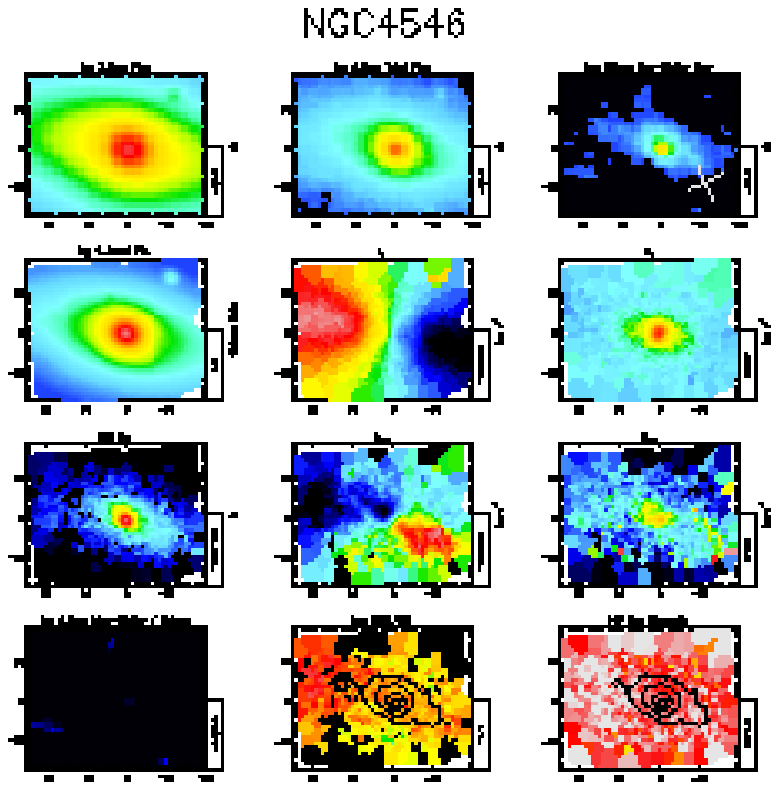}
	\begin{center}
{\bf Figure~\ref{Main-SF}.} {\it continued}
	\end{center}
\end{figure*}
\clearpage
\begin{figure*}
	\centering
	\includegraphics[width=17cm]{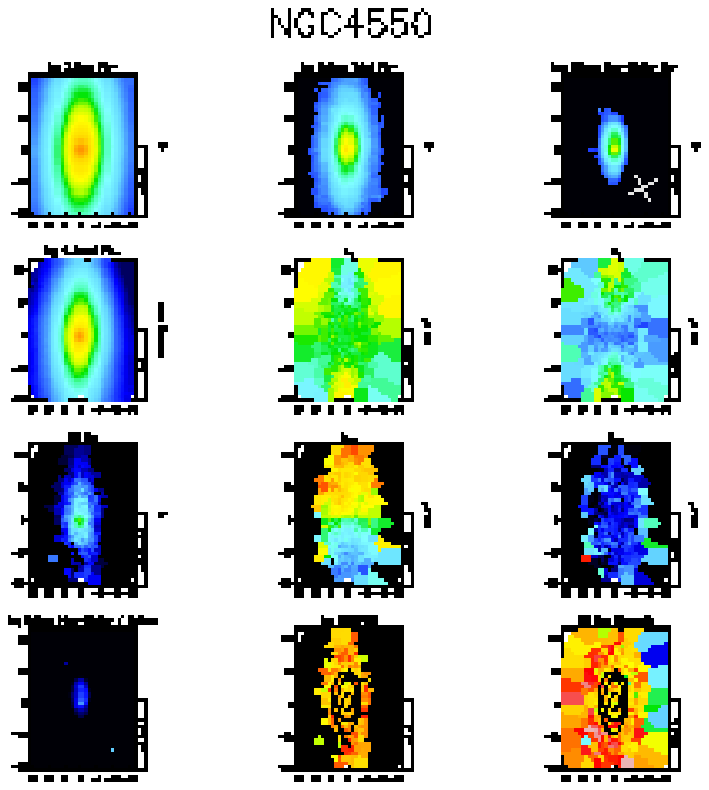}
	\begin{center}
{\bf Figure~\ref{Main-SF}.} {\it continued}
	\end{center}
\end{figure*}
\clearpage
\begin{figure*}
	\centering
	\includegraphics[width=17cm]{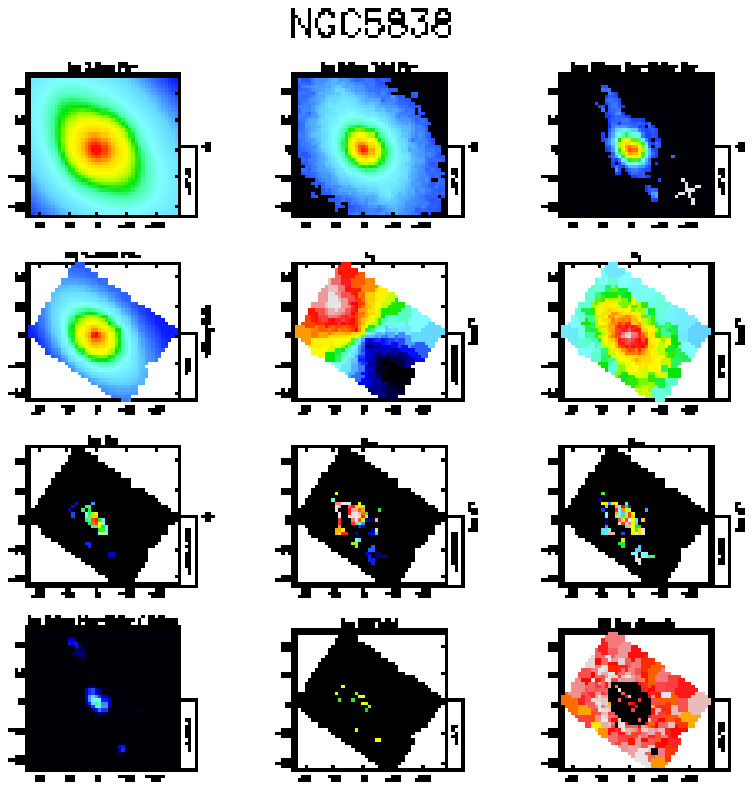}
	\begin{center}
{\bf Figure~\ref{Main-SF}.} {\it continued}
	\end{center}
\end{figure*}
\clearpage
\begin{figure*}
	\centering
	\includegraphics[width=17cm]{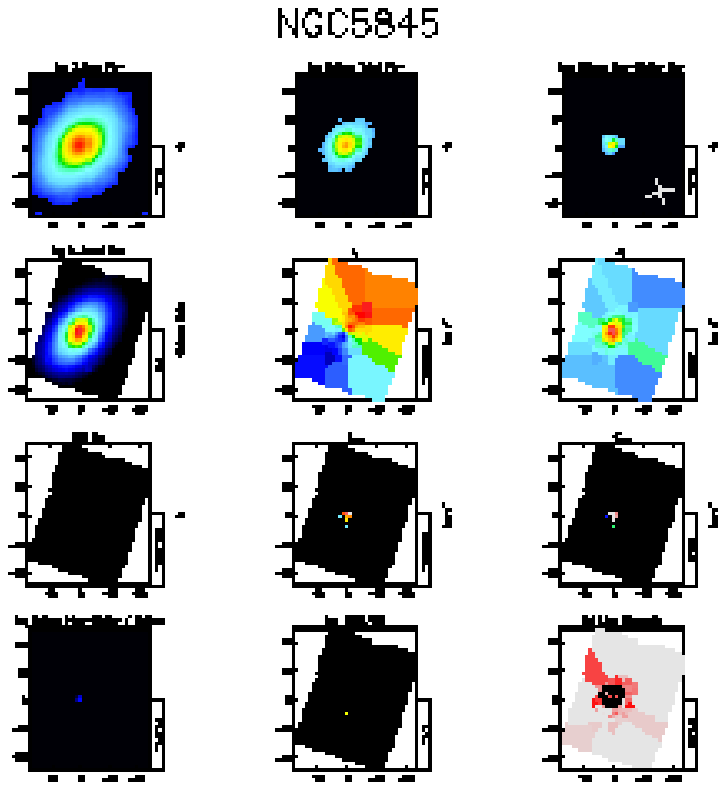}
	\begin{center}
{\bf Figure~\ref{Main-SF}.} {\it continued}
	\end{center}
\end{figure*}
\clearpage
\begin{figure*}
	\centering
	\includegraphics[width=17cm]{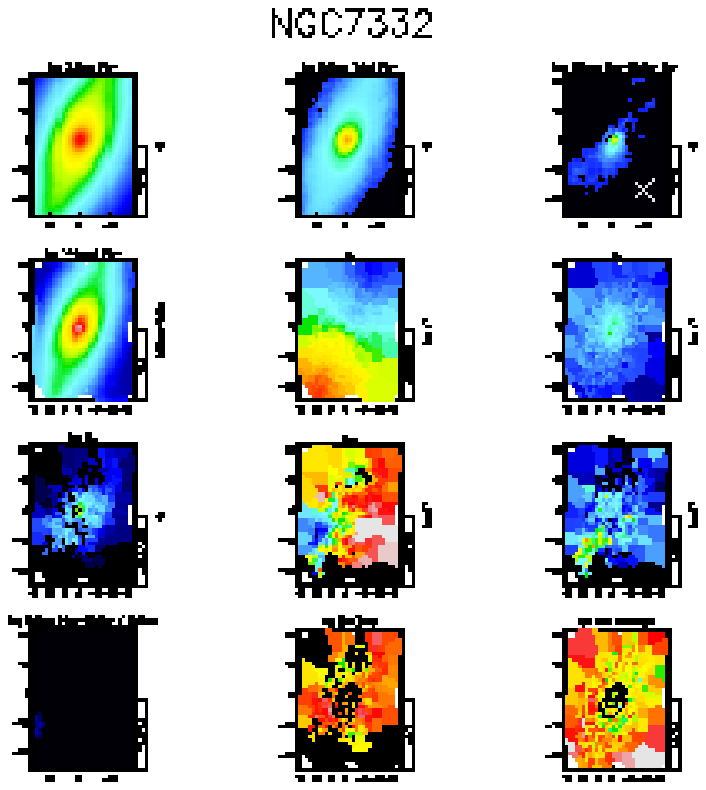}
	\begin{center}
{\bf Figure~\ref{Main-SF}.} {\it continued}
	\end{center}
\end{figure*}
\clearpage
\begin{figure*}
	\centering
	\includegraphics[width=17cm]{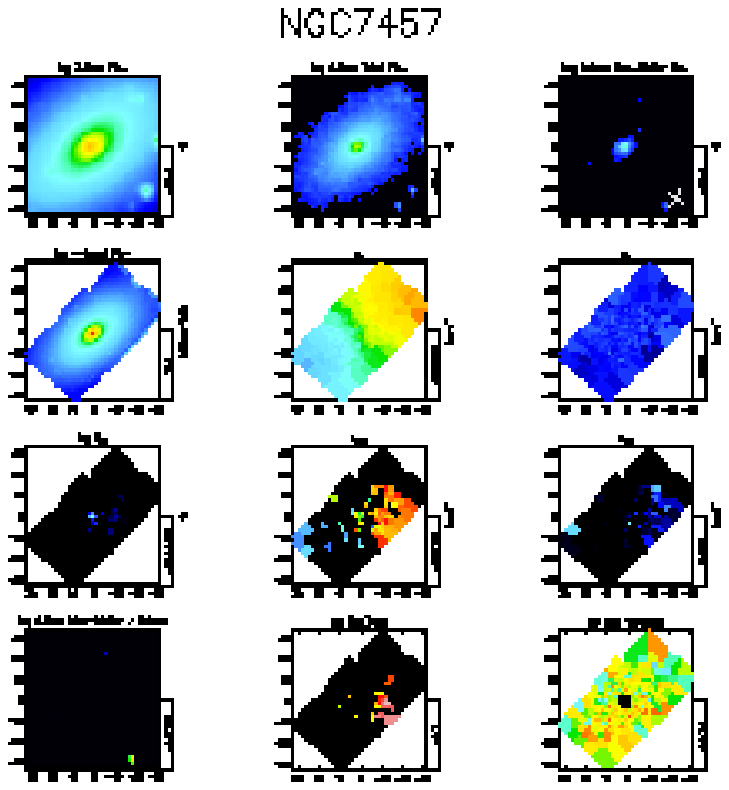}
	\begin{center}
{\bf Figure~\ref{Main-SF}.} {\it continued}
	\end{center}
\end{figure*}
\clearpage


\label{lastpage}

\end{document}